\newif\ifsingle
\singletrue % comment out for single column version

\newif\ifFullVersion
\FullVersiontrue % comment out for shortened version

\ifsingle
\documentclass[12pt,draftclsnofoot, onecolumn]{IEEEtran}		
\else		
\documentclass[10pt,final, twocolumn]{IEEEtran}
\usepackage[a4paper, total={7.3in, 10.4in}]{geometry}
\fi

\usepackage{times}
\usepackage{amsmath,dsfont}
\usepackage{amssymb,amsthm}
\usepackage{epsfig,verbatim}
\usepackage{subfigure}
\usepackage{setspace}
\usepackage{color}
\usepackage{cite}
\usepackage{epstopdf}
\usepackage{graphics}
\usepackage{accents}
\usepackage{acronym}
\usepackage{enumitem}
\usepackage[bookmarks,colorlinks]{hyperref}

\newtheorem{definition}{Definition}
\newtheorem{theorem}{Theorem}
\newtheorem*{theorem*}{Theorem}
\newtheorem{corollary}{Corollary}
\newtheorem{proposition}{Proposition}
\newtheorem{lemma}{Lemma}

\newcommand{\E}{\mathds{E}}

\newcommand{\myVec}[1]{{\boldsymbol{#1}}}
\newcommand{\myMat}[1]{\mathsf{#1}}
\newcommand{\mySet}[1]{\mathcal{#1}}
\newcommand{\Cmat}{\mathsf{C}}

\newcommand{\Ind}{{\bf 1}}

\newcommand{\Rs}{R_c}
\newcommand{\Xin}{{\myVec{X}}}
\newcommand{\xin}{{x}}
\newcommand{\Yi}{{\myVec{Y}}}

\newcommand{\Wi}{{\myVec{W}}}
\newcommand{\wi}{{\myVec{w}}}

\newcommand{\PCst}{P}
\newcommand{\opt}{^{\rm opt}}

\newcommand{\zkn}[2]{\tilde{Z}_{#1,#2}}
\newcommand{\zkeps}[1]{Z_{#1}}
\newcommand{\zk}[1]{Z_{#1}}
\newcommand{\zetan}[1]{\zeta_{#1}}

\newcommand{\Fkeps}[1]{F_{#1}}
\newcommand{\Fkn}[2]{\tilde{F}_{#1,#2}}

\newcommand{\dsM}{\mathds{M}}

\newcommand{\mN}{\mathcal{N}}
\newcommand{\mK}{\mathcal{K}}

\newcommand{\mR}{\mathcal{R}}

\newcommand{\xvec}{\mathbf{x}}

\newcommand{\yvec}{\mathbf{y}}
\newcommand{\wvec}{\mathbf{w}}

\newcommand{\talpha}{\tilde{\alpha}}
\newcommand{\tbeta}{\tilde{\beta}}

\newcommand{\Det}{{ \rm Det}}
\newcommand{\myEps}{\epsilon}
\newcommand{\eps}{\epsilon}

\newcommand{\Wc}{W_{\rm c}}
\newcommand{\Weps}{W_{\myEps}}
\newcommand{\Wepsvec}{\myVec{W}_{\myEps}}
\newcommand{\Yeps}{Y_{\myEps}}
\newcommand{\Yepsvec}{\myVec{Y}_{\myEps}}
\newcommand{\Vin}{V_n}
\newcommand{\Veps}{V_{\myEps}}
\newcommand{\Charac}{\Phi}
\newcommand{\Xscal}{X}

\newcommand{\Wn}{W_{n}}

\newcommand{\Yn}{Y_{n}}
\newcommand{\Ynvec}{{\bf Y}_{n}}

\newcommand{\Cwc}{\sigma^2_{\Wc}}
\newcommand{\Cweps}{c_{\Weps}}
\newcommand{\Sigweps}{\sigma^2_{\Weps}}

\newcommand{\Cwn}{c_{\Wn}}
\newcommand{\Sigwn}{\sigma^2_{\Wn}}
\newcommand{\Tc}{T_{\rm pw}}
\newcommand{\Tsamp}{T_{\rm s}}
\newcommand{\Td}{p}

\newcommand{\cdf}[1]{F_{#1}}
\newcommand{\pdf}[1]{p_{#1}}

\newcommand{\TDelta}{{\bar{\Delta}}}

\newcommand{\Capacity}{C}
\newcommand{\fkn}{f_{k,n}}
\newcommand{\fkeps}{f_{k,\myEps}}

\newcommand{\Xnvec}[1][n]{{\myVec{X}}_{#1}}
\newcommand{\corrmate}{\Cmat_{\myEps}}

\newcommand{\corrmatn}{\Cmat_{n}}
\newcommand{\dsE}{\mathds{E}}

\newcommand{\DC}{t_{\rm dc}}
\newcommand{\Trise}{t_{\rm rf}}

\newcommand{\ConvDist}[1]{\mathop{\longrightarrow}\limits^{(dist.)}_{#1}}

\acrodef{cdf}[CDF]{cumulative distribution function}
\acrodef{pdf}[PDF]{probability density function}
\acrodef{rv}[RV]{random variable}
\acrodef{dt}[DT]{discrete-time} 
\acrodef{ct}[CT]{continuous-time} 
\acrodef{wscs}[WSCS]{wide-sense cyclostationary}
\acrodef{wsacs}[WSACS]{wide-sense almost cyclostationary}
\acrodef{ofdm}[OFDM]{orthogonal frequency division multiplexing}
\acrodef{tdma}[TDMA]{time division multiple access}
\acrodef{cdma}[CDMA]{code division multiple access}
\acrodef{noma}[NOMA]{non-orthogonal multiple access}
\acrodef{asmcgn}[AS-MCGNC]{asynchronously-sampled memoryless cyclostationary Gaussian noise channel}
\acrodef{cmt}[CMT]{continuous mapping theorem}

\definecolor{NewColor}{rgb}{0,0,0}

\newcommand{\myFtn}[1]{
	\ifFullVersion
	\footnote{#1}
	\fi	
}

\newif\ifcomments
\definecolor{CmtColor}{rgb}{0,0.6,1}

\long\def\symbolfootnote[#1]#2{\begingroup\def\thefootnote{\fnsymbol{footnote}}\footnote[#1]{#2}\endgroup}

\IEEEoverridecommandlockouts
\title{The Capacity of Memoryless Channels with Sampled Cyclostationary Gaussian Noise
}

\vspace{-0.5cm}

\author{
	\IEEEauthorblockN{\vspace{-0.2cm} Nir Shlezinger, Emeka Abakasanga, Ron Dabora, and Yonina C. Eldar \\
	}

	\thanks{This paper was presented in part at the 2019 IEEE International Symposium on Information Theory. }
	\thanks{N. Shlezinger  and Y. C. Eldar are with the faculty of Math and CS, Weizmann Institute of Science,  Israel (e-mail: nirshlezinger1@gmail.com; yonina@weizmann.ac.il).  
		E. Abakasanga and R. Dabora  are with the department of ECE, Ben-Gurion University,  Israel (e-mail:  abakasan@post.bgu.ac.il; ron@ee.bgu.ac.il).
		This work was supported by the Israel Science Foundation under Grants 1685/16 and 0100101, and by the Israeli Ministry of Economy through the HERON 5G consortium.}
	
	\vspace{-1.0cm}
	
}

\vspace{-0.75cm}

\begin{document}
	
	\maketitle
	\pagestyle{plain}
	\thispagestyle{plain}
	%----------------------------------------------------------------------------------------
	%	ABSTRACT
	%----------------------------------------------------------------------------------------
	\vspace{-0.2cm}
	\begin{abstract}
		\vspace{-0.2cm}
		Non-orthogonal communications  play an important role in future digital communication architectures. In such  scenarios, the received signal is corrupted by an interfering communications signal, which is much stronger than the thermal noise, and is often modeled as a cyclostationary process in continuous-time. 
		To facilitate digital processing, the receiver typically samples the received signal synchronously with the symbol rate of the {\em information signal}. % resulting in a discrete-time (DT)  model in which each channel input represents a different transmitted symbol. 
		If the period of the statistics of the interference is synchronized with that of the information signal, then the sampled interference is modeled as a discrete-time (DT) {\em cyclostationary} random process. 
		\label{txt:Typo1}
		However, in the common interference scenario, the period of the statistics of the interference is not necessarily synchronized with that of the information signal. In such cases, the DT interference may be modeled as an {\em almost cyclostationary}  random process. 
		In this work we characterize the capacity of DT memoryless additive noise channels in which the noise arises from a sampled cyclostationary Gaussian process. 
		For the case of synchronous sampling, capacity can be obtained in closed form.
		When sampling is not synchronized with the symbol rate of the interference, the resulting channel is not information stable, thus classic information-theoretic tools are not applicable. 
		Using information spectrum methods, we prove that capacity can be obtained as the limit of a sequence of capacities of channels with additive {\em cyclostationary} Gaussian noise. Our results allow to characterize the effects of changes in the sampling rate and sampling time offset on the capacity of the resulting DT channel. In particular, it is demonstrated that minor variations in the sampling period, such that  the resulting noise switches from being synchronously-sampled to being asynchronously-sampled, can substantially change the capacity.    
	\end{abstract}
	
	%----------------------------------------------------------------------------------------
	%	INTRODUCTION
	%----------------------------------------------------------------------------------------
	\vspace{-0.3cm}
	\section{Introduction}
	\label{sec:Intro}
	\vspace{-0.1cm}
	In many communications scenarios, the signal which interferes with decoding at the receiver exhibits periodic characteristics. 
	An important such scenario is interference-limited communications, in which the interfering signal is a communications signal. Recent years have witnessed a growing interest  in interference-limited communications due to the transition from  orthogonal architectures, which have dominated wireless communication standards to date, to non-orthogonal schemes \cite{Andrews:14}. 
	Among the important examples of non-orthogonal communications is  \ac{noma}, which is becoming a major paradigm for 5G communications \cite{Dai:15};  DSL communications,  which is limited by  crosstalk \cite{Campbell:83}; and  cognitive radio networks, in which the primary user are the dominant source of interference for the secondary user \cite{Devroye:08,Cohen:18}. As digital communication signals are generated by a random procedure which repeats with each transmitted symbol and frame \cite[Sec. 5]{Gardner:06}, the statistics of the interference in the \ac{ct} domain is   {\em cyclostationary }  \cite[Ch. 1]{Gardner:94}. 
	The digital receiver then operates on the \ac{dt} signal obtained by sampling the \ac{ct} received signal. When  the receiver cannot decode the interference (e.g., since it has no knowledge of the interferer's codebook), then it has to treat the interference as noise. 
	\label{txt:Typo2}
	Consequently, a common received signal model for digital communications in the presence of interference consists of the transmitted signal with an additive noise corresponding to a sampled \ac{ct} cyclostationary process. 
	%In the following we consider interference scenarios in which the thermal noise is much weaker than the interference. 
	
	The capacity of channels with additive {\em stationary} white noise  was shown by Shannon in \cite{Sahnnon:49} to be invariant to the specific value of the sampling interval, as long as the sampling rate satisfies Nyquist's condition with respect to the bandwidth of the information signal. More recent works, \cite{Chen:13,Chen:14,Chen:17}, studied the effect  of different sampling mechanisms, operating below the Nyquist sampling rate, on capacity, when the additive noise is stationary. When the noise is {\em cyclostationary},  even when the sampling rate satisfies Nyquist's condition with respect to the information signal, different sampling rates result in considerably different \ac{dt} models. This indicates that in the presence of cyclostationary noise, the sampling rate can significantly affect capacity even when sampling is above the Nyquist rate.
	
	\ac{dt} communication scenarios with  additive noise obtained by {\em synchronously sampling} a \ac{ct} cyclostationary interference signal were considered in \cite{Shlezinger:16b}. 
	When sampling is synchronous with the period of the cyclostationary interference, namely, the sampling interval and  period of the statistics of the  \ac{ct} interference are commensurable,  the resulting  \ac{dt} interference signal  is cyclostationary \cite[Sec. 3.9]{Gardner:06}. This fact facilitates the analysis of \ac{dt} channels, obtained from \ac{ct} received signals via synchronous sampling, by applying  classical tools for stationary channels, thereby obtaining a characterization of the fundamental rate limits \cite{Shlezinger:15,Shlezinger:16b,Shlezinger:18}, as well as deriving signal processing schemes, e.g., for estimation of statistical moments \cite[Ch. 17.3]{Giannakis:98}, channel identification  \cite{Heath:99}, synchronization \cite{Shaked:18}, spectrum sensing \cite{Cohen:17}, and noise mitigation \cite{Shlezinger:14}. Nonetheless, in many important scenarios of interference-limited communications, the sampling rate and the symbol rate of the \ac{ct} interference are not related in any way, and thus the {synchronous sampling} assumption may not hold.
	
	When the sampling interval and the period of the \ac{ct} additive interference are incommensurable, which is referred to as {\em asynchronous sampling}, the resulting \ac{dt} interference is an {\em almost cyclostationary stochastic process}  \cite[Sec. 3.9]{Gardner:06}.  
	Such scenarios may arise due to specific settings of the sampling interval and the interference symbol period, as well as due to unintentional offsets in these values.
	Communications in the presence of additive almost cyclostationary noise was previously studied for several specific signal processing problems, including spectrum sensing for cognitive radios  \cite{Axell:12}, filter design \cite{Gardner:93}, and parameter estimation \cite{Brown:93, Napolitano:11}. A detailed survey of communications-related applications in the presence of almost cyclostationary signals can be found in \cite{Napolitano:16}.
	Nonetheless, while channels with additive almost cyclostationary noise is an important class of channels with a direct relationship  to interference-limited communications, their fundamental rate limits have not yet been characterized, which is the focus of the current work.

	%\vspace{-0.25cm}
	%\subsection *{Main Contributions}
	%\vspace{-0.1cm}
	In this paper we study the fundamental rate limits for \ac{dt} memoryless channels with additive sampled cyclostationary Gaussian noise. Such channels arise, for example, in interference-limited communications, when the interfering signal is an \ac{ofdm} modulated signal \cite{Wei:10}. 
	Unlike \cite{Chen:13,Chen:14,Chen:17}, we assume that the sampling rate satisfies Nyquist's condition with respect to the information signal, and accordingly, we consider the equivalent \ac{dt} model %where each channel input represents a different symbol 
	as in, e.g., \cite{Medard:00}, instead of studying the \ac{ct} channel. 
	In the case of synchronous sampling, capacity has already been derived in our previous work \cite{Shlezinger:16b}. Consequently, here we focus on capacity characterization for asynchronous sampling. A major benefit from this characterization is quantifying how capacity changes when the sampling rate varies along a continuous range, and in particular, when sampling switches from being synchronous to asynchronous. 

	The main difficulty associated with characterizing the capacity of asynchronously-sampled channels stems from the fact that they are not {\em information-stable}, namely, the conditional distribution of the channel output given the input does not behave ergodically \cite{Dobrushin:63}. Consequently, it is not possible to employ many of the standard information-theoretic considerations, based, e.g., on joint typicality, which, in turn, makes the characterization of capacity of interference-limited communications a very challenging problem. In the current work, we resort to information spectrum tools for characterizing the capacity of asynchronously-sampled channels, as such tools are applicable to non information-stable channels \cite{Han:03}. Although capacity characterizations obtained via information spectrum analysis tend to be difficult to compute, we are able to obtain a meaningful statement of capacity by showing that the capacity of asynchronously-sampled channels can be represented as the limit of a sequence of capacities of synchronously-sampled channels.

	Our derivation allows  to evaluate capacity {\em for any sampling rate} which satisfies Nyquist's condition with respect to the information signal. Numerically evaluating the capacities over a continuous range of sampling frequencies gives rise to some non-trivial insights: For example, we show that changing the sampling rate  changes capacity of the resulting \ac{dt} channel, which stands in contrast to the case of additive stationary noise. Furthermore, we show that {\em very small} variations in the sampling interval  can result in {\em significant changes} in the capacity of the resulting \ac{dt} channel.
	\label{txt:Insight}
	\textcolor{NewColor}{Another important insight, which arises from the cyclostationarity of the \ac{ct} interference and does not follow from the common stationary noise models, is that sampling time offsets have a notable effect on the capacity of \ac{dt} channels  when the sampling rate is synchronized with the symbol rate of the interference.}  However, when sampling is asynchronous, capacity of the \ac{dt} channel becomes invariant to sampling time offsets.
	The results of this work can be used to determine the sampling rate and the sampling time offset which maximize capacity in interference-limited communications.   
	
	%\vspace{-0.25cm}
	%\subsection *{Organization and Notations}
	%\vspace{-0.1cm}
	The rest of this paper is organized as follows: 
	Section~\ref{sec:Preliminaries}  elaborates on the cyclostationarity of communication signals, presents the problem formulation, and reviews some standard definitions.
	Section~\ref{sec:Cap_Async} derives the capacity of memoryless channels with sampled Gaussian noise.
	Numerical examples are discussed in Section~\ref{sec:Simulations}.
	Finally,  Section~\ref{sec:Conclusions}  concludes the paper. 
	\ifFullVersion
	Proofs of the  results stated in the paper are detailed in the appendices.
	\fi
	
	Throughout this paper, we use upper-case letters, e.g., $X$, to denote \acp{rv},  lower-case letters, e.g., $x$, for deterministic values,  and calligraphic letters, e.g., $\mathcal{X}$, for sets. The \ac{pdf} and the \ac{cdf} of a continuous-valued \ac{rv} $X \in\mySet{X}$ evaluated at $x \in \mySet{X}$  are denoted $\pdf{X}(x)$ and $\cdf{X}(x)$, respectively. 
	Column vectors are denoted with boldface letters, where lower-case letters denote deterministic vectors, e.g., ${\bf{x}}$, and upper-case letters are used for
	random vectors, e.g., ${\bf X}$;  the $i$-th element of ${\bf{x}}$ ($i \geq 0$) is written as $({\bf{x}})_i$.
	We use  capital Sans-Serif fonts for matrices, e.g., $\mathsf{A}$,  where the 
	\ifFullVersion
	element at the $i$-th row and $j$-th column 
	\else
	$(i,j)$-th element
	\fi
	of $\mathsf{A}$ is   $(\mathsf{A})_{i,j}$, and the $n\times n$ identity matrix is denoted with $\myMat{I}_n$.
	Complex conjugate, transpose, Hermitian transpose,  Euclidean norm, stochastic expectation, differential entropy, and mutual information are denoted by $(\cdot)^*$, $(\cdot)^T$, $(\cdot)^H$, $\left\|\cdot\right\|$, $\E\{ \cdot \}$, $h(\cdot)$,  and $I(\cdot; \cdot)$,  respectively,
	and we define $a^+ \triangleq \max\left\{0,a\right\}$.
	The Kronecker delta is written as $\delta[i]$, such that $\delta[i]\! =\! 1$ when $i\!=\!0$ and $\delta[i]\! =\! 0$ otherwise.
	We use $\ConvDist{}$ to denote convergence in distribution \cite[Pg. 103]{Kosorok:07}, and $\Ind(\cdot)$ to denote the indicator function. 
	The sets of  positive integers, integers, and real numbers are denoted by  $\mathcal{N}$, $\mathcal{Z}$, and $\mathcal{R}$, respectively. 
	All logarithms are taken to base-2. 
	Finally, for any sequence $y[i]$, $i \in \mySet{N}$, and positive integer $k$,  ${\bf y}^{(k)}$ is the column vector $\left[ { y}[1],\ldots, { y}[k]\right]^T$.

	%% TODO NIR CONTINUE FROM HERE DECEMBER 12
	
	%----------------------------------------------------------------------------------------
	%	PRELIMINARIES
	%----------------------------------------------------------------------------------------
	\vspace{-0.25cm}
	\section{Problem Formulation}
	\label{sec:Preliminaries}
	\vspace{-0.1cm}
	We first  review the cyclostationarity of communication signals  in Subsection \ref{subsec:Pre_Cycsta}. In Subsection \ref{subsec:Pre_Sampl} we present  statistical models for the sampled \ac{dt} process, leading to the  channel model detailed in Subsection \ref{subsec:Pre_Problem}. Finally, in Subsection \ref{subsec:Pre_Defs} we introduce several relevant information-theoretic definitions.
	
	%-----------------------------------
	%	Cyclostationary Signals
	%-----------------------------------
	\vspace{-0.2cm}
	\subsection{Cyclostationarity of Communication Signals}
	\label{subsec:Pre_Cycsta}
	\vspace{-0.1cm}
	As detailed in the introduction, the main motivation for our study of channels with sampled cyclostationary Gaussian noise  stems from the fact that digitally modulated  signals are typically cyclostationary processes. Consequently, the received signal in interference-limited scenarios in which the receiver cannot decode the interference, can be modeled as the sum of the sampled communications signal and sampled cyclostationary noise. To highlight the importance of the cyclostationary model for digital  communications, in the following we elaborate on the cyclostationarity of communications signals, and the resulting \ac{dt} models obtained via sampling such \ac{ct} signals. We begin by recalling the definition of wide-sense cyclostationarity \cite[Sec. 3.2]{Gardner:06}:
	\begin{definition}[Wide-sense cyclostationarity]
		\label{def:WSCS}
		A scalar stochastic process $\{X(t)\}_{t \in \mySet{T}}$, where $\mySet{T}$ is either discrete or continuous, is said to be {\em \ac{wscs}} if both its first-order and  second-order moments are periodic with respect to $t \in \mySet{T}$ with some period $T_p$. 
	\end{definition}
	For example, a real-valued process $\{X(t)\}_{t \in \mySet{T}}$ is \ac{wscs} if 
	$\E\{X(t)\} = \E\{X(t + T_p)\}$ and $\E\{X(t + \tau)X(t)\} = \E\{X(t +  T_p+ \tau)X(t+  T_p)\}$, for all $t$ and  $\tau$ in $\mySet{T}$.   
	
	It is well-established that digitally-modulated communication signals are \ac{wscs} processes in \ac{ct} \cite[Sec. 5]{Gardner:06}. 
	The periodicity of the statistical moments follows from multiple access protocols as well as from the symbol generation model. For example, when using multiple access protocols such as \ac{tdma} and \ac{cdma}, the overall signal is \ac{wscs} with a period which equals the frame duration set by the protocol \cite{Oner:04}. 
	%, digitally-modulated signals exhibit cyclostationarity induced by the symbol generation method, whose period is given by the symbol duration.   
	To demonstrate how the symbol generation scheme induces cyclostationarity, consider generalized linear modulations: 
	\label{txt:missDef}
	Let $T_{\rm sym} > 0$ denote the symbol duration, \textcolor{NewColor}{$K_{\rm f}$ be the number of data symbols in each frame,} $A_{m,k}$ denote the $k$-th data symbol at the $m$-th frame, $k \in \{1,2,\ldots,K_{\rm f}\}$, $m \in \mySet{Z}$, and $q_k(t)$ denote the pulse-shaping function of the $k$-th symbol. The resulting modulated signal in baseband is
	\begin{equation}
	\label{eqn:DigMod1}
	S(t) = \sum\limits_{m=-\infty}^{\infty}  \sum\limits_{k=1}^{K_{\rm f}} A_{m,k}   q_k\left(t - m T_{\rm sym}\right).
	\end{equation}
	
	For example, for $K_{\rm f} = 1$, \eqref{eqn:DigMod1} yields the class of pulse amplitude modulations \cite[Ch. 1]{Gardner:94}. Alternatively,  for a fixed $T_{\rm data} < T_{\rm sym}$ and pulse shaping function $\tilde{q}(t)$ such that $q_k(t)$ can be written as $q_k(t) = \tilde{q}(t)\exp\left(j \frac{2\pi \cdot k \cdot t}{ T_{\rm data}}\right) $, the model \eqref{eqn:DigMod1} represents \ac{ofdm} modulation \cite{Heath:99}. Assuming that the data symbols $\{A_{m,k}\}$ are i.i.d., it can be easily shown that $S(t)$ in \eqref{eqn:DigMod1} satisfies Def. \ref{def:WSCS}, and is thus \ac{wscs}. 
	Since zero-mean and proper complex \ac{wscs} baseband signals are also \ac{wscs} in passband \cite[Sec. II-C]{Shlezinger:14}, passband digitally modulated communication signals are also \ac{wscs}, and it follows that digital communication signals are typically modeled as \ac{wscs} signals in \ac{ct}. 
	In the current work we model the statistics of the interfering signal as Gaussian. While this is not strictly accurate for some digital modulations, it is shown  in \cite{Wei:10} that for i.i.d. data symbols, \ac{ofdm} signals approach the distribution of Gaussian processes.

	%-----------------------------------
	%	Sampling \ac{ct} \ac{wscs} Signals
	%-----------------------------------
	\vspace{-0.2cm}
	\subsection{Sampling \ac{ct} \ac{wscs} Random Processes}
	\label{subsec:Pre_Sampl}
	\vspace{-0.1cm}
	As digital receivers operate on sampled signals, we next discuss  sampling of \ac{ct} \ac{wscs} stochastic processes. Consider the \ac{dt} random signal $S_{\Tsamp, \phi}[i]$, $i \in \mySet{Z}$, obtained by uniformly sampling $S(t) $ with sampling period $\Tsamp$ and sampling time offset $\phi$, i.e., $S_{\Tsamp, \phi}[i] \triangleq S (i \cdot \Tsamp + \phi)$. In contrast to sampling of stationary signals, here the values of $\Tsamp$ and $\phi$ have a notable effect on the statistical model of the sampled signal $S_{\Tsamp, \phi}[i]$. As an example, we illustrate in Fig. \ref{fig:SampledVar} how the variance of a sampled process can vary considerably with the sampling rate and the sampling time offset: 
	\begin{figure}
		\centering
		\includegraphics[width=\columnwidth]{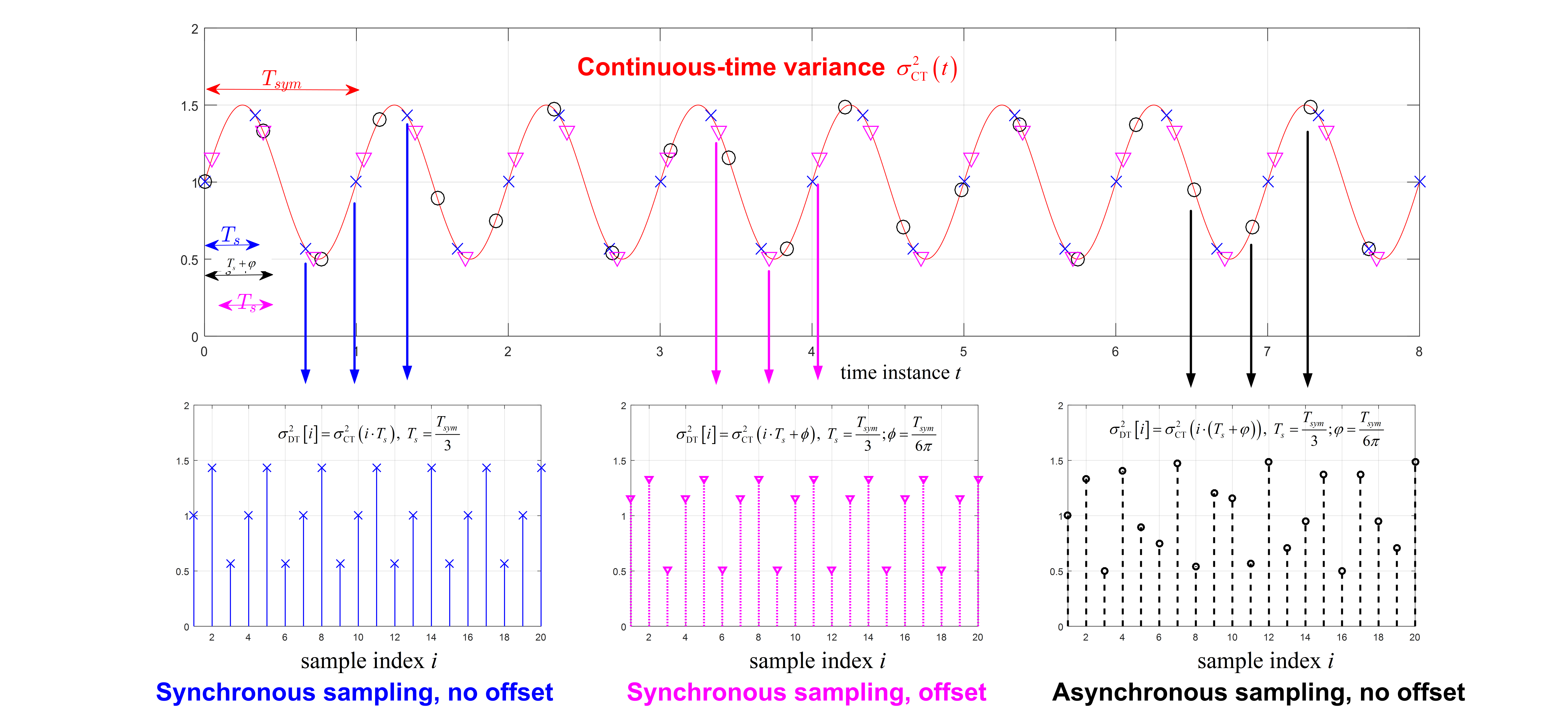}
		\vspace{-0.6cm}
		\caption{Illustration of  \ac{dt} variances obtained by sampling a \ac{ct} \ac{wscs} signal with different sampling settings.}
		\vspace{-0.8cm}
		\label{fig:SampledVar}
	\end{figure}
	The red curve in the upper plot in Fig. \ref{fig:SampledVar} depicts the periodic variance of the \ac{ct} \ac{wscs} signal $S(t)$, denoted $\sigma_{\rm CT}^2(t)$, whose period is $T_{\rm sym}$. The bottom plots in  Fig. \ref{fig:SampledVar} depict the variance of the sampled $S_{\Tsamp, \phi}[i]$, denoted $\sigma_{\rm DT}^2[i]$, for three different combinations of sampling interval and offset: 
	The bottom left blue plot depicts the variance of the sampled signal without offset when the sampling period is $\Tsamp = \frac{T_{\rm sym}}{3}$; 
	The bottom center magenta plot depicts the  variance of the sampled signal for the same sampling period $\Tsamp$ with a small sampling offset $\phi =  \frac{\Tsamp}{2\pi}$. We note that this variance, as well as the one obtained without offset, are {\em periodic} in \ac{dt}, however, their values are {\em different}. These plots, in which the periodicity of the statistics is maintained in \ac{dt}, correspond to  {\em synchronous sampling}. 
	In the bottom right black curve we depict the variance of the sampled process when there is no offset and the sampling period is $\left( 1+\frac{1}{2\pi}\right)\frac{T_{\rm sym}}{3} $, which is not an integer division of $T_{\rm sym}$ (or of an integer multiple of $T_{\rm sym}$).  
	We refer to such situations as {\em asynchronous sampling}. 
	\textcolor{NewColor}{
	Unlike the previous cases, here the \ac{dt} variance is not periodic, but is an almost periodic function, namely, it is the limit of a uniformly convergent sequence of trigonometric polynomials \cite[Ch. 1.2]{Napolitano:12}. Accordingly, as detailed in  \cite[Ch. 3.9]{Gardner:06}, the resulting \ac{dt} random process is not \ac{wscs}, but {\em \ac{wsacs}}, namely, it satisfies the following definition \cite[Sec. 3.2]{Gardner:06}}:
	\begin{definition}[Wide-sense almost cyclostationarity]
		\label{def:WSACS}
		A scalar stochastic process $\{X(t)\}_{t \in \mySet{T}}$, where $\mySet{T}$ is either discrete or continuous, is said to be {\em \acl{wsacs}} if both its first-order and second-order moments are almost periodic functions with respect to $t\in \mySet{T}$. Specifically, for a real-valued $\{X(t)\}_{t \in \mySet{T}}$ there exist a countable set $\mySet{A} \subset \mySet{R}$ and coefficients $\{\mu_\alpha\}_{\alpha \in \mySet{A}}$ such that 	for all $ t, \tau \in \mySet{T}$, these moments can  be written as
		\begin{equation*}
		\E\{X(t)\} = \sum\limits_{\alpha \in \mySet{A}}\mu_\alpha  e^{j 2\pi \alpha t}; \qquad \E\{X(t +   \tau)X(t)\}  = \sum\limits_{\alpha \in \mySet{A}}c_\alpha(\tau) e^{j 2\pi \alpha t}.
		\end{equation*} 
	\end{definition} 
	
	The simple example presented here demonstrates how the statistical properties of a sampled \ac{wscs} process can change considerably with minor variations in the sampling period and sampling time offset. Consequently, in communication channels where the noise corresponds to a sampled \ac{wscs} \ac{ct} process, e.g., as in interference-limited communications, capacity can vary significantly as the sampling rate changes. This motivates the need to characterize capacity for {\em any given sampling rate}, as mathematically formulated in the next subsection.
	
	%-----------------------------------
	%	Problem Formulation
	%-----------------------------------
	\vspace{-0.2cm}
	\subsection{Problem Formulation}
	\label{subsec:Pre_Problem}
	\vspace{-0.1cm}
	Consider a \ac{ct} real-valued zero-mean \ac{wscs}  Gaussian random process $\Wc(t)$ with period $\Tc$, i.e., the variance $\Cwc(t) \triangleq \E\big\{\left( \Wc(t)\right) ^2\big\}$ satisfies  $\Cwc(t)  =  \Cwc(t + \Tc)$, for all $t \in \mySet{R}$.
	Furthermore, the variance function $\Cwc(t)$ is  continuous with respect to time $t$, and is strictly positive. Note that since $\Cwc(t)$ is periodic and continuous, it holds that it is also bounded and uniformly continuous \cite[Thm. 3.13]{Amann:05}.
	Let the signal $\Wc(t)$ be uniformly sampled with a sampling interval of $\Tsamp$ such that $\Tc = (\Td + \myEps)\cdot \Tsamp$ for some fixed $\Td \in \mySet{N}$ and $\myEps \in [0,1)$, resulting in the \ac{dt} signal $\Weps[i] = \Wc(i \cdot  \Tsamp)$. 
	In this work we assume that the span of the temporal correlation of the \ac{ct} signal $\Wc(t)$ is sufficiently shorter than the sampling period, and in particular, $\E\{\Wc(t + \lambda)\Wc(t) \} = 0$ for all real $\lambda \ge \Tsamp$, and $t \in \mySet{R}$. 
	The resulting \ac{dt} process $\Weps[i]$ is clearly a {\em memoryless} zero-mean Gaussian process with  autocorrelation function
	\vspace{-0.1cm}
	\begin{align}
	\Cweps[i,\tau]
	&= \E\left\{\Weps[i+\tau] \Weps[i]\right\} \notag \\
	&= \E\left\{\Wc\left(\frac{(i+\tau)\cdot \Tc}{\Td + \myEps}\right) \cdot\Wc\left(\frac{i \cdot \Tc}{\Td + \myEps}\right)\right\}
	=\Bigg(  \Cwc\left(\frac{i \cdot\Tc}{\Td + \myEps} \right)\Bigg) \cdot\delta[\tau].
	\label{eqn:AnsycAutocorr}
	\vspace{-0.1cm}
	\end{align}
	%It follows from \eqref{eqn:AnsycAutocorr} that $\Cweps[i,\tau] = 0$ for all $|\tau| >\left\lceil \frac{(\Td + 1) \cdot\Memc}{\Tc} \right\rceil\triangleq \Memd < \infty$, hence $\Weps[i]$ has a finite memory $\Memd<\infty$.
	The variance of $\Weps[i]$  is thus given by $\Sigweps[i] = \Cwc\left(\frac{i \cdot\Tc}{\Td + \myEps}\right)$. 
	While we do not explicitly account for sampling time offsets in our definition of the sampled process $\Weps[i]$, it can be incorporated by replacing $\Cwc(t)$ with its time-shifted version, i.e., $\Cwc(t-\phi)$. 
	
	By treating the sampled \ac{wscs} interfering signal as additive noise assumed to be much stronger than the thermal noise, we arrive at the following \ac{dt} channel model:   
	Consider 
	%The sampled signal represents the additive noise in the considered communications channel, which is therefore modeled as 
	a \ac{dt} memoryless channel with  additive sampled \ac{wscs} Gaussian noise $\Weps[i]$. We keep the subscript  $\myEps$ to emphasize the dependence of the noise statistics on the synchronization mismatch between the sampling interval and the noise period.
	Let $\mySet{U}$ denote the set of messages,  $\Xscal[i]$ be the real channel input and $\Yeps[i]$ denote the output, both at time index $i \in \mySet{N}$. The input-output relationship of this channel for the transmission of $l \in \mySet{N}$ symbols is given by
	\vspace{-0.2cm}
	\begin{equation}
	\label{eqn:AsnycModel1}
	\Yeps[i] =   \Xscal[i ] + \Weps[i], \qquad i \in \{1,2,\ldots,l\}.
	\vspace{-0.1cm}
	\end{equation}
	%Note that the initial state of the channel is  ${\bf S}_0 = \big[\Xscal[-\Mem],\ldots,\Xscal[-1],\Weps[-\Mem],\ldots,\Weps[-1]\big]^T$.
	%Let $\mySet{U}$ denote the set of messages. 
	The channel input sequence $\left\{\Xscal[i]\right\}_{i \in \mySet{N}}$ is assumed to be independent of the noise process $\left\{\Weps[i]\right\}_{i \in \mySet{N}}$, % and of ${\bf S}_0$,
	and is subject to an average power constraint $P$,
	i.e., for each message $u \in \mathcal{U}$, the corresponding codeword $\{\xin_{(u)}\left[ i \right]\}_{i=1}^{l}$ satisfies
	\vspace{-0.2cm}
	\begin{equation}
	\label{eqn:AsyncConst1}
	\frac{1}{{{l}}}\sum\limits_{i = 1}^{l  } \left| \xin_{(u)}\left[ i \right] \right|^2 \le P.
	\vspace{-0.1cm}
	\end{equation}
	% TODO NIR - move to discussion section?
	The channel \eqref{eqn:AsnycModel1} represents a sampled \ac{ct} channel, and we assume that the sampling rate satisfies Nyquist theorem with respect to the information signal, see  \cite[Sec. II]{Medard:00} for discussion on sampled time-varying channels. Hence, unlike \cite{Chen:13,Chen:14,Chen:17} which considered sub-Nyquist sampling, here we  carry out the capacity analysis by considering the \ac{dt} sampled channel and not the  \ac{ct} channel. 
	
	As follows from our discussion above and in the introduction, the channel model in \eqref{eqn:AsnycModel1} is particularly relevant for interference-limited communications, as well as to cognitive radio communications.  In these cases, $\Weps[i]$ is the sampled version of $\Wc(t)$, which represents a  digitally-modulated  interfering  signal. Accordingly, $\Wc(t)$ is a \ac{ct} \ac{wscs} process, as discussed in Subsection~~\ref{subsec:Pre_Cycsta}. %In particular, when $\Wc(t)$ corresponds to an \ac{ofdm}  signal with a sufficiently large number of subcarriers, then it can be modeled as a \ac{wscs} Gaussian process \cite{Wei:10}. 
	Our objective is to characterize the capacity of the real channel defined in \eqref{eqn:AsnycModel1} subject to the power constraint \eqref{eqn:AsyncConst1} {\em for any value of $\myEps \in [0,1)$}.
	
	In general, the interfering signal may have memory. Thus,  the general sampled interference-limited setup has two major aspects of the noise statistics that need to be addressed: The non-stationary behavior and the memory.  As channel memory has been extensively addressed for stationary channels, in this work we focus on the new aspect which is the {\em non-stationary nature of the noise statistics}, leaving its combination with channel memory to future work. 
	
	We note from \eqref{eqn:AnsycAutocorr} that when $\myEps$ is a rational number, i.e., there exist $ u,v \in \mySet{N}$ such that $\myEps = \frac{u}{v}$, then  $\Weps[i]$ is a \ac{wscs} process with period $\Td \cdot v + u \in \mySet{N}$.  Recall that we refer to this situation as synchronous sampling. 
	As we discuss in Subsection \ref{subsec:Cap_Sync}, for such channel models capacity was derived in \cite{Shlezinger:16b}. 
	However, when $\myEps$ is irrational,  $\Weps[i]$ is a \ac{wsacs} process, as defined in Def. \ref{def:WSACS}.
	We refer to 
	\ifFullVersion
	the scenario when $\myEps$ is irrational 
	\else
	such scenarios 
	\fi 
	as {\em asynchronous sampling}. 
	In order to understand how capacity varies with continuous variations in the sampling rate, due to, e.g., hardware impairments, capacity with asynchronous sampling has to be characterized.  
	Additionally,  in interference-limited setups, there is no reason to assume that the sampling rate is synchronized with the 
	\ifFullVersion
	symbol rate of the interference,
	\else
	interference period,
	\fi
	which further motivates the characterization of capacity with asynchronous sampling. 
	
	By characterizing the capacity of the channel \eqref{eqn:AsnycModel1} subject to a power constraint \eqref{eqn:AsyncConst1} for each $\myEps \in [0,1)$, we are able to rigorously quantify the effect of variations in the sampling rate and sampling offset on capacity. In our numerical study in Section \ref{sec:Simulations}, and particular, in Figs. \ref{fig:CnTs_Offset0}-\ref{fig:CnTs_Offset05}, we demonstrate how capacity varies as $\myEps$ changes, noting that for different synchronous sampling rates, capacity exhibits dependence on sampling offset, 
	which can result in either an increase or a decrease with respect to zero offset, while for asynchronous sampling a relatively constant capacity value is obtained. We conjecture that the asynchronously-sampled capacity represents the capacity of the analog channel, which is invariant to the sampling mechanism, but we leave the rigorous investigation of this  for future work. 
	Another non-trivial insight which follows from our analysis is that capacity can change dramatically with minor variations in the sampling rate. For example, in our numerical study, and  specifically in Fig. \ref{fig:CnP_Offset0}, we demonstrate that a  variation of $0.2\%$ in the sampling interval can result in significant variations of $30 \%$ in capacity. This result is consistent with the fundamentally different statistical models observed heuristically in Fig. \ref{fig:SampledVar} induced by small variations in the sampling rate.

	%We will address channel memory in our future work. Accordingly, as stated above, we model the additive noise  as a memoryless \ac{wsacs} process. 
	
	%%%%%%%%%%%%%%%%%%%%%%%%%%%%%%%%%%%%%%%%%%%%%%%%%%%%%%%%%%%%%%%%%%%%%%%%%%%%%%%%%%%%%%%%%%%%%%%%%%%%%%%%%%%%%%%%%%%%%%%%%%%%%%%%%%%%%%%%%%%%%%%%%%%%%%%%%%%%%%%%%%%%%%%%%%%%%%%%%%%%%%%%%%%%%%%%%%%%%%%%%%%%%%%%%%%%%%%%%
	%%%%%%%%%%%%%%%%%%%%%%%%%%%%%%%%%%%%%%%%%%%%%%%%%%%%%%%%%%%%%%%%%%%%%%%%%%%%%%%%%%%%%%%%%%%%%%%%%%%%%%%%%%%%%%%%%%%%%%%%%%%%%%%%%%%%%%%%%%%%%%%%%%%%%%%%%%%%%%%%%%%%%%%%%%%%%%%%%%%%%%%%%%%%%%%%%%%%%%%%%%%%%%%%%%%%%%%%%

	%-----------------------------------
	%	Definitions
	%-----------------------------------
	\vspace{-0.2cm}
	\subsection{Definitions}
	\label{subsec:Pre_Defs} 
	\vspace{-0.1cm}
	We end this section by introducing the set of definitions used in our capacity analysis, beginning with information spectrum quantities. 
	As mentioned in the introduction, we utilize the information spectrum approach for defining the capacity, since it can be applied for arbitrary channels. Standard information-theoretic methods, which are based on the law of large numbers, require the conditional distribution of the channel output given its input to be ergodic, i.e., these methods hold for {\em information-stable channels} \cite{Verdu:94}, and thus are not applicable to the non-ergodic \ac{dt} channel which arises from asynchronous sampling. 
	In the following we review the basic information spectrum quantities, following their definitions in \cite[Defs. 1.3.1-2]{Han:03}:
	\begin{definition}
		\label{def:pliminf}
		The {\em limit-inferior in probability} of a sequence of real \acp{rv} $\{\zk{k}\}_{k \in \mySet{N}}$ is defined as
		\vspace{-0.1cm}
		\begin{equation}
		\label{eqn:pliminf}
		{\rm p-}\mathop{\lim \inf}\limits_{k \rightarrow \infty} \zk{k} \triangleq \sup\left\{\alpha \in \mySet{R} \big| \mathop{\lim}\limits_{k \rightarrow \infty}\Pr \left(\zk{k} < \alpha \right) = 0   \right\}
		\triangleq \alpha_0.
		\vspace{-0.1cm}
		\end{equation}
	\end{definition}
	Hence, $\alpha_0$ is the largest real number satisfying that  $\forall \talpha < \alpha_0$ and  $\forall \delta>0$ there exists $k_0(\delta,\talpha) \in \mN$ such that $\Pr(Z_k<\talpha)<\delta$, $\forall k>k_0(\delta,\talpha)$.
	\begin{definition}
		\label{def:plimsup}
		The {\em limit-superior in probability} of a sequence of real \acp{rv} $\{\zk{k}\}_{k \in \mySet{N}}$ is defined as
		\vspace{-0.1cm}
		\begin{equation}
		\label{eqn:plimsup}
		{\rm p-}\mathop{\lim \sup}\limits_{k \rightarrow \infty} \zk{k} \triangleq \inf\left\{\beta  \in \mySet{R} \big| \mathop{\lim}\limits_{k \rightarrow \infty}\Pr \left(\zk{k} > \beta \right) = 0   \right\}
		\triangleq \beta_0.
		\vspace{-0.1cm}
		\end{equation}
	\end{definition}
	Hence,  $\beta_0$ is the smallest real number satisfying that $\forall \tbeta > \beta_0$ and $\forall \delta>0$, there exists $k_0(\delta,\tbeta)\in\mN$, such that $\Pr(Z_k>\tbeta)<\delta$, $\forall k>k_0(\delta,\tbeta)$.
	
	The above quantities are well-defined even when the sequence of \acp{rv} $\{\zk{k}\}_{k \in \mySet{N}}$ does not converge in distribution \cite[Pg. VIII]{Han:03}, \cite[Sec. II]{HanVerdu:IT93}. Consequently, these quantities play an important role in  information-theoretic  analysis when methods based on the law of large numbers cannot be applied, e.g., when non-stationary and non-ergodic signals are considered \cite[Sec. I]{Verdu:94}.
	The main difficulty in the application of Defs. \ref{def:pliminf}-\ref{def:plimsup} to characterize information-theoretic quantities is that, except for very specific scenarios, they are quite difficult to compute \cite[Pg. XIV]{Han:03}. 
	%\footnote{In fact, in \cite[Pg. XIV]{Han:03}, the author states that {\em "Information-spectrum methods are still very immature and need to be cultivated much further. For example, specific nonstationary or nonergodic examples that we were able to demonstrate in this book are, in most cases,	mixed sources and mixed channels. This is because mixed sources and mixed 	channels are very simple but typically nonergodic, while, regrettably enough, we are not yet successful in demonstrating another type of much more substantially	nonstationary and/or nonergodic sources and/or channels. This problem remains to be exploited further also along the information-spectrum methods."}}. 
	In  Subsection \ref{subsec:Cap_InfSpec} we prove an identity which allows us to obtain a meaningful characterization of the capacity of the channel \eqref{eqn:AsnycModel1} with asynchronous sampling using  Defs. \ref{def:pliminf}-\ref{def:plimsup}. % result in a meaningful characterization.
	
	We next introduce three additional standard definitions used in the capacity derivation:
	%
	%\begin{definition}[Memoryless channel]
	%	\label{def:Channel}
	%	A real, memoryless, scalar channel without feedback
	%	%, where the transmitter has antennas and the receiver has $n_r$ antennas,
	%	consists of an input stream $\Xscal[i] \in \mySet{R}$, $i \in \mySet{N}$, an output stream  $Y[i] \in \mySet{R}$, $i \in \mySet{N}$,   and a sequence of \acp{pdf} $\left\{p\left(\Yi^{(n) }|\Xin^{(n)}\right) \right\}_{n=1}^{\infty}$ satisfying 
	%	\begin{equation*}
	%	p\left(\Yi^{(n) }|\Xin^{(n) }\right)  = \prod\limits_{i=1}^{n}p(Y[i] | \Xscal[i]).
	%	\end{equation*}.
	%\end{definition}
	
	% Define - code
	\begin{definition}[Channel code]
		\label{def:Code}
		An $\left[R, l \right]$ code with rate $R$ and blocklength $l \in \mySet{N}$ consists of:
		{\em 1)} A message set $\mathcal{U} \triangleq \{1,2,\ldots,2^{lR}\}$.
		{\em 2)} An  encoder $e_l$ which maps a message $u \in \mySet{U}$ into a codeword $\myVec{x}_{(u)}^{(l)} = \Big[\xin_{(u)}\left[1\right],\xin_{(u)}\left[2\right],\ldots,\xin_{(u)}\left[l\right]\Big]$.
		{\em 3)}  A decoder $d_l$ which maps the channel output  $\myVec{y}^{(l)}$ into a message $\hat{u} \in \mySet{U}$.
		%
		%	The encoder operates independently of the initial state ${\bf S}_0$.
	\end{definition}
	The set $\{{\bf x}_{(u)}^{(l)}\}_{u =1}^{2^{lR}}$ is referred to as the {\em codebook} of the $\left[R, l \right]$ code.
	Letting the message $U$  be  selected uniformly from  $\mySet{U}$, the average probability of error 
	%For a given initial state $\svec_0$, the average probability of error is defined as\footnote{NIR COMMENT: Note that I added a distribution to the initial state. This allowed me not to keep the term of the previous noise samples in my expressions which I found pretty difficult to work with as now the initial state also depends on $n$.}:
	\ifFullVersion
	can be expressed as
	\begin{equation*}
	\label{eqn:def_avgError}
	%P_e^l \left( {{\bf s} }_0 \right) \!= \! \frac{1}{2^{lR}}\sum\limits_{u \!= \! 1}^{2^{lR}} \Pr \left( {\left. {{d_l }\left( \Yi^{l}  \right) \ne u} \right|U \!= \! u,{{{\bf S} }_0} \!= \! {{{\bf s} }_0}} \right).
	P_e^{(l)}  \!= \! \frac{1}{2^{lR}}\sum\limits_{u \!= \! 1}^{2^{lR}} \Pr \left( \left. {{d_l }\big( \Yi^{(l) }\big) \ne u} \right|U \!= \! u  \right).
	\end{equation*}
	\else
	is
	$P_e^{(l)}  \!= \! \frac{1}{2^{lR}}\sum\limits_{u \!= \! 1}^{2^{lR}} \Pr \left( \left. {{d_l }\big( \Yi^{(l) }\big) \ne u} \right|U \!= \! u  \right)$.
	\fi
	
	% Define - rate
	\begin{definition}[Achievable rate]
		\label{def:Rate}
		A rate $\Rs$ is achievable if for every $\eta_1,\eta_2 > 0$, $\exists l_0\left( \eta_1,\eta_2\right)  \in\mN$ s.t. $\forall l > l_0\left( \eta_1,\eta_2\right)$ there exists an $\left[R, l \right]$ code which satisfies %\cite[Sec. III]{Goldsmith:01}
		\ifFullVersion
		\vspace{-0.2cm}
		\begin{subequations}
			\label{eqn:def_Rs}
			\begin{equation}
			\label{eqn:def_Rs1}
			%		\mathop {\sup }\limits_{{{{\bf s} }_0} \in \mathcal{S}_0} P_e^l \left( {{{{\bf s} }_0}} \right) < {\eta_1},
			P_e^{(l)} < \eta_1,
			\end{equation}
			\vspace{-0.2cm}
			and
			\begin{equation}
			\label{eqn:def_Rs2}
			R \geq \Rs - \eta_2.
			\end{equation}
		\end{subequations}
		\else
		$P_e^{(l)} < \eta_1$ and $	R \geq \Rs - \eta_2$.
		\fi
	\end{definition}
	
	\begin{definition}[Capacity]
		{\em Capacity} is defined as the supremum over all achievable rates.
	\end{definition}
	
	% TODO NIR CONTINUE FROM HERE FEB 7
	
	%-----------------------------------
	%	Capacity of  Sampled Cyclostationary Noise Channels
	%-----------------------------------
	\vspace{-0.2cm}
	\section{Capacity of  Sampled \ac{wscs} Additive Gaussian Noise Channels}
	\label{sec:Cap_Async}
	\vspace{-0.1cm}
	In order to characterize the capacity of the channel  \eqref{eqn:AsnycModel1} subject to \eqref{eqn:AsyncConst1}, we first  present a theorem in Subsection \ref{subsec:Cap_InfSpec} which relates the information spectrum quantities of a set of sequences of \acp{rv} to the information spectrum quantities of its limit sequence of \acp{rv}. 
	Next, we recall in Subsection \ref{subsec:Cap_Sync} the capacity with synchronous sampling, as a preliminary step to our derivation of the capacity with asynchronous sampling. 
	In Subsection \ref{subsec:Cap_Main} we use the relationship established in Subsection \ref{subsec:Cap_InfSpec} to derive the capacity with asynchronous sampling  as the limit of a sequence of capacities of channels with \ac{dt} {\em \ac{wscs}} Gaussian noise, where each element in the sequence can be evaluated as a closed form expression detailed in Subsection \ref{subsec:Cap_Sync}. Finally, in Subsection \ref{subsec:Cap_Discussion} we discuss our results and point out some insights which arise from them.
	
	%-----------------------------------
	%	Information Spectrum Limit
	%-----------------------------------
	\vspace{-0.2cm}
	\subsection{Information Spectrum Limits}
	\label{subsec:Cap_InfSpec} 
	\vspace{-0.1cm}
	In our capacity derivation, we utilize the following new theorem for random sequences:
	\begin{theorem}
		\label{thm:plim}
		Let $\big\{\zkn{k}{n}\big\}_{n,k \in \mySet{N}}$ be a set of real scalar  \acp{rv} satisfying two assumptions:
		\begin{enumerate}[label={\em AS\arabic*}]
			\item \label{itm:assm1} For every fixed $n \in \mySet{N}$, every convergent subsequence of $\big\{\zkn{k}{n}\big\}_{k \in \mySet{N}}$ converges in distribution as $k \rightarrow \infty$ to a finite deterministic scalar. Each subsequence may converge to a different scalar.
			\item \label{itm:assm2}  For every fixed $k \in \mySet{N}$, as $n \rightarrow \infty$ the sequence $\big\{\zkn{k}{n}\big\}_{n\in \mySet{N}}$ converges uniformly in distribution to a scalar real-valued \ac{rv} $\zkeps{k}$. Specifically,  letting $\Fkn{k}{n}(\alpha)$ and $\Fkeps{k}(\alpha)$, $\alpha \in \mySet{R}$, denote the \acp{cdf} of $\zkn{k}{n}$ and of $\zkeps{k}$, respectively, then  $\forall \eta>0$, there exists $n_0(\eta)$ such that for every  
			\ifFullVersion	
			$ n > n_0(\eta)$ and 	for each $\alpha \in \mySet{R}$  and $ k \in \mySet{N}$,	
			\begin{equation*} 
			\left|\Fkn{k}{n}(\alpha) - \Fkeps{k}(\alpha)\right| < \eta. 
			\end{equation*}
			\else 
			$ n > n_0(\eta)$, $\left|\Fkn{k}{n}(\alpha) - \Fkeps{k}(\alpha)\right| < \eta$ 	for each $\alpha \in \mySet{R}$, $ k \in \mySet{N}$.
			%		it holds that $|\Fkn{k}{n}(\alpha) - \Fkeps{k}(\alpha)| < \eta$ for all $\alpha \in \mySet{R}$ and  $k \in \mySet{N}$. 
			\fi
		\end{enumerate}
		
		When  $\big\{\zkn{k}{n}\big\}_{n,k \in \mySet{N}}$ satisfies \ref{itm:assm1}-\ref{itm:assm2}, it holds that 
		\begin{subequations}
			\label{eqn:plim}
			\begin{eqnarray}
			\label{eqn:plima}
			{\rm p-}\mathop{\lim \inf}\limits_{k \rightarrow \infty} \zkeps{k} &=&\mathop{\lim}\limits_{n \rightarrow \infty} \left({\rm p-}\mathop{\lim \inf}\limits_{k \rightarrow \infty} \zkn{k}{n} \right), \\
			%		\end{eqnarray}
			%		and
			%		\begin{eqnarray}
			\label{eqn:plimb}
			{\rm p-}\mathop{\lim \sup}\limits_{k \rightarrow \infty} \zkeps{k} &=&\mathop{\lim}\limits_{n \rightarrow \infty} \left({\rm p-}\mathop{\lim \sup}\limits_{k \rightarrow \infty} \zkn{k}{n} \right).
			\end{eqnarray}
		\end{subequations}
	\end{theorem}
	
	\begin{IEEEproof}
		The proof is given in Appendix \ref{app:Proof1}. 	
	\end{IEEEproof}
	
	\smallskip
	For various information-theoretic problems, the terms $	{\rm p-}\mathop{\lim \inf}\limits_{k \rightarrow \infty} \zkeps{k}$ and / or ${\rm p-}\mathop{\lim \sup}\limits_{k \rightarrow \infty} \zkeps{k} $ represent {\em unknown quantities}, i.e., quantities for which it is not possible to obtain {\em meaningful expressions} using current tools, while ${\rm p-}\mathop{\lim \inf}\limits_{k \rightarrow \infty} \zkn{k}{n}$ and / or ${\rm p-}\mathop{\lim \sup}\limits_{k \rightarrow \infty} \zkn{k}{n}$ correspond to {\em known quantities}  for which  {\em meaningful expressions} can be established. Consequently, Theorem~\ref{thm:plim} facilitates deriving meaningful characterizations of the unknown quantities. In Subsection~\ref{subsec:Cap_Main} we use Theorem~\ref{thm:plim} to characterize the capacity of asynchronously-sampled memoryless cyclostationary Gaussian noise channels.

	%-----------------------------------
	%	Capacity Characterization with Synchronous Sampling
	%-----------------------------------
	\vspace{-0.2cm}
	\subsection{Capacity Characterization for Synchronous Sampling}
	\label{subsec:Cap_Sync} 
	\vspace{-0.1cm}
	As a preliminary step to our capacity characterization for the  channel \eqref{eqn:AsnycModel1} subject to the constraint \eqref{eqn:AsyncConst1}, resulting from asynchronous sampling, we present here the capacity for the model resulting from synchronous sampling. In this case, the synchronization mismatch $\myEps$ can be written as $\myEps = \frac{u}{v}$ for some positive integers $u,v$. As discussed in Subsection \ref{subsec:Pre_Problem}, the resulting $\Weps[i] = \Wc\Big(\frac{i \Tc}{\Td + \frac{u}{v}}\Big)$  is a \ac{wscs} process with  period $\bar{\Td}_{u,v} = \Td \cdot  v + u$. 
	Consequently, the channel  \eqref{eqn:AsnycModel1} is a \ac{dt} memoryless channel with additive {\em \ac{wscs}} Gaussian noise, whose capacity can be obtained from \cite[Thm. 1]{Shlezinger:16b}, which is recalled in the following proposition:
	
	\begin{proposition}
		\label{cor:SynSample}
		Let $\Sigweps[i] \triangleq \Cweps[i,0]$,  and let $\TDelta_{u,v}$ be the unique solution to
		\begin{equation}
		\label{eqn:MainThmCst}
		\frac{1}{  \bar{\Td}_{u,v}}\sum\limits_{i=0}^{\bar{\Td}_{u,v}-1}   \left(\TDelta_{u,v} - \Sigweps[i] \right)^+  =  \PCst.
		\vspace{-0.1cm}
		\end{equation}
		The capacity of the channel  \eqref{eqn:AsnycModel1} subject to \eqref{eqn:AsyncConst1} when $\myEps = \frac{u}{v}$, denoted $\bar{\Capacity}_{u,v}$, is given by
		\vspace{-0.1cm}
		\begin{equation}
		\label{eqn:MainThm2}
		\bar{\Capacity}_{u,v} = \frac{1}{2 \cdot \bar{\Td}_{u,v}}\sum\limits_{i=0}^{\bar{\Td}_{u,v} -1}   \left(\log   \left(  \frac{\TDelta_{u,v}}{\Sigweps[i]}\right)  \right)^+  .
		\vspace{-0.1cm}
		\end{equation}
	\end{proposition} 
	
	\begin{IEEEproof}
		The proposition is obtained by specializing \cite[Thm. 1]{Shlezinger:16b}, which characterizes the capacity of finite-memory \ac{dt} multivariate channels with additive \ac{wscs} noise, to memoryless \ac{dt} scalar channels with additive \ac{wscs} noise. 
	\end{IEEEproof}
	
	\smallskip
	Proposition \ref{cor:SynSample} expresses the capacity in closed-form for 
	\ifFullVersion
	channels in which the noise is synchronously-sampled. 
	\else
	synchronously-sampled channels.
	\fi
	In the next subsection we show that the limit inferior of a specific sequence whose elements are capacities of the form of \eqref{eqn:MainThm2}, characterizes the capacity for 
	\ifFullVersion
	channels in which the noise is asynchronously-sampled. 
	\else
	asynchronously-sampled channels. 
	\fi

	%-----------------------------------
	%	Capacity Characterization with Asynchronous Sampling
	%-----------------------------------
	\vspace{-0.2cm}
	\subsection{Capacity Characterization for Asynchronous Sampling}
	\label{subsec:Cap_Main} 
	\vspace{-0.1cm}
	%In the following we  characterize the capacity of asynchronously-sampled memoryless cyclostationary Gaussian noise channels.
	To characterize the capacity of the channel \eqref{eqn:AsnycModel1} subject to \eqref{eqn:AsyncConst1} for asynchronous sampling, we first define a sequence of channels with synchronous sampling such that in the limit, the sampling interval approaches the asynchronous sampling interval. Then, we relate the capacities of these channels to the capacity  with asynchronous sampling in Theorem \ref{thm:AsycCap}. % and discuss the insights which arise from this result. 
	
	We begin by defining for each $n \in \mySet{N}$, $\myEps_n \triangleq \frac{\lfloor n \cdot \myEps \rfloor}{n}$, for which we define the \ac{dt} zero-mean Gaussian process $\Wn[i] \triangleq \Wc \left(\frac{i \cdot \Tc}{\Td + \myEps_n} \right)$. 
	%It directly follows that the \ac{dt} process $\Wn[i]$  is {\em a zero-mean Gaussian process}. 
	Since $\myEps_n$ is a rational number, it follows from the discussion in Subsection \ref{subsec:Pre_Problem}   that $\Wn[i]$ is a \ac{wscs} process with period $\Td_n = \Td \cdot n + \lfloor n \cdot \myEps \rfloor$. The autocorrelation function of the \ac{dt} process $\Wn[i]$ is given by
	\vspace{-0.1cm}
	\begin{align}
	\Cwn[i,\tau]
	&= \E\Big\{\Wn[i+\tau]\cdot\Wn[i]\Big\}  
	%&= \E\left\{\Wc\left(\frac{(i+\tau)\Tc}{\Td + \myEps}\right)\Wc\left(\frac{i \cdot \Tc}{\Td + \myEps}\right)\right\}
	= \left( \Cwc\left(\frac{i \cdot\Tc}{\Td + \myEps_n} \right)\right)  \cdot \delta[\tau],
	\label{eqn:CSAutocorr}
	\vspace{-0.1cm}
	\end{align}
	and by its periodic nature, we have that $\Cwn[i,\tau] = \Cwn[i + \Td_n,\tau]$, for all $i, \tau \in \mySet{Z}$. 
	
	Next, we define a channel with input $X[i]$ and output $\Yn[i]$, whose input-output relationship for the transmission of $l \in \mySet{N}$ symbols is given by
	\vspace{-0.2cm}
	\begin{equation}
	\label{eqn:AsnycModel2}
	\Yn[i] =   X[i] + \Wn[i],  \qquad i \in \{1,2,\ldots,l\},
	\vspace{-0.1cm}
	\end{equation}
	where the channel input is subject to the constraint \eqref{eqn:AsyncConst1}.
	The channel  \eqref{eqn:AsnycModel2} corresponds to synchronous sampling, therefore, its capacity can  be obtained via Proposition \ref{cor:SynSample}. In particular, by letting $\Capacity_n$ denote the capacity of \eqref{eqn:AsnycModel2}, it holds that 
	\vspace{-0.2cm}
	\begin{equation}
	\label{eqn:SyncCap1}
	\Capacity_n = \bar{\Capacity}_{\lfloor n \cdot \myEps \rfloor, n},
	\vspace{-0.1cm}
	\end{equation}
	where $\bar{\Capacity}_{u,v}$ is given in \eqref{eqn:MainThm2}.
	Now, applying Theorem \ref{thm:plim}, we can  characterize the capacity of the asynchronously-sampled channel \eqref{eqn:AsnycModel1}, denoted with $\Capacity_\myEps$, as stated in the following Theorem~\ref{thm:AsycCap}:
	\begin{theorem}
		\label{thm:AsycCap}
		Consider the channel \eqref{eqn:AsnycModel1} subject to the power constraint \eqref{eqn:AsyncConst1}.
		%	 if the noise sequence satisfies
		%	
		%	\textcolor{red}{NIR COMMENT - The condition must be adapted for memoryless noise.}
		%	
		%    \begin{equation}
		%    \label{eqn:ConditionThmUnifConv}
		%        \Cwdel[k+1,0] - \big(\cvec_{\delta}[k+1]\big)^T \cdot \left(\corrmatd^{k} \right)^{-1} \cdot \cvec_{\delta}[k+1] > 1,
		%        \quad \forall \delta\in[0,1)
		%    \end{equation}
		%    then,
		Then, for any fixed irrational value of $\eps\in [0,1)$, $\Capacity_\myEps$ is obtained as
		\vspace{-0.2cm}
		\begin{equation}
		\label{eqn:capacity-lim}
		\Capacity_\myEps = \mathop{\lim \inf}\limits_{n \rightarrow \infty} \Capacity_n,
		\vspace{-0.1cm}
		\end{equation}
		where $\Capacity_n$ is given in \eqref{eqn:SyncCap1}. Furthermore, Gaussian inputs are optimal.
	\end{theorem}
	
	\begin{IEEEproof} 
		The  proof is given in Appendix \ref{app:Proof2}, and here we only provide a brief outline.  Recall that for arbitrary channels, capacity is given by the supremum over input distributions of the limit-inferior in probability (see Def. \ref{def:pliminf}) of the mutual information density rate \cite{Verdu:94}. Therefore, to prove the theorem, we  show that, when the distribution of the input to the channel  \eqref{eqn:AsnycModel2} converges uniformly to that of the input to \eqref{eqn:AsnycModel1}, then  the sequences of mutual information densities of the channels \eqref{eqn:AsnycModel1} and \eqref{eqn:AsnycModel2} satisfy the conditions of Theorem \ref{thm:plim}. This allows us to relate the achievable rates of the channels for input distributions which satisfy the uniform convergence requirement.
		We then use the fact that the optimal input to \eqref{eqn:AsnycModel2} is temporally independent and Gaussian \cite{Shlezinger:16b}, to identify its convergent subsequence, which leads to the proof of \eqref{eqn:capacity-lim}.   
	\end{IEEEproof}
	
	%-----------------------------------
	%	Discussion
	%-----------------------------------
	\vspace{-0.2cm}
	\subsection{Discussion}
	\label{subsec:Cap_Discussion}
	\vspace{-0.1cm}
	We note that unlike previous capacity characterizations derived for   memoryless time-varying channels, such as %\cite[Thm. 10]{Verdu:94} and 
	\cite[Remark 3.2.3]{Han:03}, our expression is {\em not restricted to finite alphabets} and {\em accounts for average power constraints}. Moreover, the fact that we focus specifically on asynchronously-sampled \ac{wscs} noise leads to an expression which is relatively simple to compute, as a limit inferior of a sequence with closed-form elements. 
	
	Note that the sampling period used for obtaining the channel in \eqref{eqn:AsnycModel2} is $\frac{\Tc}{\Td + \myEps_n}$ and that $\myEps_n$ is a convergent Cauchy sequence of rational numbers. Therefore,  the sequence $\{\Capacity_n\}_{n  \in \mySet{N}}$ represents the sequence of capacities of channels with additive sampled \ac{ct} \ac{wscs}  Gaussian noise, where for sufficiently large $n$, the sampling rate varies only slightly as $n$ increases.  
	%
	%It is therefore concluded from Thm. \ref{thm:AsycCap} that light variations in the sampling rate, due to which the scenario switches from asynchronous sampling to synchronous sampling, may lead to an increase in the capacity of the \ac{dt} channel. 
	%
	%We also emphasize that, in general, there is no guarantee that the sequence of capacities $\{\Capacity_n\}_{n \in \mySet{N}}$ is convergent. 
	%The fact that $\Capacity_\myEps$ equals the limit inferior of the sequence  $\{\Capacity_n\}_{n \in \mySet{N}}$ implies it may have subsequences with larger limits \cite[Ch. 3]{Rudin:76}. 
	
	For small values of $n$, Theorem \ref{thm:AsycCap} does not indicate whether $\Capacity_n$ is larger than or smaller than $\Capacity_\myEps$.  However, using Theorem \ref{thm:AsycCap}, these values can be computed numerically, and  in Section \ref{sec:Simulations} we demonstrate that the difference between  $\Capacity_n$ and $\Capacity_\myEps$ can be notable for small $n$. 
	Combining this with that fact that when the sampling period is sufficiently small, i.e., $\Td \gg 1$, then $\Capacity_n$ and $\Capacity_\myEps$ correspond to channels sampled at roughly the same sampling rate for each $n \in \mySet{N}$, we conclude that in some scenarios, relatively small variations in the sampling rate may result in relatively large variations in capacity. Theorem \ref{thm:AsycCap} allows to precisely compute these variations, and consequently to properly set the sampling rate such that capacity is maximized.
	
	\label{txt:Insight2}
	Another insight which arises from Theorem \ref{thm:AsycCap}, compared to the synchronous sampling scenario in Proposition \ref{cor:SynSample}, is related to the dependence of capacity on the sampling time offset: 
	\textcolor{NewColor}{Note that for synchronous sampling in which the numerator and denominator of $\myEps$ are relatively small integers, e.g., for $\myEps_n$ with relatively small $n$, replacing the \ac{ct} variance $\Cwc(t)$ with its time-shifted version $\Cwc(t-\phi)$, results in a different variance function of the sampled \ac{dt} noise. Consequently, the variance of the sampled noise depends on $\phi$, as also numerically illustrated in Fig. \ref{fig:SampledVar}, and hence, capacity of the \ac{dt} channel can vary, possibly notably, between different values of the sampling time offset $\phi$.} However, as $n$ increases, the number of sampling points within a period of the \ac{ct} variance increases, and consequently the difference between the sets of values of the respective sampled variances within a single period of the \ac{ct} variance obtained with different time offsets decreases as $n$ increases. For sufficiently large $n$, they become approximately identical up to a permutation due to the time shift, implying that, by Theorem \ref{thm:AsycCap}, capacity with asynchronous sampling is invariant to sampling offsets. This behavior is also observed in the numerical study, presented in the following section. 
	
	\label{txt:NewDisc}
	\color{NewColor}
	Finally, we note that the aforementioned insights which arise from our capacity analysis, i.e., the dependence of capacity on the sampling rate and the sampling time offset, may not reflect in systems operating with short blocklenghts,  due to the asymptotic nature of capacity analysis. For example, when the duration of a codeword is shorter than the period of the statistics of the interfering signal, performance is clearly affected by sampling time offset, regardless of whether sampling is synchronous or asynchronous, as the codeword may be subject to different noise power levels for different offsets. These limitations of the insights which arise from our capacity analysis stem from the fact that the fundamental performance limit of capacity for a given channel requires asymptotically large blocklengths to facilitate decreasing the probability of error. For the considered channel model, which represents practical scenarios of communications in the presence of interference, transmissions of short blocks and large blocks may undergo channels with substantially different characteristics, due to the non-stationary nature of the channel. Hence, the insights associated with the capacity expression in Theorem \ref{thm:AsycCap} may not reflect the behavior of systems communicating with short blocklengths.

	The insights discussed above are directly relevant to communications in which the codeword duration is sufficiently larger than the period of the statistics of the interference. In such scenarios, each codeword spans over a large number of periods of the statistics of $W_c(t)$, and the properties of capacity reflect in systems with finite blocklengths.  Here, these  insights can be translated into practical code design guidelines. 
	For example, since capacity of synchronously sampled channels depends on the sampling offset, it may seem attractive to design the communications scheme based on the sampling time offset which maximizes capacity. The insight that this property does not hold for asynchronous sampling indicates that such an approach may result in outage, i.e., using code rates higher than capacity. This follows since hardware impairments and limitations of symbol rate estimation result in non-intentional jitter in the sampling rate clock, which may in turn cause a system designed with synchronous sampling to experience an asynchronously sampled received signal. Consequently, we suggest to design coding schemes with rates up to the asynchronous sampling capacity, even when the system is designed to sample synchronously.
	\color{black}

	\vspace{-0.25cm}
	\section{Numerical Examples and Discussion}
	\label{sec:Simulations}
	\vspace{-0.1cm}
	In this section we numerically evaluate the capacity of \ac{dt} memoryless channels with sampled \ac{wscs} Gaussian noise. Since the capacity of such channels with asynchronous sampling, denoted $\Capacity_{\myEps}$, was derived in Theorem \ref{thm:AsycCap} to be equal to the limit inferior of a sequence of capacities of \ac{dt} memoryless channels with additive \ac{wscs} Gaussian noise, denoted $\{\Capacity_n\}_{n \in \mySet{N}}$, we first empirically study the convergence properties of $\{\Capacity_n\}_{n \in \mySet{N}}$ in Subsection \ref{subsec:Sim_capCS}. Then, in Subsection \ref{subsec:Sim_capSYNC} we study how variations in the sampling rate and different sampling time offsets affect the capacity of \ac{dt} channels with additive noise corresponding to the sampling of a \ac{ct} \ac{wscs} Gaussian noise. 
	
	Let $\Pi_{\DC,\Trise}(t)$ be a periodic continuous pulse  function with rise / fall time $\Trise = 0.01$, duty cycle $\DC \in [0,0.98]$, and period of $1$, i.e., $ \Pi_{\DC,\Trise}(t+1) = \Pi_{\DC,\Trise}(t)$ for all $t \in \mySet{R}$. Specifically, for $t \in [0,1)$ the function $\Pi_{\DC,\Trise}(t)$ is given by 
	%
	%In the following, we 
	%we set the input power constraint to $\PCst = 1$, and the period of the \ac{ct} \ac{wscs} process $\Wc(t)$ to $\Tc = 5$ microseconds. We consider the time-varying variance of the noise to be given by a periodic continuous pulse  function with rise / fall time $\Trise = 0.01$ and duty cycle $\DC \in [0,0.98]$. 
	%Specifically, the periodic continuous pulse function satisfies $ \Pi_{\DC,\Trise}(t+1) = \Pi_{\DC,\Trise}(t)$ for all $t \in \mySet{R}$, and for $t \in [0,1)$ is given by
	\ifFullVersion 
	\begin{equation}
	\label{eqn:PerWaveFunc}
	\Pi_{\DC,\Trise}(t) = \begin{cases}
	\frac{t}{\Trise} & t \in [0,\Trise] \\
	1	&				t \in (\Trise, \DC+\Trise ) \\
	1 - \frac{t-\DC-\Trise}{\Trise} & t \in[\DC +\Trise, \DC +2\cdot \Trise] \\
	0 & t\in (\DC +2\cdot \Trise, 1).
	\end{cases}
	\end{equation}
	\else
	\vspace{-0.1cm}
	\begin{align}
	\Pi_{\DC,\Trise}(t) 
	&= 
	\frac{t}{\Trise} \Ind\left( t \in [0,\Trise]\right)  + \Ind\left( 	t \in (\Trise, \DC+\Trise ) \right)  \notag \\
	&+ \left( 1 - \frac{t-\DC-\Trise}{\Trise}\right) \Ind\left(  t \in[\DC +\Trise, \DC +2\cdot \Trise] \right).  
	\label{eqn:PerWaveFunc}
	\vspace{-0.1cm}
	\end{align}
	\fi
	In the following we consider the time-varying variance of the noise, $\Cwc(t)$, to be   a periodic and continuous pulse  function. To formulate $\Cwc(t)$, 
	let $\phi \in [0,1)$ represent the offset between the first sample and the rise start time of the periodic continuous pulse function, corresponding to the sampling time offset normalized to the period $\Tc$. The variance of  $\Wc(t)$ is given by
	\vspace{-0.1cm}
	\begin{equation}
	\label{eqn:WcVar}
	\Cwc(t) = 0.2 + 4.8 \cdot \Pi_{\DC,\Trise}\left(\frac{t}{\Tc} - \phi\right),
	\vspace{-0.1cm}
	\end{equation}
	with period of $\Tc = 5$ $\mu$secs.
	Such periodic variance profiles arise, e.g., when the digitally-modulated interfering signal obeys a \ac{tdma} protocol, see \cite[Ch. 14.2]{Goldsmith:05}.  In such cases, when the interfering signal is present, the noise variance is high, while when it is absent, only weak background noise impairs communications.
	
	%-----------------------------------
	%	Convergence Properties of $\{\Capacity_n\}_{n \in \mySet{N}}$
	%-----------------------------------
	\vspace{-0.2cm}
	\subsection{Convergence Properties of $\{\Capacity_n\}_{n \in \mySet{N}}$}
	\label{subsec:Sim_capCS}
\begin{figure}
		\centering
		\begin{minipage}{0.45\textwidth}
			\centering
			{\includegraphics[width=1.1\linewidth]{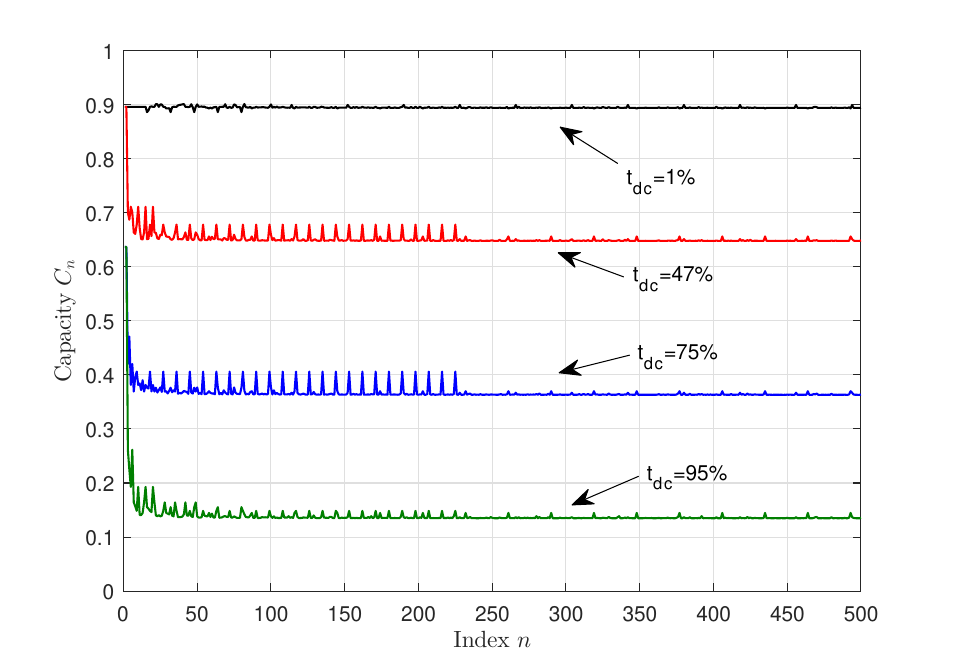}}
			\vspace{-0.8cm}
			\caption{$\Capacity_n$ versus $n$ for offset $\phi=0$.
			}
			\label{fig:CnDc_Offset0}		
		\end{minipage}
		$\quad$
		\begin{minipage}{0.45\textwidth}
			\centering
			{\includegraphics[width=1.1\linewidth]{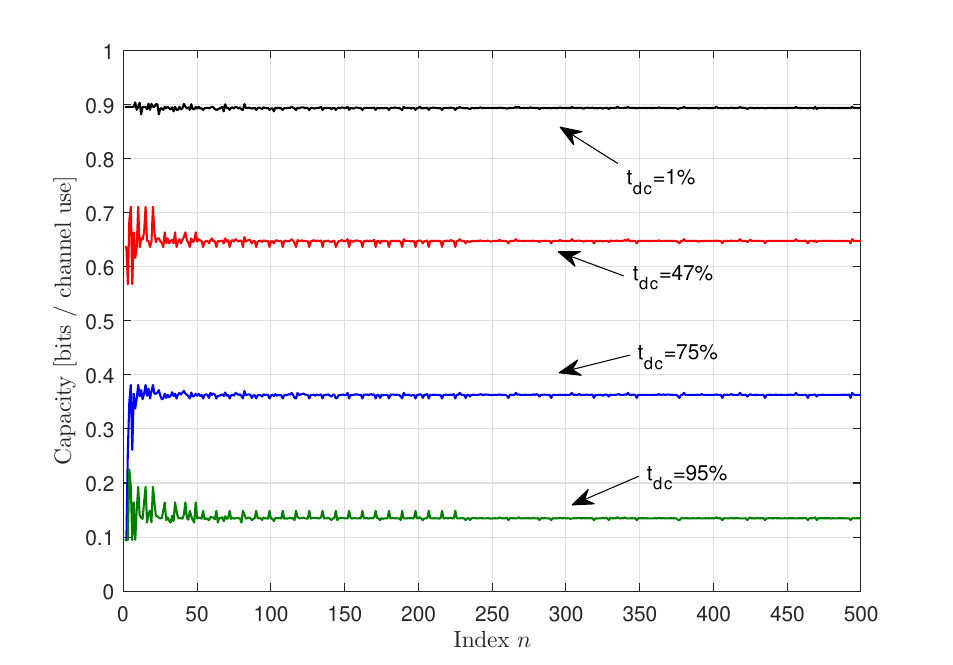}}
			\vspace{-0.8cm}
			\caption{$\Capacity_n$ versus $n$ for offset $\phi=\frac{1}{4}$.
			}
			\label{fig:CnDc_Offset05}
		\end{minipage}
		\vspace{-0.8cm}
	\end{figure}

	\vspace{-0.1cm}
	By Theorem \ref{thm:AsycCap}, capacity with asynchronous sampling $\Capacity_{\myEps}$ is the limit inferior of the sequence $\{\Capacity_n\}_{n \in \mySet{N}}$. 
	Hence, in the following we examine the behavior of the sequence of capacities  $\{\Capacity_n\}_{n \in \mySet{N}}$ defined in \eqref{eqn:SyncCap1} as $n$ increases. 
	In the first study, we fix the input power constraint to $\PCst = 1$ and set $\myEps = \frac{\pi}{7}$, $\Td = 2$. For this setting, we evaluate the capacities $\Capacity_n$ via \eqref{eqn:MainThm2} with the sampling period given by $\Tsamp(n) = \frac{\Tc}{\Td + \myEps_n}$ where  $\myEps_n = \frac{\lfloor n \cdot \myEps \rfloor}{n}$ is a rational number which approaches $\myEps$ as $n \rightarrow \infty$. 
	\textcolor{NewColor}{The reason for selecting a relatively small value of $\Td = 2$ is that for this value the variations in $\Tsamp(n)$ as $n$ changes are more pronounced than with larger values of $\Td$, allowing to better visualize the variation properties of the sequence $\{\Capacity_n\}_{n \in \mySet{N}}$.}
	%Specifically, for every $n \in [1,500]$  we compute the capacity $\Capacity_n$. 
	%Accordingly, the ratio between the period of the \ac{ct} \ac{wscs} noise and the sampling interval is set to $\frac{\Tc}{\Tsamp(n)} = \Td + \myEps_n$. 
	Figs. \ref{fig:CnDc_Offset0} and \ref{fig:CnDc_Offset05} present $\Capacity_n$ for $n \in [1,500]$ and for duty cycles $\DC = \{1,47,75,95\} \%$, where  in Fig. \ref{fig:CnDc_Offset0} there is no sampling time offset, i.e., $\phi = 0$, and in Fig. \ref{fig:CnDc_Offset05} the sampling time offset is set to $\phi = \frac{1}{4}$. 
	We observe in both figures that capacity values are larger for smaller $\DC$. This can be explained by noting that the time-averaged noise variance decreases as $\DC$ decreases. 
	Furthermore,   for all considered configurations, $\Capacity_n$ exhibits notable variations for small values of $n$, i.e., when an  increase in $n$ induces a relatively significant change in the sampling frequency. Comparing Fig. \ref{fig:CnDc_Offset0} and Fig. \ref{fig:CnDc_Offset05}, we conclude that the nature of these variations depends on the sampling offset $\phi$. For example, for $\DC = 95 \%$ at $n \in [5, 15]$, then  for $\phi =0$ capacity varies in the range $[0.1407,0.2615]$ bits per channel use, while  for $\phi =\frac{1}{4}$  capacity varies in the range $[0.0946,0.1929]$  bits per channel use. However, as $n$ increases beyond $250$, the variations in $\Capacity_n$ become smaller and are less dependent on the sampling offset, as the resulting values of $\Capacity_n$ are approximately in the same range in both Figs. \ref{fig:CnDc_Offset0} and \ref{fig:CnDc_Offset05} for $n \ge 250$. 
	%, yet, empirical convergence is not illustrated. For example, observing the behavior of $\Capacity_n$ with $\DC = 47\%$, $\phi = \frac{1}{2}$, and $n > 250$, we note that $\Capacity_n$ still varies in the range $[1.387,1.408]$. Consequently, the capacity of the asynchronously-sampled channel here is $\Capacity_{\myEps} = \mathop{\lim \inf}\limits_{n \rightarrow \infty} \Capacity_n = 1.387$ bits per channel use, and thus, by adding a negligible variation to the sampling rate, the resulting capacity can increase to $1.408$ bits per channel use. 
	% Furthermore ,while the sampling offset was shown to have a notable effect on capacity for small values of $n$,  as $n$ increases, the resulting values of $\Capacity_n$ are approximately the same in both Figs. \ref{fig:CnDc_Offset0} and \ref{fig:CnDc_Offset05}. 
	These variations are in agreement with the discussion following Theorem \ref{thm:AsycCap} in Subsection \ref{subsec:Cap_Main}, where it was noted that capacity with synchronous sampling depends on the sampling offset, yet 
	%   When $\Tsamp(n)$ is set such that the \ac{dt} period $\Td_n$ is alternatively small, i.e., when the denominator of $\myEps_n$ is a small integer, then sampling of the variance at $\phi = 0$ and at $\phi = \frac{1}{4}$ can result in substantially different sampled variances with different average values. These differences become smaller as $\myEpsn$ increases towards $\myEps$ as  the resulting \ac{dt} noise essentially captures the complete variance profile of the \ac{ct} noise,  and capacity becomes less sensitive to sampling offset.  
	%  Consequently, 
	when the sampling rate approaches being asynchronous, the effect of sampling offset on capacity becomes negligible. 
	%This can explained by noting that for small values of $n$, the value of the equivalent \ac{dt} period $\Td_n$ is smaller, and thus the resulting \ac{dt} exhibits less different values of the variance $\sigma_{w_{n}}[i]$, taken as sampled from $\Cwc(t)$ in the range $t \in [0, \Tc]$. Consequently, introducing sampling offset results in different variance profiles. However, as the sampling rate approaches being asynchronous, the equivalent period $\Td_n$ increases, and thus the resulting \ac{dt} noise essentially captures the complete variance profile of the \ac{ct} noise,  and capacity becomes less sensitive to sampling offset.  

	%-----------------------------------
	%	The Dependence of Capacity on Sampling Rate
	%-----------------------------------
	\vspace{-0.2cm}
	\subsection{The Dependence of Capacity on the Sampling Rate}
	\label{subsec:Sim_capSYNC}
\begin{figure}
		\centering
		\begin{minipage}{0.45\textwidth}
			\centering
			{\includegraphics[width=1.1\linewidth]{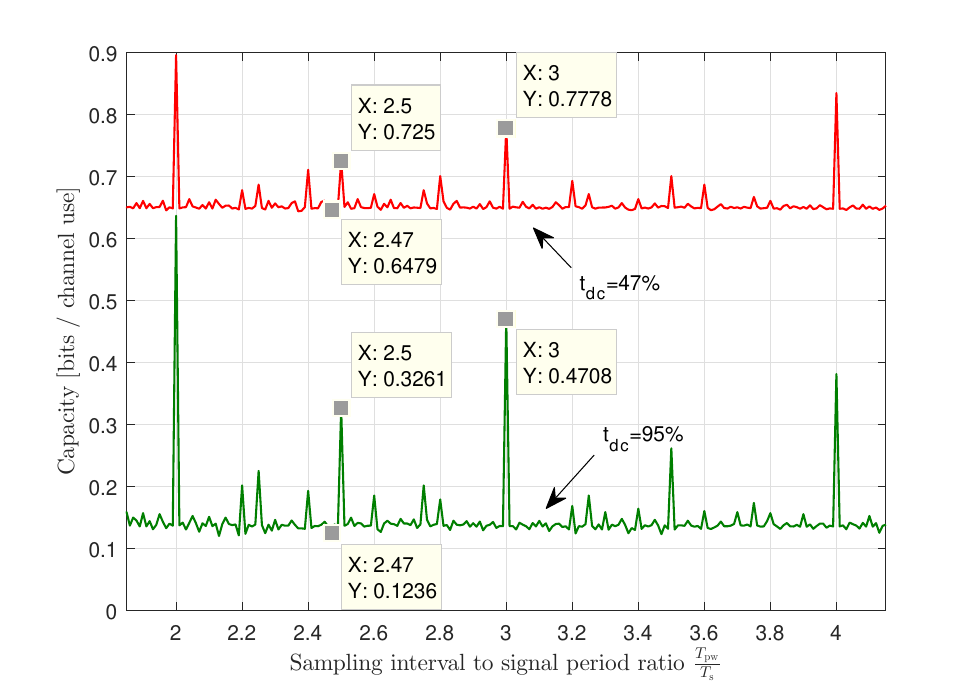}}
			\vspace{-0.8cm}
			\caption{$\Capacity_n$ versus $\frac{\Tc}{\Tsamp}$ for offset $\phi=0$.
			}
			\label{fig:CnTs_Offset0}		
		\end{minipage}
		$\quad$
		\begin{minipage}{0.45\textwidth}
			\centering
			{\includegraphics[width=1.1\linewidth]{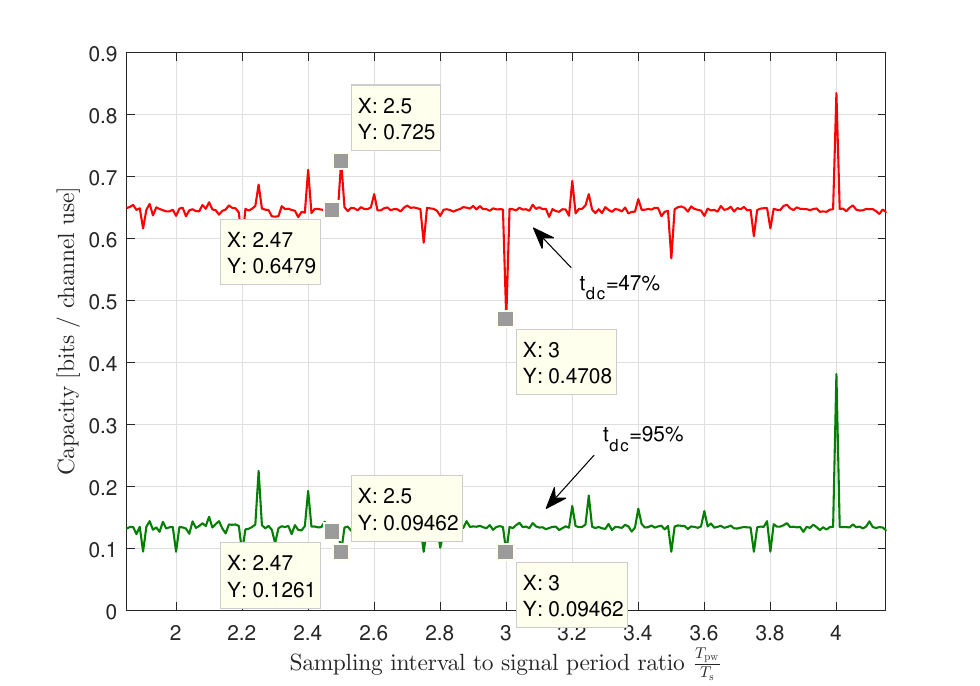}}
			\vspace{-0.8cm}
			\caption{$\Capacity_n$ versus $\frac{\Tc}{\Tsamp}$  for offset $\phi=\frac{1}{4}$.
			}
			\label{fig:CnTs_Offset05}
		\end{minipage}
		\vspace{-0.8cm}
	\end{figure}

	\vspace{-0.1cm}
	Next, we numerically evaluate the dependence of the capacity of sampled memoryless channels with  additive \ac{wscs} Gaussian noise on the specific selection of the sampling interval $\Tsamp$. To that aim, we  first set the transmit power constraint to $\PCst = 1$ and set the duty cycle in the  noise model  \eqref{eqn:WcVar} to $\DC \in \{47, 95\} \%$. In Figs. \ref{fig:CnTs_Offset0}-\ref{fig:CnTs_Offset05} we depict the numerically computed capacity values for sampling intervals satisfying $2 < \frac{\Tc}{\Tsamp} < 4 $ with sampling time offsets $\phi = 0 $ and $\phi = \frac{1}{4}$, respectively.  
	\label{txt:ObsFigs}
	Observing Figs. \ref{fig:CnTs_Offset0}-\ref{fig:CnTs_Offset05} we note when $\frac{\Tc}{\Tsamp}$ has a fractional part with a relatively small integer denominator, notable variations in capacity are observed, which depend on the sampling offset. \textcolor{NewColor}{The denominator of the fractional part of $\frac{\Tc}{\Tsamp}$ determines the number of periods of the \ac{ct} noise which correspond to a single period of the \ac{dt} sampled process, hence, a smaller denominator results in more pronounced periodicity while a larger denominator resembles asynchronous sampling scenarios.} It follows that, when $\frac{\Tc}{\Tsamp}$ approaches an irrational number, the period of the sampled variance function becomes very long, and consequently, capacity is a constant which is independent of the sampling offset. For example, for $\frac{\Tc}{\Tsamp} = 3$ and $\DC = 47\%$, then for sampling time offset $\phi = 0$ capacity is as high as  $0.7778$ bits per channel use, while for sampling offset $\phi = \frac{1}{4}$  capacity is as low as $0.4708$ bits per channel use. However, when approaching asynchronous sampling, capacity is fixed at approximately $0.64$ bits per channel use for all considered values of $\frac{\Tc}{\Tsamp}$ and both offsets of $\phi$. 
	This again follows as when the denominator of the fractional part of $\frac{\Tc}{\Tsamp}$ increases, the \ac{dt} period of the sampled variance increases and practically captures the entire set of values of the \ac{ct} variance regardless of the sampling offset. 
	\textcolor{NewColor}{It is emphasized that capacity is not continuous in $\frac{\Tc}{\Tsamp}$, and notable singularities are observed for synchronous sampling when the fractional part of $\frac{\Tc}{\Tsamp}$ has a relatively small denominator.}
	We conjecture that the fact that asynchronous sampling captures the entire set of values of the \ac{ct} variance implies that it represents the capacity of the analog channel, which does not depend on the specific sampling rate and offset. We leave the investigation of this conjecture to future work. 
	
	%
	%This again follows since when  the sampling period approaches the asynchronous sampling, the period of the sampled \ac{dt} channel increases, and thus the variance of the sampled noise $\Sigwn[i]$ contains more discrete samples within the period of the \ac{ct} variance $\Cwc(t)$.
	%
	%It is emphasized for the scenario considered here, and more generally for $\Td \ge 2$, the variance of the sampled noise $\Sigwn[i]$ is obtained by sampling the variance of the \ac{ct} noise $\Cwc(t)$ above Nyquist rate. However, as shown here, the fact that $\Cwc(t)$ can be perfectly restored from $\Sigwn[i]$ for all considered sampling rates does not imply that the resulting capacity is independent of the sampling rate. 
	%
	
	Figs. \ref{fig:CnTs_Offset0}-\ref{fig:CnTs_Offset05} demonstrate how minor variations in the sampling rate can result in significant changes in capacity. For example, for sampling offset $\phi = 0$ it is observed in Fig.  \ref{fig:CnTs_Offset0} that when the sampling rate switches from $\Tsamp = 2.47 \cdot \Tc$ to $\Tsamp = 2.5 \cdot \Tc$, i.e., the sampling rate switches from being nearly asynchronous to being synchronous,  capacity increases from  $0.647$  bits per channel use to $0.725$  bits per channel use for $\DC=47\%$, and increases from $0.123$  bits per channel use to $0.326$  bits per channel use for $\DC=95\%$.

\begin{figure}
		\centering
		\begin{minipage}{0.45\textwidth}
			\centering
			{\includegraphics[width=1.1\linewidth]{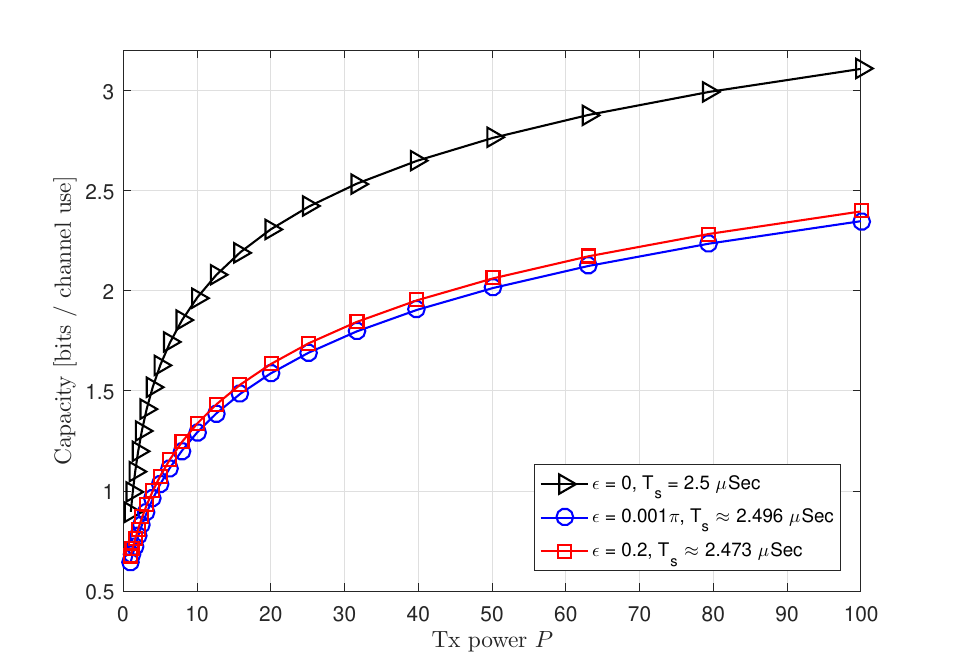}}
			\vspace{-0.8cm}
			\caption{$\Capacity$ versus $\PCst$ for $\DC =47\%$ and  offset $\phi=0$.
			}
			\label{fig:CnP_Offset0}		
		\end{minipage}
		$\quad$
		\begin{minipage}{0.45\textwidth}
			\centering
			{\includegraphics[width=1.1\linewidth]{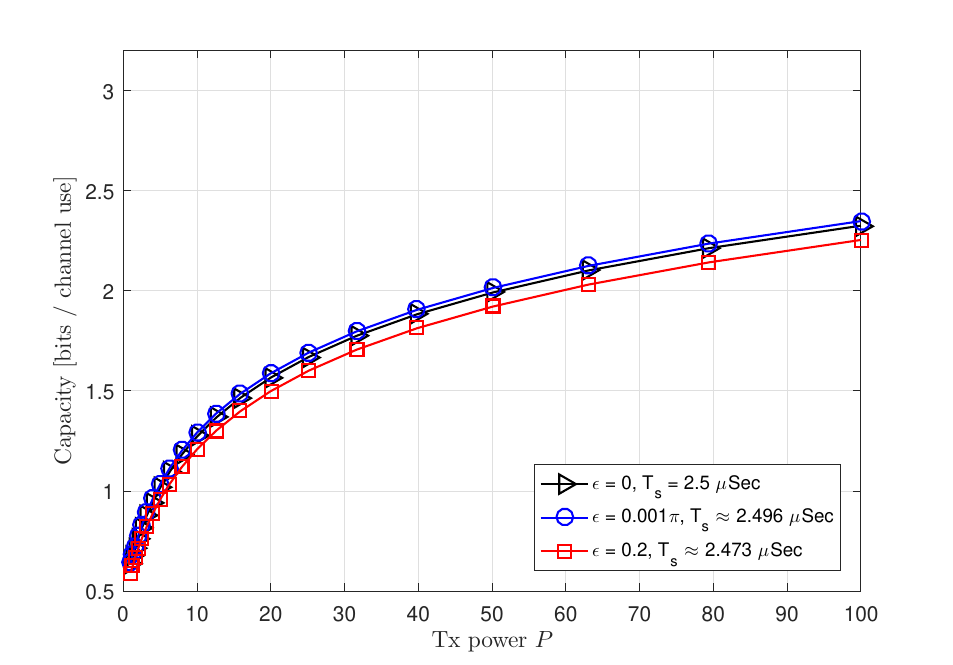}}
			\vspace{-0.8cm}
			\caption{$\Capacity$ versus  $\PCst$ for $\DC =47\%$ and  offset $\phi=\frac{1}{4}$.
			}
			\label{fig:CnP_Offset05}
		\end{minipage}
		\vspace{-0.8cm}
	\end{figure}

	To further demonstrate the  assertion that minor variations in the sampling interval can lead to notable variations in capacity, we numerically evaluate the capacity versus the  transmit power constraint $\PCst$ for different values of synchronization mismatch $\myEps$. In particular, we set $\DC = 47\%$, fix $\Td = 2$ and evaluate the capacity versus  $\PCst \in [1,100]$  for $\myEps \in \{0, \frac{\pi}{1000}, 0.2\}$. Note that only $\myEps = \frac{\pi}{1000}$ corresponds to asynchronous sampling, and that its sampling interval is approximately $2.496$ $\mu$secs, namely, a negligible variation from the sampling intervals corresponding to $\myEps \in \{0, 0.2\}$, which are  $2.5$ $\mu$secs and $2.473$ $\mu$secs, respectively. 
	The results of this numerical evaluation are depicted Figs. \ref{fig:CnP_Offset0}-\ref{fig:CnP_Offset05} for sampling offsets $\phi = 0$ and $\phi = \frac{1}{4}$, respectively. Observing Figs.  \ref{fig:CnP_Offset0}-\ref{fig:CnP_Offset05}, we note that a change of less than $0.2\%$ in the sampling interval, corresponding to the synchronization mismatch $\myEps$ changing from $\myEps = 0$ to $\myEps = \frac{\pi}{1000}$, has a notable effect on capacity: At sampling offset $\phi = 0$ such a change results in a dramatic decrease in capacity, e.g., at $\PCst = 10$ capacity decreases by roughly $30 \%$. 
	For $\phi = \frac{1}{4}$ such a change in the sampling rate slightly increases capacity. A similar behavior of a much smaller magnitude is observed comparing the curves corresponding to $\myEps = 0.2$ and $\myEps = \frac{\pi}{1000}$. It is also noted that the capacity curve for the asynchronous sampling mismatch $\myEps = \frac{\pi}{1000}$ is identical for both sampling offsets, indicating once again that when the sampling is asynchronous, capacity is invariant to sampling offsets.

	In summary, the results presented above demonstrate that for \ac{dt} memoryless channels with additive sampled \ac{ct} \ac{wscs} noise, capacity can vary significantly between different sampling rates and sampling offsets. In particular, it is shown that when the sampling rate is not synchronized with the period of the  \ac{wscs} noise, {\em differently} from the synchronized sampling case, capacity is {\em not} sensitive to sampling time offsets. It is also shown that capacity can change significantly due to minor variations in the sampling rate, especially when the variations cause the sampling rate to switch between synchronous and asynchronous sampling.

	\vspace{-0.3cm}
	\section{Conclusions}
	\label{sec:Conclusions}
	\vspace{-0.1cm}
	In this work we characterized the capacity of \ac{dt} memoryless communication channels with additive sampled \ac{wscs} Gaussian noise, a model which represents important  scenarios, including interference-limited communications and cognitive communications. This model can be analyzed using common information-theoretic tools, e.g., methods based on the law of large numbers, only when the sampling rate is synchronized with the period of the noise statistics. To characterize its capacity with asynchronous sampling, we first derived a new relationship between the information spectrum quantities for uniformly convergent sequences of \acp{rv}. We then used this relationship to express the capacity of  asynchronously-sampled  memoryless additive \ac{wscs} Gaussian noise channels as the limit of a sequence of capacities of synchronously-sampled memoryless additive \ac{wscs} Gaussian noise channels.  
	Our numerical analysis demonstrates how  variations in the sampling rate, which switch the resulting model from  synchronous sampling to  asynchronous sampling, can significantly change capacity. 
	In particular, it was shown that a small change of $0.2\%$ in the sampling rate caused a decrease of $30 \%$ in capacity. 
	Our characterization can be used, for example, to properly determine the sampling period in interference-limited communications such that capacity is maximized.

	%----------------------------------------------------------------------------------------
	%	APPENDICES
	%----------------------------------------------------------------------------------------
	%%\newpage
	%
	%%\renewcommand{\thesubsection}{\thesection.\alph{subsection}}
	%\begin{appendices}
	%\numberwithin{proposition}{subsection} 
	%\numberwithin{lemma}{subsection} 
	%\numberwithin{corollary}{subsection} 
	%\numberwithin{remark}{subsection} 
	%\numberwithin{equation}{subsection}	
	
	%\newpage
	\numberwithin{proposition}{section} 
	\numberwithin{lemma}{section} 
	\numberwithin{corollary}{section} 
	\numberwithin{remark}{section} 
	\numberwithin{equation}{section}

	\vspace{-0.2cm}
	\begin{appendices}
		
		%----------------------------------------------------------------------------------------
		%	APPENDIX A
		%----------------------------------------------------------------------------------------
		\vspace{-0.2cm}
		\section{Proof of Theorem \ref{thm:plim}}
		\label{app:Proof1}
		\vspace{-0.1cm}
		In the following we prove only \eqref{eqn:plima}, as the proof of \eqref{eqn:plimb} is obtained by following similar arguments. 
\ifFullVersion
\else
		A more detailed version of the proof presented in this appendix can be found in \cite{Shlezinger:19}. 
\fi
		First, we note that  Def. \ref{def:pliminf} can be written as 
		\vspace{-0.2cm}
		\begin{align}
		\hspace{-0.2cm}
		{\rm p-}\mathop{\lim \inf}\limits_{k \rightarrow \infty} \zk{k} 
		%&=   \sup\left\{\alpha \in \mySet{R} \big|\mathop{\lim}\limits_{k \rightarrow \infty} \Pr \left(\zk{k} < \alpha \right) =0   \right\} \notag \\
		\!\stackrel{(a)}{=} \!  \sup\left\{\!\alpha\! \in\! \mySet{R} \big|\mathop{\lim \sup}\limits_{k \rightarrow \infty} \Pr \left(\zk{k}\!< \!\alpha \right) \!\!=0   \right\} 
		%\notag \\
		%&
		\!=\!   \sup\left\{\!\alpha \!\in \!\mySet{R} \big|\mathop{\lim \sup}\limits_{k \rightarrow \infty} \Fkeps{k}(\alpha) \!=\!0   \right\}.
		\label{eqn:pliminfEq}
		\vspace{-0.1cm}
		\end{align}
		For $(a)$ 
		\ifFullVersion
		we first note that for a given $\alpha \in \mySet{R}$, the non-negative sequence $\{\Pr \left(\zk{k} < \alpha \right)\}_{k \in \mySet{N}}$ may not converge. Nonetheless, 
		%if for some $\alpha \in \mySet{R}$ this sequence converges to zero, then its limit superior converges to zero. To see this equivalence, we note that 
		if for a given $\alpha$ it holds that $\mathop{\lim}\limits_{k \rightarrow \infty} \Pr \left(\zk{k} < \alpha \right) =0$, then for this value of $\alpha$ the limit exists and thus $\mathop{\lim \sup}\limits_{k \rightarrow \infty} \Pr \left(\zk{k} < \alpha \right) =0  $. On the other hand, if for some $\alpha \in \mySet{R}$ it holds that $\mathop{\lim \sup}\limits_{k \rightarrow \infty} \Pr \left(\zk{k} < \alpha \right) =0  $, then, since  $\{\Pr \left(\zk{k} < \alpha \right)\}_{k \in \mySet{N}}$ is non negative, it holds that  $\mathop{\lim \inf}\limits_{k \rightarrow \infty} \Pr \left(\zk{k} < \alpha \right)  \ge 0  $. 
		Therefore by \cite[Thm. 3.17]{Rudin:76} the limit exists and is equal to $\mathop{\lim}\limits_{k \rightarrow \infty} \Pr \left(\zk{k} < \alpha \right) =0$.  We 
		\else
		we note that $\{\Pr \left(\zk{k}\! < \!\alpha \right)\}_{k \in \mySet{N}}$ are all non-negative, thus, for any $\alpha \in \mySet{R}$ for which it holds that $\mathop{\lim \sup}\limits_{k \rightarrow \infty} \Pr \left(\zk{k}\! < \!\alpha \right) =0$, it follows from \cite[Thm. 3.17]{Rudin:76} that $\mathop{\lim}\limits_{k \rightarrow \infty} \Pr \left(\zk{k} < \alpha \right)$ exists and is equal to $0$,  
		$\mathop{\lim}\limits_{k \rightarrow \infty} \Pr \left(\zk{k} < \alpha \right) =0$.  We  also
		\fi
		%
		%\ifFullVersion 
		note that since $\Fkeps{k}(\alpha) \in [0,1]$, then,   $\mathop{\lim \sup}\limits_{k \rightarrow \infty} \Fkeps{k}(\alpha) $ exists and is finite 
		\cite[Thm. 3.17]{Rudin:76}, even if  $\mathop{\lim}\limits_{k \rightarrow \infty} \Fkeps{k}(\alpha) $ does not exist. 
		%\else 
		%  \cite[Thm. 3.17]{Rudin:76}. 
		%  \fi

		The proof of  Theorem \ref{thm:plim} uses the following lemma:
		\begin{lemma}
			\label{lem:AidLemma2}
			Given assumption \ref{itm:assm2},  for all $\alpha \in \mySet{R}$ it holds that
			\vspace{-0.1cm}
			\begin{equation}
			\label{eqn:AidLemma2}
			\mathop{\lim \sup}\limits_{k \rightarrow \infty} \Fkeps{k}(\alpha) = \mathop{\lim }\limits_{n \rightarrow \infty}\mathop{\lim \sup}\limits_{k \rightarrow \infty} \Fkn{k}{n}(\alpha).
			\vspace{-0.1cm}
			\end{equation}  
		\end{lemma}
		
		\begin{IEEEproof}
			To prove the lemma we first show that $\mathop{\lim \sup}\limits_{k \rightarrow \infty} \Fkeps{k}(\alpha) \le \mathop{\lim }\limits_{n \rightarrow \infty}\mathop{\lim \sup}\limits_{k \rightarrow \infty} \Fkn{k}{n}(\alpha)$, and then we show 	$\mathop{\lim \sup}\limits_{k \rightarrow \infty} \Fkeps{k}(\alpha) \ge \mathop{\lim }\limits_{n \rightarrow \infty}\mathop{\lim \sup}\limits_{k \rightarrow \infty} \Fkn{k}{n}(\alpha)$.
			Recall that by \ref{itm:assm2}, for all $\alpha \in \mySet{R}$ and $k \in \mySet{N}$, $ \Fkn{k}{n}(\alpha)$ converges as $n \rightarrow \infty$ to $\Fkeps{k}(\alpha)$, uniformly over $k$ and $\alpha$, i.e., for all $\eta >  0$ there exists $n_0(\eta) \in \mySet{N}$, $k_0\big(n_0(\eta), \eta\big) \in \mySet{N}$ such that for every $n > n_0(\eta)$, $\alpha \in \mySet{R}$ and  $k  > k_0\big(n_0(\eta), \eta\big)$, it holds that $\big|\Fkn{k}{n}(\alpha) -  \Fkeps{k}(\alpha)\big| < \eta$. 
			Consequently, for every subsequence $k_1, k_2, \ldots$ such that $\mathop{\lim}\limits_{l \rightarrow \infty} \Fkn{k_l}{n}(\alpha)$  exists for any $n > n_0(\eta)$, it follows from \cite[Thm. 7.11]{Rudin:76} that, as the convergence over $k$ is uniform, the limits over $n$ and $l$ are interchangeable:
			\vspace{-0.4cm}
			\begin{equation}
			\mathop{\lim}\limits_{n \rightarrow \infty}\mathop{\lim}\limits_{l \rightarrow \infty} \Fkn{k_l}{n}(\alpha) 
			=\mathop{\lim}\limits_{l \rightarrow \infty}  \mathop{\lim}\limits_{n \rightarrow \infty}\Fkn{k_l}{n}(\alpha)
			% \notag \\
			%&
			= \mathop{\lim}\limits_{l \rightarrow \infty}  \Fkeps{k_l}(\alpha). 
			\label{eqn:AidLemma21}
			\vspace{-0.1cm}
			\end{equation}	
			The existence of such a convergent subsequence is guaranteed by the Bolzano-Weierstrass Theorem \cite[Thm. 2.42]{Rudin:76} as $\Fkn{k}{n}(\alpha) \in [0,1]$.    
			
			From 
			\ifFullVersion
			the properties of the limit superior 
			\fi 
			\cite[Thm. 3.17]{Rudin:76} if follows that there exists a subsequence of $\big\{\Fkeps{k}(\alpha)\big\}_{k \in \mySet{N}}$, denoted $\big\{\Fkeps{k_m}(\alpha)\big\}_{m \in \mySet{N}}$, such that $\mathop{\lim}\limits_{m \rightarrow \infty}  \Fkeps{k_m}(\alpha) = \mathop{\lim\sup}\limits_{k \rightarrow \infty}  \Fkeps{k}(\alpha)$. Consequently, 
			\ifFullVersion
			\begin{align}
			\mathop{\lim\sup}\limits_{k \rightarrow \infty}  \Fkeps{k}(\alpha) 
			&= \mathop{\lim}\limits_{m \rightarrow \infty}  \Fkeps{k_m}(\alpha) 
			%\notag \\
			%&
			\stackrel{(a)}{=} \mathop{\lim}\limits_{n \rightarrow \infty}\mathop{\lim}\limits_{m \rightarrow \infty} \Fkn{k_m}{n}(\alpha) \notag \\
			&\stackrel{(b)}{\le} \mathop{\lim}\limits_{n \rightarrow \infty}\mathop{\lim\sup}\limits_{k \rightarrow \infty} \Fkn{k}{n}(\alpha),
			\label{eqn:AidLemma23}
			\end{align}
			where $(a)$ follows from \eqref{eqn:AidLemma21}, and $(b)$ follows from the definition of the limit superior \cite[Def. 3.16]{Rudin:76}. 
			\else
			\vspace{-0.2cm}
			\begin{align}
			\mathop{\lim\sup}\limits_{k \rightarrow \infty}  \Fkeps{k}(\alpha) 
			&= \mathop{\lim}\limits_{m \rightarrow \infty}  \Fkeps{k_m}(\alpha)  
			\stackrel{(a)}{\le} \mathop{\lim}\limits_{n \rightarrow \infty}\mathop{\lim\sup}\limits_{k \rightarrow \infty} \Fkn{k}{n}(\alpha),
			\label{eqn:AidLemma23}
			\vspace{-0.1cm}
			\end{align}
			where $(a)$ follows from \eqref{eqn:AidLemma21} combined with the definition of the limit superior \cite[Def. 3.16]{Rudin:76}. 
			\fi
			%Therefore, $\mathop{\lim \sup}\limits_{k \rightarrow \infty} \Fkeps{k}(\alpha) \le \mathop{\lim }\limits_{n \rightarrow \infty}\mathop{\lim \sup}\limits_{k \rightarrow \infty} \Fkn{k}{n}(\alpha)$. 
			
			\ifFullVersion
			Similarly,  by  \cite[Thm. 3.17]{Rudin:76}, for any $n \in \mySet{N}$  there exists a subsequence $\big\{\Fkn{m_l}{n}(\alpha)\big\}_{l \in \mySet{N}}$ where  $\{m_l\}_{l \in \mySet{N}}$ satisfy  $0<m_1< m_2< \ldots$, 
			such that $\mathop{\lim}\limits_{l \rightarrow \infty}  \Fkn{m_l}{n}(\alpha) = \mathop{\lim\sup}\limits_{k \rightarrow \infty}  \Fkn{k}{n}(\alpha)$. Therefore, 
			\begin{align}
			\mathop{\lim}\limits_{n \rightarrow \infty} \mathop{\lim\sup}\limits_{k \rightarrow \infty}  \Fkn{k}{n}(\alpha)
			&=  \mathop{\lim}\limits_{n \rightarrow \infty} \mathop{\lim}\limits_{l \rightarrow \infty}  \Fkn{m_l}{n}(\alpha)   \notag \\ 
			&\stackrel{(a)}{=} \mathop{\lim}\limits_{l \rightarrow \infty}  \Fkeps{m_l}(\alpha)
			%\notag \\
			%&
			\stackrel{(b)}{\le} \mathop{\lim\sup}\limits_{k \rightarrow \infty}  \Fkeps{k}(\alpha),
			\label{eqn:AidLemma24}
			\end{align}
			where $(a)$ follows  
			from \eqref{eqn:AidLemma21}, and $(b)$ follows from the definition of the limit superior \cite[Def. 3.16]{Rudin:76}. 
			Therefore, $\mathop{\lim \sup}\limits_{k \rightarrow \infty} \Fkeps{k}(\alpha) \ge \mathop{\lim }\limits_{n \rightarrow \infty}\mathop{\lim \sup}\limits_{k \rightarrow \infty} \Fkn{k}{n}(\alpha)$. Combining \eqref{eqn:AidLemma23} and \eqref{eqn:AidLemma24} proves \eqref{eqn:AidLemma2} in the statement of the lemma. 
			\else
			Using similar arguments, it can be shown that $\mathop{\lim}\limits_{n \rightarrow \infty} \mathop{\lim\sup}\limits_{k \rightarrow \infty}  \Fkn{k}{n}(\alpha) \le \mathop{\lim\sup}\limits_{k \rightarrow \infty}  \Fkeps{k}(\alpha)$.   Combining this with \eqref{eqn:AidLemma23} proves \eqref{eqn:AidLemma2} in the statement of the lemma.
			\fi
		\end{IEEEproof}	
		
		\begin{lemma}
			\label{lem:AidLemma3} 
			Under assumptions \ref{itm:assm1}-\ref{itm:assm2},  the sequence of \acp{rv} $\big\{\zkn{k}{n} \big\}_{k,n \in \mySet{N}}$ satisfies
			\vspace{-0.2cm}
			\begin{align}
			\mathop{\lim }\limits_{n \rightarrow \infty}\left( {\rm p-}\mathop{\lim \inf}\limits_{k \rightarrow \infty} \zkn{k}{n} \right)  
			&=   \sup\left\{\alpha \in \mySet{R} \Big|\mathop{\lim }\limits_{n \rightarrow \infty}\mathop{\lim \sup}\limits_{k \rightarrow \infty} \Fkn{k}{n}(\alpha) =0   \right\}.
			\label{eqn:pliminfEq4}
			\vspace{-0.1cm}
			\end{align}
		\end{lemma}

		\begin{IEEEproof}
			Since by assumption \ref{itm:assm1}, for every $n \in \mySet{N}$,  every convergent subsequence of 	 $\big\{\zkn{k}{n} \big\}_{k\in \mySet{N}}$ converges in distribution as $k \rightarrow \infty$ to a deterministic scalar, it follows that every convergent subsequence of $\Fkn{k}{n}(\alpha)$ converges as $k \rightarrow \infty$ to a step function, which is the \ac{cdf} of the corresponding sublimit of $\zkn{k}{n}$. In particular, 
			$\mathop{\lim \sup}\limits_{k \rightarrow \infty} \Fkn{k}{n}(\alpha)$ is a step function representing the \ac{cdf} of the deterministic scalar $\zetan{n}$, i.e.,
			\ifFullVersion
			\begin{equation}
			\mathop{\lim \sup}\limits_{k \rightarrow \infty} \Fkn{k}{n}(\alpha) = 
			\begin{cases}
			0 & \alpha \le\zetan{n} \\
			1 & \alpha > \zetan{n}.
			\end{cases}
			\label{eqn:IntProof1}
			\end{equation}
			\else
			$\mathop{\lim \sup}\limits_{k \rightarrow \infty} \Fkn{k}{n}(\alpha) = \Ind\left( {\alpha > \zetan{n}}\right) $.
			\fi
			Since, by Lemma \ref{lem:AidLemma2},  \ref{itm:assm2} implies that the limit $\mathop{\lim }\limits_{n \rightarrow \infty}\mathop{\lim \sup}\limits_{k \rightarrow \infty} \Fkn{k}{n}(\alpha)$ exists\footnote{The convergence to a discontinuous function is in the sense of \cite[Ex. 7.3]{Rudin:76}}, then $\mathop{\lim }\limits_{n \rightarrow \infty} \zetan{n} $ exists. Hence, we obtain that 
			\ifFullVersion
			\begin{equation*}
			\mathop{\lim }\limits_{n \rightarrow \infty}\mathop{\lim \sup}\limits_{k \rightarrow \infty} \Fkn{k}{n}(\alpha) = 
			\begin{cases}
			0 & \alpha \le \mathop{\lim }\limits_{n \rightarrow \infty} \zetan{n} \\
			1 & \alpha > \mathop{\lim }\limits_{n \rightarrow \infty} \zetan{n}.
			\end{cases}
			\end{equation*}
			It is emphasized that the equality in the above expression is arbitrary and does not affect the proof, i.e., while we wrote that $ \mathop{\lim }\limits_{n \rightarrow \infty}\mathop{\lim \sup}\limits_{k \rightarrow \infty} \Fkn{k}{n}(\alpha) = 0$ for  $\alpha = \mathop{\lim }\limits_{n \rightarrow \infty} \zetan{n}$, the proof holds also when $ \mathop{\lim }\limits_{n \rightarrow \infty}\mathop{\lim \sup}\limits_{k \rightarrow \infty} \Fkn{k}{n}(\alpha) = 1$ for  $\alpha = \mathop{\lim }\limits_{n \rightarrow \infty} \zetan{n}$.  Consequently,
			\else
			$\mathop{\lim }\limits_{n \rightarrow \infty}\mathop{\lim \sup}\limits_{k \rightarrow \infty} \Fkn{k}{n}(\alpha) =  \Ind \big(\alpha > \mathop{\lim }\limits_{n \rightarrow \infty} \zetan{n} \big) $, and 
			\fi
			the right-hand side of \eqref{eqn:pliminfEq4} equals to $\mathop{\lim }\limits_{n \rightarrow \infty} \zetan{n} $. 
			
			Next, we note that 
			\ifFullVersion	 
			\begin{align*}
			{\rm p-}\mathop{\lim \inf}\limits_{k \rightarrow \infty} \zkn{k}{n}
			\stackrel{(a)}{=}  \sup\left\{\alpha \in \mySet{R} \Big|\mathop{\lim \sup}\limits_{k \rightarrow \infty} \Fkn{k}{n}(\alpha) =0   \right\} 	  =  \zetan{n},
			\end{align*} 
			where $(a)$ follows from \eqref{eqn:pliminfEq}. % and $(b)$ follows from \eqref{eqn:IntProof1}.
			\else 
			\eqref{eqn:pliminfEq} implies that ${\rm p-}\mathop{\lim \inf}\limits_{k \rightarrow \infty} \zkn{k}{n} = \zetan{n}$. 
			\fi	 
			Consequently, the left-hand side of \eqref{eqn:pliminfEq4} is equal to $\mathop{\lim }\limits_{n \rightarrow \infty} \zetan{n} $, thus proving equality \eqref{eqn:pliminfEq4} in the statement of the lemma.  
		\end{IEEEproof}

		%\textcolor{red}{TODO NIR - I need to prove this, it is not trivial, and I am not sure that I can just replace the function with its limit. Perhaps there is something related to uniform convergence?...}
		%
		%\begin{align}
		%\mathop{\lim }\limits_{n \rightarrow \infty}\left( {\rm p-}\mathop{\lim \inf}\limits_{k \rightarrow \infty} \zkn{k}{n} \right) 
		%&=   \sup\left\{\alpha \big| G(\alpha) =0   \right\} \notag \\
		%&=   \sup\left\{\alpha \big|\mathop{\lim }\limits_{n \rightarrow \infty}\mathop{\lim \sup}\limits_{k \rightarrow \infty} \Fkn{k}{n}(\alpha) =0   \right\}.
		%\label{eqn:pliminfEq4}
		%\end{align}
		
		Substituting \eqref{eqn:AidLemma2} into \eqref{eqn:pliminfEq} results in 
		\vspace{-0.2cm}
		\begin{align}
		{\rm p-}\mathop{\lim \inf}\limits_{k \rightarrow \infty} \zk{k}  
		% &=   \sup\left\{\alpha \in \mySet{R} \Big|\mathop{\lim \sup}\limits_{k \rightarrow \infty} \Fkeps{k}(\alpha) =0   \right\}  \notag \\
		&=\sup\left\{\alpha\in \mySet{R} \Big|\mathop{\lim }\limits_{n \rightarrow \infty}\mathop{\lim \sup}\limits_{k \rightarrow \infty} \Fkn{k}{n}(\alpha) =0   \right\} 
		% \notag \\
		% &
		\stackrel{(a)}{=}\mathop{\lim }\limits_{n \rightarrow \infty}\left( {\rm p-}\mathop{\lim \inf}\limits_{k \rightarrow \infty} \zkn{k}{n} \right), 
		\label{eqn:pliminfEq3}
		\vspace{-0.1cm}
		\end{align}
		where $(a)$ follows from \eqref{eqn:pliminfEq4}. Eq. \eqref{eqn:pliminfEq3} proves \eqref{eqn:plima}. Following  similar arguments, we can prove \eqref{eqn:plimb}, thus concluding the proof of the theorem.
		%
		%\textcolor{red}{TODO NIR - I need to prove this, it is not trivial, and nobody says that the limit even exists...}
		%
		%
		%
		%
		\qed

		\vspace{-0.2cm}
		\section{Proof of Theorem \ref{thm:AsycCap}}
		\label{app:Proof2} 
		\vspace{-0.1cm}
		%To detail the proof of Theorem \ref{thm:AsycCap}, let us first recall the following notation, which is  frequently used henceforth: For any sequence, $\{{y}[i]\}_{i \in \mySet{N}}$, and positive integer $k$,  ${\bf y}^{(k)}$ denotes the column vector obtained by stacking $\big[ { y}[1],\ldots, { y}[k]\big]^T$. 
		The outline of the proof of  Theorem \ref{thm:AsycCap} is as follows:
		\begin{itemize}
			\item First, we show in Subsection \ref{app:Proof2a} that for any  $k\in \mySet{N}$, in the limit of $n \rightarrow \infty$, the \ac{pdf} of  $\Wi_n^{(k)}$ converges to the  \ac{pdf} of  $\Wi_{\eps}^{(k)}$, and that  convergence is uniform with respect to $k\in\mN$ and to the realization $\myVec{w}^{(k)} \in \mySet{R}^k$. This is stated in Lemma~\ref{Lem:PDF-convergence}.
			\item Next, in Subsection \ref{app:proof2b} we use Theorem \ref{thm:plim} to relate the mutual information density rates of the channels \eqref{eqn:AsnycModel1} and \eqref{eqn:AsnycModel2}. To properly state the relationship proved in Subsection \ref{app:proof2b}, let  $\cdf{\Xin}$ denote the distribution of the stochastic process $\{\Xscal[i]\}_{i \in \mySet{N}}$, i.e.,  $\cdf{\Xin} \equiv \left\{\cdf{\Xin^{(k)}}  \right\}_{k=1}^{\infty}$ is the set of \acp{cdf} of all random vectors $\Xin^{(k)}$ whose entries are the elements of $\{\Xscal[i]\}_{i \in \mySet{N}}$ \cite[Ch. 10.1]{Papoulis:91}. We define the following random functions:
			\ifFullVersion	
			\begin{subequations}
				\label{eqn:zkdefs}
				\begin{equation}
				\label{eqn:zkdefs2}
				\zkeps{k}'\left( \cdf{\Xin} \right) \triangleq \frac{1}{k}\log \frac{\pdf{\Yi_{\eps}^{(k)} | \Xin^{(k)} }\left( \Yi_{\eps}^{(k)} \big| \Xin^{(k)} \right) }{\pdf{\Yi_{\eps}^{(k)}}\left( \Yi_{\eps}^{(k)} \right) },
				\end{equation}
				and
				\begin{equation}
				\label{eqn:zkdefs1}
				\zkn{k}{n}'\left( \cdf{\Xin} \right)  \triangleq \frac{1}{k}\log \frac{\pdf{\Yi_n^{(k)} | \Xin^{(k)} }\left( \Yi_n^{(k)} \big|\Xin^{(k)} \right) }{\pdf{\Yi_n^{(k)} }\left( \Yi_n^{(k)}  \right) },
				\end{equation}		
			\end{subequations}
			\else
			\vspace{-0.1cm}
			\begin{equation}
			\label{eqn:zkdefs}
			\hspace{-0.2cm}
			\zkeps{k}'\!\left( \cdf{\Xin} \right) \!\triangleq\! \frac{1}{k}\log \frac{\pdf{\Yi_{\eps}^{(k)} | \Xin^{(k)} }\!\big( \Yi_{\eps}^{(k)} \big| \Xin^{(k)} \big) }{\pdf{\Yi_{\eps}^{(k)}}\big( \Yi_{\eps}^{(k)} \big) }; \!\quad \!
			\zkn{k}{n}'\!\left( \cdf{\Xin} \right)\!  \triangleq\! \frac{1}{k}\log \frac{\pdf{\Yi_n^{(k)} | \Xin^{(k)} }\!\big( \Yi_n^{(k)} \big|\Xin^{(k)} \big) }{\pdf{\Yi_n^{(k)} }\big( \Yi_n^{(k)}  \big) },
			\vspace{-0.1cm}
			\end{equation}
			\fi
			$k,n \in \mySet{N}$.
			%  and prove  in Lemmas \ref{lem:AsyncZk}-\ref{lem:AsyncZk2} that these sequences of \acp{rv} satisfy the conditions of Theorem \ref{thm:AsycCap}. In particular, 
			Note that the \acp{rv} in \eqref{eqn:zkdefs} represent the  mutual information density rates \cite[Def. 3.2.1]{Han:03}  for the sampled channel \eqref{eqn:AsnycModel1} and for the additive \ac{wscs} noise channel \eqref{eqn:AsnycModel2}, respectively, with a given input distribution. 
			In Lemma \ref{lem:AsyncZk} we show that   if the Gaussian  random vectors $\Xin_n^{(k)}$ and $\Xin^{(k)}$ satisfy that $\Xin_n^{(k)} \ConvDist{n \rightarrow \infty} \Xin^{(k)}$ uniformly with respect to $k$, then %\footnote{Note that when convergence occurs in distribution, then this type of convergence occurs uniformly in $k$, see proof of Corollary \ref{lem:AsyncConvDist}}  
			$\zkn{k}{n}'\left( \cdf{\Xin_n} \right)  \ConvDist{n \rightarrow \infty} \zkeps{k}'\left( \cdf{\Xin} \right)$ uniformly in $k \in \mySet{N}$. 
			Subsequently,  Lemma  \ref{lem:AsyncZk2} proves that every subsequence of $\zkn{k}{n}'\left( \cdf{\Xin} \right)$ converges in distribution to a  deterministic scalar. %, where each subsequence may converge to a different scalar.
			\item Finally, in Subsection \ref{app:proof2c}, we combine the above results and show  in Lemmas \ref{lem:CapLowBound} and \ref{lem:CapUpBound} that 	$\Capacity_\myEps \ge \mathop{\lim \inf}\limits_{n \rightarrow \infty} \Capacity_n$ and 	$\Capacity_\myEps \le \mathop{\lim \inf}\limits_{n \rightarrow \infty} \Capacity_n$, respectively, concluding that $	\Capacity_\myEps = \mathop{\lim \inf}\limits_{n \rightarrow \infty} \Capacity_n$.
		\end{itemize}
	\ifFullVersion
	\else
	 	A more  detailed version  of the proof presented in this appendix  can be found in \cite{Shlezinger:19}. 
 	\fi
		
		We henceforth assume that $\frac{1}{2\pi} < \Cwc(t) < \infty $ for all $t \in \mySet{R}$. 
		The motivation for this assumption is that it allows us to show that $\Wi_n^{(k)}$ converges uniformly to $\Wepsvec^{(k)}$, without having to consider the power of the information signal.
		Note that this assumption has no effect on the generality of our capacity derivation, since multiplying $\{\Yeps[i]\}_{i \in \mySet{N}}$ by some positive constant $\kappa$ is an invertible transformation hence it does not affect capacity. Consequently,   the capacity of the channel \eqref{eqn:AsnycModel1} subject to an average power constraint $\PCst$ is identical to the capacity of a channel whose output is given by $\Xscal[i ] + \kappa \cdot \Weps[i]$ subject to an average power constraint $\kappa^2 \PCst$. Therefore, if there exists $t_0 \in \mySet{R}$ for which $\Cwc(t_0) \le \frac{1}{2\pi}$, then one can obtain a channel with the same capacity  
		which satisfies the assumption above by properly scaling the output signal and the power constraint.   %Therefore, by noting that this condition does not restrict our generality, we are able to characterize capacity without introducing new conditions.

		%-----------------------------------
		%	Uniform convergence of noise pdf
		%-----------------------------------
		\vspace{-0.2cm}
		\subsection{Convergence in Distribution of $\Wi_n^{(k)}$ to  $\Wi_{\eps}^{(k)}$ Uniformly with respect to $k \in \mySet{N}$}
		\label{app:Proof2a}
		\vspace{-0.1cm}
		To prove that $\Wi_n^{(k)}$ converges in distribution to  $\Wi_{\eps}^{(k)}$ as $n \rightarrow \infty$ uniformly with respect to $k\in \mySet{N}$, we first prove in Lemma \ref{Lem:PDF-convergence} that the \ac{pdf} of $\Wi_n^{(k)}$ converges to the \ac{pdf} of  $\Wi_{\eps}^{(k)}$ uniformly in $k$. We then conclude in Corollary \ref{lem:AsyncConvDist} that   $\Wi_n^{(k)} \ConvDist{n \rightarrow \infty} \Wi_{\eps}^{(k)}$ uniformly in $k \in \mySet{N}$. 
		
		Define the set $\mK \triangleq \{1, 2,..., k\}$,
		\ifFullVersion
		and consider the zero-mean random vectors of dimension $k$: $\Wi_{n}^{(k)}$ and $\Wi_{\myEps}^{(k)}$. Let $\corrmate^k$ and $\corrmatn^k$ denote the correlation matrices:
		\begin{subequations}
			\label{eqn:corrmatdef}
			\begin{eqnarray}
			\corrmate^k  & \triangleq & \dsE\left\{(\Wi_{\myEps}^{(k)}) (\Wi_{\myEps}^{(k)})^T\right\} \equiv  {\rm diag}\left(\Sigweps[1], \ldots, \Sigweps[k] \right) \\
			\corrmatn^k  & \triangleq & \dsE\left\{(\Wi_{n}^{(k)}) (\Wi_{n}^{(k)})^T\right\} \equiv  {\rm diag}\left(\Sigwn[1], \ldots, \Sigwn[k] \right) ,
			\end{eqnarray}
		\end{subequations}
		%for $(u, v) \in \mK \times \mK$. Note that since $\dsE\left\{W_{n}[u]\cdot W_{n}[v]\right\} = \dsE\left\{W_{n}[v]\cdot W_{n}[u]\right\}$, then
		%$\disccorrn[v,u-v] = \disccorrn[u,v-u]$, and $ \left(\corrmatn^k\right)_{u,v} =  \left(\corrmatn^k\right)_{v,u}$. Similarly, $\disccorr[v,u-v] = \disccorr[u,v-u]$, and $ \left(\corrmate^k\right)_{u,v} =  \left(\corrmate^k\right)_{v,u}$. 
		where ${\rm diag}(m_1, m_2, \ldots, m_l)$ denotes an $l \times l$ diagonal matrix with the specified elements, i.e., letting $\myMat{M} = {\rm diag}(m_1, m_2, \ldots, m_l)$ then $\left( \myMat{M}\right)_{i,i} = m_i$. 
		\else
		and let $\corrmate^k$ and $\corrmatn^k$ denote the correlation matrices of  $\Wi_{n}^{(k)}$ and $\Wi_{\myEps}^{(k)}$, respectively. 
		\fi 
		We can now state the following lemma:

		\begin{lemma}
			\label{Lem:PDF-convergence}
			As $n \rightarrow \infty$, the \ac{pdf} of $\Wi_{n}^{(k)}$ converges uniformly in $ \wi^{(k)} \in \mySet{R}^k$ and in $k \in \mySet{N}$ to the \ac{pdf} of $\Wi_{\myEps}^{(k)}$: 
\ifFullVersion			
			\vspace{-0.3cm}
			\begin{equation*}
			\lim_{n\rightarrow\infty} \pdf{\Wi_{n}^{(k)}}\big(\wi^{(k)}\big) = \pdf{\Wi_{\myEps}^{(k)}}\big(\wi^{(k)}\big), \qquad \forall \wi^{(k)} \in \mySet{R}^k, \forall k \in \mySet{N}.
			\vspace{-0.1cm}
			\end{equation*}
\else
$	\lim_{n\rightarrow\infty} \pdf{\Wi_{n}^{(k)}}\big(\wi^{(k)}\big) = \pdf{\Wi_{\myEps}^{(k)}}\big(\wi^{(k)}\big)$ for all $\wi^{(k)} \in \mySet{R}^k$ and $ k \in \mySet{N}$.
\fi
		\end{lemma}
		
		\begin{IEEEproof}
			To prove the lemma, we first fix $k \in \mySet{N}$, and show that $ \pdf{\Wi_{n}^{(k)}}\big(\wi^{(k)}\big)$ converges to $\pdf{\Wi_{\myEps}^{(k)}}\big(\wi^{(k)}\big)$ uniformly in  $\wi^{(k)} \in \mySet{R}^k$. Then, we prove that this convergence is uniform in $k$.

			We start by recalling that $\Wi_{n}^{(k)}$ and $\Wi_{\myEps}^{(k)}$ have independent entries, and by  noting that since $\eps_n\triangleq \frac{1}{n}\cdot\lfloor n\cdot \eps\rfloor$ it holds that $\frac{n\myEps-1}{n}\le \eps_n \le \frac{n\myEps}{n}$, hence, 
			\ifFullVersion
			\begin{equation}
			\label{eqn:limeps}
			\lim_{n\rightarrow \infty}\eps_n = \myEps.
			\end{equation}
			\else
			$\mathop{\lim}\limits_{n\rightarrow \infty}\eps_n = \myEps$.
			\fi
			Note that  since $\Cwc (t)$ is a uniformly continuous function, then by the definition of a uniformly continuous function, for each $i \in \mySet{N}$ 
			%\footnote{
			%From \cite[Def. 11.1.5]{Rankin:Book}: A function $f$  defined on a general interval $\mA$ is said to be continuous on $\mA$, if it is continuous at every point $c \in \mA$
			%
			%From \cite[Def. 11.1.1]{Rankin:Book}: The function $f$ is said to be continuous at the point $c$ if $f(x) \rightarrow f(c)$ as $x \rightarrow c$. This means that $f$ is defined on some interval $(c-\delta_0, c+\delta_0)$ and for every $\myEps > 0$ there exists a positive $\delta(\myEps)$ such that
			%\[
			%    \big| f(x)-f(c) \big| < \myEps, \qquad \forall x\in \big(c-\delta(\myEps), c+\delta(\myEps)\big).
			%\]},
			\ifFullVersion
			\eqref{eqn:limeps} implies that 
			\begin{equation}
			\label{eqn:limcorr}
			\lim_{n\rightarrow \infty} \Sigwn[i]= \lim_{n\rightarrow \infty} \Cwc\left(i\cdot \frac{\Tc}{p+\eps_n} \right) =  \Cwc\left(i\cdot \frac{\Tc}{p+\eps} \right) \equiv \Sigweps[i].
			\end{equation}
			\else
			it follows that $   \mathop{\lim}\limits_{n\rightarrow \infty} \Sigwn[i]=  \mathop{\lim}\limits_{n\rightarrow \infty} \Cwc\left(i\cdot \frac{\Tc}{p+\eps_n} \right) =  \Cwc\left(i\cdot \frac{\Tc}{p+\eps} \right) \equiv \Sigweps[i]$.
			\fi
			Now, from the definitions of the correlation matrices $\corrmate^k$ and $\corrmatn^k$, we  have that 
			\ifFullVersion
			\vspace{-0.1cm}
			\begin{equation}
			\label{eqn:Matn2e}
			\lim_{n\rightarrow \infty} \max_{(u,v)\in \mK \times \mK} \left\{ \left| \left(\corrmatn^k\right)_{u,v}  - \left(\corrmate^k\right)_{u,v}\right| \right\} =     \lim_{n\rightarrow \infty} \max_{i\in \mK} \left\{ \left|  \Sigwn[i] -  \Sigweps[i] \right| \right\}  = 0. 
			\vspace{-0.1cm}
			\end{equation} 
			\else 
			since $\mathop{\lim}\limits_{n\rightarrow \infty} \mathop{\max}\limits_{i\in \mK} \left\{ \left|  \Sigwn[i] -  \Sigweps[i] \right| \right\}  = 0$,
			the matrix $\corrmatn^k$ converges to $\corrmate^k$.
			\fi
			%where, due  the uniform convergence of each element in the matrix, convergence in \eqref{eqn:Matn2e} is uniform over $k \in \mySet{N}$. 
			%
			%\color{red}
			%where, due to the uniform convergence of each element in the matrix, convergence of\\ $\max_{(u,v)\in \mK \times \mK} \left\{ \left| \left(\corrmatn^k\right)_{u,v}  - \left(\corrmate^k\right)_{u,v}\right| \right\}$ to $0$ is uniform in $n$.\footnote{\textcolor{red}{COMMENT NIR - I think this sentence is inaccurate. What do you mean by uniform convergence in $n$? One can write $\alpha_n \triangleq \max_{(u,v)\in \mK \times \mK} \left\{ \left| \left(\corrmatn^k\right)_{u,v}  - \left(\corrmate^k\right)_{u,v}\right| \right\}$, and thus the only mode of convergence is in $n$, so there is cannot be uniform convergence in $n$. Do you mean that the convergence of $  \left| \left(\corrmatn^k\right)_{u,v}  - \left(\corrmate^k\right)_{u,v}\right| $ is uniform over $u,v$? because that statement is correct, though I am not sure it is relevant...  }}
			%\color{black}
			\ifFullVersion
			
			Next, define the $k \times k$ diagonal matrix $\Cmat^k$, the real vector $\wi^{(k)} \in \mySet{R}^k$, and the mappings $\dsM^{(1)}_k$, $\dsM^{(2)}_k$, and $\dsM_k$ as follows: 
			The mapping
			\[
			\dsM^{(1)}_k: \mR^{k^2} \mapsto \mR
			\]
			is defined as
			\[
			\dsM^{(1)}_k\big(\Cmat^k\big) = \Det\big(\Cmat^k\big) \stackrel{(a)}{=} \prod_{i=1}^k \left(\Cmat^k\right)_{i,i},
			\]
			where $(a)$ follows from the fact that $\Cmat^k$ is diagonal \cite[Ch. 6.1]{Meyer:Book}. 
			%Since the determinant of a matrix $\Cmat$ can be expressed using the definition \cite[Sec. 6.1]{Meyer:Book} as the polynomial
			%\[
			%    \Det\big(\Cmat^k\big) = \sum_{\sigma \in \mS_k} \mbox{sgn}(\sigma)\prod_{i=1}^k \left(\Cmat^k\right)_{i,\sigma(i)},
			%\]
			%where $\mS_k$ is the set of all permutations of the set $\mK$ and $\sigma\in\mS_k$ is a permutation of the elements in $\mK$, and $\mbox{sgn}(\sigma)$ is the sign of the permutation $\sigma\in\mS_k$\footnote{For a permutation $\sigma$ of the set $\mK$, $\mbox{sgn}(\sigma)=1$ if the number of interchanges required to obtain $\sigma$ from $\mK$ is even, otherwise $\mbox{sgn}(\sigma)=-1$.}.
			Obviously, the function $\dsM_k^{(1)}\big(\Cmat^k\big)$ is continuous in $\Cmat^k$.
			
			The mapping
			\[
			\dsM^{(2)}_k: \mR^{k^2} \mapsto \mR^{k^2}
			\]
			is next defined via
			\[
			\dsM^{(2)}_k\big(\Cmat^k\big) = \big(\Cmat^k\big)^{-1}.
			\]
			Note that $ \dsM^{(2)}_k\big(\Cmat^k\big)$ is a continuous mapping, see, e.g., \cite{Stewart:69}.
			
			Finally, consider the mapping
			\[
			\dsM_k: \mR^{k^2} \times \mR^{k} \mapsto \mR
			\]
			via
			\[
			\dsM_k\big(\Cmat^k, \wi^{(k)}\big) = (2\pi)^{-k/2}\Big(\Det\big(\Cmat^k\big)\Big)^{-1/2}\cdot \exp\left(-\frac{1}{2}\big(\wi^{(k)}\big)^T \big(\Cmat^k\big)^{-1}\wi^{(k)}\right).
			\] 
			Since the composition of continuous functions is a continuous function \cite[Thm. 11.2.3]{Rankin:Book}, it follows that 
			$\exp \Big(-\frac{1}{2}\big(\wi^{(k)}\big)^T \big(\Cmat^k\big)^{-1}\wi^{(k)} \Big)$ is a continuous function in $\Cmat^k$ and $\wi^{(k)}$, and hence
			$\dsM_k$ is the product of two continuous functions, from which it follows that 
			$\dsM_k$ is a continuous mapping  in $\Cmat^k$ and $\wi^{(k)}$.
			Furthermore, we note that for diagonal non-singular $\Cmat^k$, it holds that   
			\begin{equation*}
			\exp \Big(-\frac{1}{2}\big(\wi^{(k)}\big)^T \big(\Cmat^k\big)^{-1}\wi^{(k)} \Big) = \prod\limits_{i=1}^{k} \exp\left(-\frac{\left(\wi^{(k)} \right)_i^2}{2 \left(\Cmat^k \right)_{i,i}}   \right), 
			\end{equation*}  
			is uniformly continuous in $\wi^{(k)} \in \mySet{R}^k$ \cite[Thm. 5.10]{Rudin:76}.
			It thus follows from  \eqref{eqn:Matn2e} and from \cite[Thm. 11.2.3]{Rankin:Book} that
			$\mathop{\lim}\limits_{n\rightarrow \infty} \dsM_k\big(\corrmatn^k,\wi^{(k)}\big) = \dsM_k\big(\corrmate^k,\wi^{(k)}\big)$. 
			
			\else
			Next,  define the mapping 
			%$\dsM_k: \mR^{k^2} \times \mR^{k} \mapsto \mR$ 
			%via
			%\[
			$\dsM_k\big(\Cmat^k, \wi^{(k)}\big)\! = \!\Big(\Det\big(2\pi\Cmat^k\big)\Big)^{-1/2}\cdot \exp\left(-\frac{1}{2}\big(\wi^{(k)}\big)^T \big(\Cmat^k\big)^{-1}\wi^{(k)}\right)$. 
			%\] 
			The uniform continuity and positivity of $\Cwc(t)$, combined with the  continuity of  $\dsM_k\big(\Cmat^k, \wi^{(k)}\big)$  with respect to $\{\Sigwn[i]\}_{i \in \mySet{K}}$ imply that    $\dsM_k\big(\Cmat^k, \wi^{(k)}\big)$ is uniformly continuous in $\wi^{(k)}$, thus 
			$\mathop{\lim}\limits_{n\rightarrow \infty} \dsM_k\big(\corrmatn^k,\wi^{(k)}\big) \!=\! \dsM_k\big(\corrmate^k,\wi^{(k)}\big)$ pointwise $\forall \wi^{(k)} \in \mySet{R}^k$.
			\fi
			Noting that  $\pdf{\Wi_{n}^{(k)}}\big(\wi^{(k)}\big) =\dsM_k\big(\corrmatn^k,\wi^{(k)}\big)$ and   $\pdf{\Wi_{\myEps}^{(k)}}\big(\wi^{(k)}\big) =\dsM_k\big(\corrmate^k,\wi^{(k)}\big)$, proves that for a given $k \in \mySet{N}$, the sequence of \acp{pdf} $ \pdf{\Wi_{n}^{(k)}}\big(\wi^{(k)}\big)$ converges pointwise as 
			\ifFullVersion
			$n \rightarrow \infty$, for each $\wi^{(k)} \in \mySet{R}^k$ . 
			
			To see that this convergence is uniform in  $\wi^{(k)} \in \mySet{R}^{k}$, we note that the sequence of zero-mean multivariate  Gaussian \acp{pdf} satisfy that $\forall \eta >0$ and $\forall n_0(k) \in \mySet{N}$ there exists some $r\big(\eta,n_0(k)\big) > 0$ such that $\max \left\{\pdf{\Wi_{n}^{(k)}}\big(\wi^{(k)}\big), \pdf{\Wi_{\myEps}^{(k)}}\big(\wi^{(k)}\big) \right\} < \eta$, for all $\wi^{(k)} \in \mySet{R}^k$ satisfying $\| \wi^{(k)}\| > r\big(\eta,n_0(k)\big)$ and for all $n > n_0(k)$. 
			Thus $ \Big|\pdf{\Wi_{n}^{(k)}}\big(\wi^{(k)}\big)- \pdf{\Wi_{\myEps}^{(k)}}\big(\wi^{(k)}\big) \Big| < \eta$, for all $\wi^{(k)} \in \mySet{R}^k$ satisfying $\| \wi^{(k)}\| > r\big(\eta,n_0(k)\big)$ and for all $n > n_0(k)$. 
			Since $\pdf{\Wi_{n}^{(k)}}\big(\wi^{(k)}\big)$ converges pointwise to $\pdf{\Wi_{\myEps}^{(k)}}\big(\wi^{(k)}\big)$ as $n \rightarrow \infty$, it follows from the uniform continuity of  $\pdf{\Wi_{n}^{(k)}}\big(\wi^{(k)}\big)$ and  $\pdf{\Wi_{\myEps}^{(k)}}\big(\wi^{(k)}\big)$ that $\exists \tilde{n}_0(\eta,k) > 0$ (independent of $\wi^{(k)} \in \mySet{R}^k$) such that for all $n > \tilde{n}_0(\eta,k)$ then $ \left|\pdf{\Wi_{n}^{(k)}}\big(\wi^{(k)}\big)- \pdf{\Wi_{\myEps}^{(k)}}\big(\wi^{(k)}\big) \right| < \eta$ for all $\wi^{(k)} \in \mySet{R}^k$ in the closed set $\| \wi^{(k)}\| \le r\big(\eta,n_0(k)\big)$. 
			This is obtained by noting that for $\| \wi^{(k)}\| \le r\big(\eta,n_0(k)\big)$, the difference $\left|\pdf{\Wi_{n}^{(k)}}\big(\wi^{(k)}\big)- \pdf{\Wi_{\myEps}^{(k)}}\big(\wi^{(k)}\big) \right|$ attains a maximal values. This value can be made arbitrarily small do to the continuity of the \acp{pdf} in $\wi^{(k)} \in \mySet{R}^k$. 
			Note also that due to convergence, the difference  $\left|\pdf{\Wi_{n}^{(k)}}\big(\wi^{(k)}\big)- \pdf{\Wi_{\myEps}^{(k)}}\big(\wi^{(k)}\big) \right|$ decreases also for  $\| \wi^{(k)}\|> r\big(\eta,n_0(k)\big)$.
			Consequently,  $ \left|\pdf{\Wi_{n}^{(k)}}\big(\wi^{(k)}\big)- \pdf{\Wi_{\myEps}^{(k)}}\big(\wi^{(k)}\big) \right| < \eta$ for all $n > \max \big\{\tilde{n}_0(\eta,k),n_0(k)\big\}$ and for each $\wi^{(k)} \in \mySet{R}^k$, and thus  convergence is uniform in $\wi^{(k)} \in \mySet{R}^{k}$. 
			\else
			$n \rightarrow \infty$. 
			Using the fact that both $\pdf{\Wi_{n}^{(k)}}\big(\wi^{(k)}\big)$ and $\pdf{\Wi_{\myEps}^{(k)}}\big(\wi^{(k)}\big)$ vanish as $\|\wi^{(k)}\| \rightarrow \infty$, it  can be shown that convergence is uniform  in $\wi^{(k)} \in \mySet{R}^{k}$. 
			\fi
			
			Next, we prove that the convergence is uniform in $k$. To that aim, we fix $\eta > 0$ and $k_0 \in \mySet{N}$, and prove that $\exists n_0 (\eta, k_0)$ such that for all $n > n_0(\eta, k_0)$ and for all sufficiently large $k$, it holds that $\big|\pdf{\Wi_{n}^{(k)}}\big(\wi^{(k)}\big) - \pdf{\Wi_{\myEps}^{(k)}}\big(\wi^{(k)}\big)\big| < \eta$ for every $\wi^{(k)} \in \mySet{R}^k$. Since  $n_0(\eta, k_0)$ does not depend on $k$ (only on the fixed $k_0$), this implies that the convergence is uniform with respect to $k \in \mySet{N}$. 
			
			To that aim we first note that since  the sequence of \acp{pdf} $\pdf{\Wi_{n}^{(k_0)}}\big(\wi^{(k_0)}\big) $  converges as $n \rightarrow \infty$ to $\pdf{\Wi_{\eps}^{(k_0)}}\big(\wi^{(k_0)}\big) $  uniformly in $\wi^{(k_0)} \in \mySet{R}^{k_0}$, it follows that $\exists n_0 (\eta, k_0) \in \mySet{N}$ such that for all $n > n_0(\eta, k_0)$ and for all $\wi^{(k_0)} \in \mySet{R}^{k_0}$, it holds that 
			\ifFullVersion
			\begin{equation*}
			\big|\pdf{\Wi_{n}^{(k_0)}}\big(\wi^{(k_0)}\big)  - \pdf{\Wi_{\eps}^{(k_0)}}\big(\wi^{(k_0)}\big)  \big| < \frac{\eta}{2}.
			\end{equation*}
			\else 
			$ \big|\pdf{\Wi_{n}^{(k_0)}}\big(\wi^{(k_0)}\big)  - \pdf{\Wi_{\eps}^{(k_0)}}\big(\wi^{(k_0)}\big)  \big| < \frac{\eta}{2}$.
			\fi
			Thus, for all $k > k_0$ and for all $\wi^{(k)} \in \mySet{R}^{k}$, using the notation $w_i \triangleq \big(\wi^{(k)}\big)_i$, we can write
			\ifFullVersion
			\begin{align}
			\left|\pdf{\Wi_{n}^{(k)}}\Big(\wi^{(k)}\Big)\!  -\! \pdf{\Wi_{\eps}^{(k)}}\Big(\wi^{(k)}\Big)  \right| 
			&= \left|\pdf{\Wi_{n}^{(k_0)}}\Big(\wi^{(k_0)}\Big) \prod\limits_{i=k_0+1}^k \pdf{W_n[i]}\left(w_i \right)   \! - \! \pdf{\Wi_{\eps}^{(k_0)}}\Big(\wi^{(k_0)}\Big) \prod\limits_{i=k_0+1}^k \pdf{W_{\eps}[i]}\left(w_i \right)   \right| \notag \\
			&= \Bigg| 
			\pdf{\Wi_{n}^{(k_0)}}\Big(\wi^{(k_0)}\Big) \left( \prod\limits_{i=k_0+1}^k \pdf{W_{n}[i]}\left(w_i \right)\! - \!  \prod\limits_{i=k_0+1}^k \pdf{W_{\eps}[i]}\left(w_i\right) \right)  
			\notag \\
			&\quad 
			+ \left(\pdf{\Wi_{n}^{(k_0)}}\Big(\wi^{(k_0)}\Big) \! - \!  \pdf{\Wi_{\eps}^{(k_0)}}\Big(\wi^{(k_0)}\Big) \right)\prod\limits_{i=k_0+1}^k \pdf{W_{\eps}[i]}\left(w_i \right)  \Bigg| \notag \\
			&\leq  \pdf{\Wi_{n}^{(k_0)}}\Big(\wi^{(k_0)}\Big) \left|  \prod\limits_{i=k_0+1}^k \pdf{W_{n}[i]}\left(w_i \right)\! - \!  \prod\limits_{i=k_0+1}^k \pdf{W_{\eps}[i]}\left(w_i \right)\right| 
			\notag \\
			&\quad   
			+\prod\limits_{i=k_0+1}^k \pdf{W_{\eps}[i]}\left(w_i \right) \left|\pdf{\Wi_{n}^{(k_0)}}\Big(\wi^{(k_0)}\Big) \! - \! \pdf{\Wi_{\eps}^{(k_0)}}\Big(\wi^{(k_0)}\Big)\right|.  
			\label{eqn:LimAid1}
			\end{align} 
			\else
			\begin{align}
			\left|\pdf{\Wi_{n}^{(k)}}\Big(\wi^{(k)}\Big)\!  -\! \pdf{\Wi_{\eps}^{(k)}}\Big(\wi^{(k)}\Big)  \right| 
			&= \left|\pdf{\Wi_{n}^{(k_0)}}\!\Big(\wi^{(k_0)}\Big) \prod\limits_{i=k_0\!+\!1}^k \pdf{W_n[i]}\!\left(w_i \right)   \! - \! \pdf{\Wi_{\eps}^{(k_0)}}\!\Big(\wi^{(k_0)}\Big) \prod\limits_{i=k_0\!+\!1}^k \pdf{W_{\eps}[i]}\!\left(w_i \right)   \right| \notag \\
			%&= \Bigg| 
			%\pdf{\Wi_{n}^{(k_0)}}\Big(\wi^{(k_0)}\Big) \left( \prod\limits_{i=k_0+1}^k \pdf{W_{n}[i]}\left(w_i \right)\! - \!  \prod\limits_{i=k_0+1}^k \pdf{W_{\eps}[i]}\left(w_i\right) \right)  
			%\notag \\
			%&\quad 
			%+ \left(\pdf{\Wi_{n}^{(k_0)}}\Big(\wi^{(k_0)}\Big) \! - \!  \pdf{\Wi_{\eps}^{(k_0)}}\Big(\wi^{(k_0)}\Big) \right)\prod\limits_{i=k_0+1}^k \pdf{W_{\eps}[i]}\left(w_i \right)  \Bigg| \notag \\
			&\leq  \pdf{\Wi_{n}^{(k_0)}}\Big(\wi^{(k_0)}\Big) \left|  \prod\limits_{i=k_0+1}^k \pdf{W_{n}[i]}\left(w_i \right)\! - \!  \prod\limits_{i=k_0+1}^k \pdf{W_{\eps}[i]}\left(w_i \right)\right| 
			\notag \\
			&\quad   
			+\prod\limits_{i=k_0+1}^k \pdf{W_{\eps}[i]}\left(w_i \right) \left|\pdf{\Wi_{n}^{(k_0)}}\Big(\wi^{(k_0)}\Big) \! - \! \pdf{\Wi_{\eps}^{(k_0)}}\Big(\wi^{(k_0)}\Big)\right|.  
			\label{eqn:LimAid1}
			\end{align} 
			\fi
			Next, by defining the subset  $\mySet{P} \triangleq [0, \Tc] \subset \mR$, we note that the Gaussian \ac{pdf} satisfies
			\begin{align}
			\pdf{W_{n}[i]} (w_i) 
			&\le \sqrt{\frac{1}{2 \pi \cdot \Sigwn[i]}} 
			%\notag \\
			%& 
			\stackrel{(a)}{\le} \sqrt{\frac{1}{2 \pi \cdot \mathop{\min}\limits_{t \in \mySet{R}} \Cwc(t)}} 
			\stackrel{(b)}{=} \sqrt{\frac{1}{2 \pi \cdot \mathop{\min}\limits_{t \in \mySet{P}} \Cwc(t)}}
			\stackrel{(c)}{<} 1, 
			\label{eqn:LimAid2}
			\end{align} 
			$\forall w_i \in \mySet{R}$. Here, 
			$(a)$ follows from \eqref{eqn:CSAutocorr}, $(b)$ follows since $\Wc(t)$ is \ac{wscs} with period $\Tc$, and $(c)$ follows from the assumption $\Cwc(t) > \frac{1}{2\pi}$. Similarly,   $\pdf{W_{\myEps}[i]} (w_i)  < 1$ for all $w_i \in \mySet{R}$. 		
			It follows from \eqref{eqn:LimAid2} that $\exists k_1(\eta) > 0$ (independent of $n$) such that for all $k > k_1(\eta)$, $\prod\limits_{i=k_0+1}^k \pdf{W_{n}[i]}\left(w_i \right) < \frac{\eta}{2}$ and $\prod\limits_{i=k_0+1}^k \pdf{W_{\eps}[i]}\left(w_i \right) < \frac{\eta}{2}$, for all $\wi^{(k)} \in \mySet{R}^k$. Furthermore, \eqref{eqn:LimAid2}  also implies that $\pdf{\Wi_{n}^{(k_0)}}\big(\wi^{(k_0)}\big) < 1$. Plugging these inequalities into \eqref{eqn:LimAid1} results in 
			\begin{align}
			\left|\pdf{\Wi_{n}^{(k)}}\big(\wi^{(k)}\big)  - \pdf{\Wi_{\eps}^{(k)}}\big(\wi^{(k)}\big)  \right| 
			&\le \frac{\eta}{2} + \frac{\eta}{2} \left|\pdf{\Wi_{n}^{(k_0)}}\big(\wi^{(k_0)}\big) - \pdf{\Wi_{\eps}^{(k_0)}}\big(\wi^{(k_0)}\big)\right| 
			% \notag \\
			%&
			\le \frac{\eta}{2}\left(\frac{\eta}{2} + 1\right),  
			\label{eqn:LimAid3}
			\end{align}	
			$\forall \wi^{(k)} \in \mySet{R}^k$. 	
			Eqn. \eqref{eqn:LimAid3} implies that for all sufficiently small $\eta<1$, if $n > n_0(\eta, k_0)$, then  $\big|\pdf{\Wi_{n}^{(k)}}\big(\wi^{(k)}\big)  - \pdf{\Wi_{\eps}^{(k)}}\big(\wi^{(k)}\big)  \big| < \eta$ for all $\wi^{(k)} \in \mySet{R}^k$ and for all sufficiently large $k \in \mySet{N}$, thus concluding the proof of the lemma. 
		\end{IEEEproof}

		\begin{corollary}
			\label{lem:AsyncConvDist}
			For any $k \in \mySet{N}$ it holds that $\Wi_n^{(k)} \ConvDist{n \rightarrow \infty} \Wi_{\eps}^{(k)}$, uniformly over $k$.
		\end{corollary}
		\begin{IEEEproof}
			Since the continuous \ac{pdf} of the continuous random vector $\Wi_n^{(k)}$ converges to the continuous \ac{pdf} of continuous random vector $\Wi_{\eps}^{(k)}$, it follows from \cite[Thm. 1]{Schefe:49} that $\Wi_n^{(k)} \ConvDist{n \rightarrow \infty} \Wi_{\eps}^{(k)}$. Since the convergence of the \acp{pdf} is uniform in $k\in\mySet{N}$,  the convergence of the \acp{cdf} is also uniform by \cite[Thm. 1]{Schefe:49}.
		\end{IEEEproof}
		
		% 
		
		%-----------------------------------
		%	Uniform convergence of mutual information density
		%-----------------------------------
		\vspace{-0.2cm}
		\subsection[text]{Showing that $\tilde{Z}_{k,n}'\left( \cdf{\Xin_n}\right)$ and $\zkeps{k}'\left(\cdf{\Xin}\right)$ Satisfy the Conditions of Thm. \ref{thm:plim}}
		\label{app:proof2b}
		\vspace{-0.1cm}
		Let $\cdf{\Xnvec}\opt$ denote the optimal input distribution for the channel \eqref{eqn:AsnycModel2} subject to the input power constraint \eqref{eqn:AsyncConst1}.
		We next prove that $\tilde{Z}_{k,n}'\left( \cdf{\Xnvec}\right)$ and $\zkeps{k}'\left(\cdf{\Xin}\right)$ satisfy \ref{itm:assm1}-\ref{itm:assm2}. In particular, Lemma \ref{lem:AsyncZk} proves  that $\tilde{Z}_{k,n}'\left( \cdf{\Xnvec}\right)  \ConvDist{n \rightarrow \infty} \zkeps{k}'\left(\cdf{\Xin}\right)$ uniformly in 
		\ifFullVersion
		$k \in \mySet{N}$ for  zero-mean Gaussian input vectors
		\else
		$k $ for  Gaussian inputs
		\fi
		with independent entries. Lemma \ref{lem:AsyncZk2} proves that for  $k \rightarrow \infty$, $\tilde{Z}_{k,n}'\left( \cdf{\Xnvec}\opt\right)$ converges
		\ifFullVersion
		in distribution
		\fi
		to a deterministic scalar. 
		\begin{lemma}
			\label{lem:AsyncZk}
			Consider a sequence of $k\times 1$ zero-mean Gaussian random vectors with independent entries $\{\Xin_n^{(k)}\}_{n \in \mySet{N}}$ and a zero-mean Gaussian random vector  with independent entries $\Xin^{(k)}$, such that  $\Xin_n^{(k)} \ConvDist{n \rightarrow \infty} \Xin^{(k)}$ uniformly with respect to $k \in \mySet{N}$. 
			%	\ifFullVersion 
			%	Specifically, letting $\sigma_{\Xscal_n}^2[i]$ and $\sigma_{\Xscal}^2[i]$ denote the variances of $\big( \Xin_n^{(k)}\big)_i $ and of  $\big( \Xin^{(k)}\big)_i $, respectively, it holds that for all $\eta > 0$ there exists $n_0 (\eta)$ such that for each $n > n_0(\eta)$, $\big|\sigma_{\Xscal}^2[i] - \sigma_{\Xscal_n}^2[i]\big| < \eta$ for all $i \in \mySet{N}$. 
			%	\fi
			Then, the \acp{rv} $\zkn{k}{n}'\left( \cdf{\Xnvec}\right)$ and $\zkeps{k}'\left( \cdf{\Xin}\right)$ defined in \eqref{eqn:zkdefs} satisfy $\zkn{k}{n}'\left( \cdf{\Xnvec}\right)  \ConvDist{n \rightarrow \infty} \zkeps{k}'\left( \cdf{\Xin}\right)$ uniformly over $k \in \mySet{N}$.
		\end{lemma}
		\begin{IEEEproof}
			%	To prove the lemma, we need to show that  for all $\eta > 0$ there exists $n_0(\eta) \in \mySet{N}$ such that $\big|\cdf{\Xin_n^{(k)}}\left(\xvec^{(k)} \right) -\cdf{\Xin^{(k)}}\left(\xvec^{(k)} \right) \big| < \eta$ for all $\xvec^{(k)} \in \mySet{R}^k$ and for all sufficiently large $k \in \mySet{N}$.
			For ${\bf y}^{(k)}, {\bf x}^{(k)} \in \mySet{R}^{k}$, define
			\begin{equation}
			\label{eqn:fkdef}
			\fkn\left(\yvec^{(k)}, {\bf x}^{(k)} \right)  \triangleq \frac{\pdf{\Yi_n^{(k)} | \Xin_n^{(k)}}\left( \yvec^{(k)} \big| \xvec^{(k)}\right) }{\pdf{\Yi_n^{(k)}}\left( \yvec^{(k)} \right) },
			\qquad
			\fkeps\left(\yvec^{(k)}, \xvec ^{(k)}\right)  \triangleq \frac{\pdf{\Yi_{\eps}^{(k)} | \Xin^{(k)}}\left( \yvec^{(k)} \big| \xvec^{(k)}\right) }{\pdf{\Yi_{\eps}^{(k)}}\left( \yvec^{(k)} \right) }.	
			\end{equation}
			\ifFullVersion
			To prove the lemma, 
			we first show that $\left[\big( \Yi_n^{(k)}\big) ^T , \big( \Xin_n^{(k)}\big) ^T \right] \ConvDist{n \rightarrow \infty} \left[\big( \Yi_{\eps}^{(k)}\big) ^T , \big( \Xin^{(k)}\big) ^T \right]$ uniformly with respect to $k$;
			Then, we
			\else Note that Corollary \ref{lem:AsyncConvDist} implies that  $\left[\big( \Yi_n^{(k)}\big) ^T , \big( \Xin_n^{(k)}\big) ^T \right] \ConvDist{n \rightarrow \infty} \left[\big( \Yi_{\eps}^{(k)}\big) ^T , \big( \Xin^{(k)}\big) ^T \right]$ uniformly in $k$. Thus, to prove the lemma,  we first  
			\fi
			use the extended \ac{cmt} \cite[Thm 7.24]{Kosorok:07} to prove that
			$\fkn\big(\Yi_n^{(k)} , \Xin_n^{(k)}  \big)  \ConvDist{n \rightarrow \infty}  \fkeps\big(\Yi_{\eps}^{(k)} , \Xin^{(k)}  \big)$ for each $k \in \mySet{N}$. 
			Since $\zkn{k}{n}'\left( \cdf{\Xnvec}\right) = \frac{1}{k}\log\fkn\big(\Yi_n^{(k)}  , \Xin_n^{(k)} \big)$ and $\zkeps{k}'\left( \cdf{\Xin}\right) =  \frac{1}{k}\log\fkeps\big(\Yi_{\eps}^{(k)}  , \Xin^{(k)} \big)$, we conclude that $\zkn{k}{n}'\left( \cdf{\Xnvec}\right)  \ConvDist{n \rightarrow \infty} \zkeps{k}'\left( \cdf{\Xin}\right)$ for each  $k \in \mySet{N}$. 
			Finally, we prove that  convergence is uniform in $k$.
			
			\ifFullVersion
			\begin{sloppypar}	
				%	For ${\bf w}^{(k)}, {\bf x}^{(k)} \in \mySet{R}^{k}$, let $\cdf{\Wi_{\eps}^{(k)}}\left( \wvec^{(k)}\right)$, $\cdf{\Wi_n^{(k)}}\left(\wvec^{(k)} \right)$, $\cdf{\Xin^{(k)}}\left(  \xvec^{(k)}\right)$, and $\cdf{\Xnvec^{(k)}}\left(   \xvec^{(k)} \right)$
				%%	, and $\cdf{\Xz}\left(   \xz\right)$
				%be the \acp{cdf} of $ \Wi_{\eps}^{(k)}$, $\Wi_n^{(k)}$, $\Xin^{(k)}$, and $\Xin_n^{(k)}$,
				%%and $\Xz$,
				%respectively. 
				Since $\Wi_n^{(k)}$ and $\Xin_n^{(k)}$ are mutually independent, it follows that the joint \ac{cdf} of the  vector $\left[\big( \Wi_n^{(k)}\big)^T,\big(\Xin_n^{(k)}\big)^T \right]^T$ evaluated at $\left[\left( \wvec^{(k)}\right) ^T, \left( \xvec^{(k)}\right) ^T  \right]^T$ is given by $\cdf{\Wi_n^{(k)}}\left( \wvec^{(k)}\right)\cdot \cdf{\Xnvec^{(k)}}\left(  \xvec^{(k)}\right) $. By Corollary \ref{lem:AsyncConvDist} and since by assumption $\Xin_n^{(k)} \ConvDist{n \rightarrow \infty} \Xin^{(k)}$, this joint \ac{cdf} converges  to $\cdf{\Wi_{\eps}^{(k)}}\left( \wvec^{(k)} \right)\cdot \cdf{\Xin^{(k)}}\left( \xvec ^{(k)}\right) $ as $n \rightarrow \infty$, which is the joint \ac{cdf} of  $\left[\big( \Wi_{\eps}^{(k)}\big)^T, \big(\Xin^{(k)}\big)^T \right]^T$. Furthermore, since the convergence in distribution of $\Xin_n^{(k)}$ to $\Xin^{(k)}$ is uniform in $k$ and the convergence in distribution $\Wi_n^{(k)}$ to $\Wi_{\eps}^{(k)}$ is also uniform in $k$, it follows that the convergence of the joint \ac{cdf} of $\left[\big( \Wi_n^{(k)}\big)^T, \big(\Xin_n^{(k)}\big)^T \right]^T$ to that of $\left[\big( \Wi_{\eps}^{(k)}\big)^T, \big(\Xin^{(k)}\big)^T \right]^T$ is also uniform in $k$.
				We thus conclude that  $\left[\big( \Wi_n^{(k)}\big)^T, \big(\Xin_n^{(k)}\big)^T \right]^T \ConvDist{n \rightarrow \infty} \left[\big( \Wi_{\eps}^{(k)}\big)^T, \big(\Xin^{(k)}\big)^T \right]^T$, uniformly in $k\in\mN$.  
				
				Next, we note that from \eqref{eqn:AsnycModel1} and \eqref{eqn:AsnycModel2},  $\Yi_n^{(k)}$ and $\Yi_{\myEps}^{(k)}$ can be written as $\Yi_n^{(k)} = \Xin_n^{(k)} + \Wi_n^{(k)}$ and $\Yi_{\eps}^{(k)} = \Xin^{(k)} + \Wi_{\eps}^{(k)}$, respectively. Thus, $\left[\big( \Yi_n^{(k)}\big)^T , \big( \Xin_n^{(k)}\big)^T \right]^T$ and $\left[\big( \Yi_{\eps}^{(k)}\big)^T , \big( \Xin^{(k)}\big)^T \right]^T$ can be obtained by applying the same linear transformation to  $\left[\big( \Wi_n^{(k)}\big)^T, \big(\Xin_n^{(k)}\big)^T \right]^T$ and to $\left[\big( \Wi_{\eps}^{(k)}\big)^T, \big(\Xin^{(k)}\big)^T \right]^T$, respectively.
				Consequently, it follows from the \ac{cmt} \cite[Thm. 7.7]{Kosorok:07} that  $\left[\big( \Yi_n^{(k)}\big)^T , \big( \Xin_n^{(k)}\big)^T \right] \ConvDist{n \rightarrow \infty} \left[\big( \Yi_{\eps}^{(k)}\big)^T , \big(\Xin^{(k)}\big)^T \right]$. Since this  linear transformation, i.e., $\Yi_n^{(k)} = \Xin_n^{(k)} + \Wi_n^{(k)}$ and $\Yi_{\eps}^{(k)} = \Xin^{(k)} + \Wi_{\eps}^{(k)}$, is  Lipschitz continuous, it follows from \cite{Kasy:15} 
				that convergence is uniform in $k$.
			\end{sloppypar}	
			
			Next, we 
			\else
			We first 
			\fi 
			apply the extended \ac{cmt} to prove that $\fkn\big(\Yi_n^{(k)} , \Xin_n^{(k)}  \big) \ConvDist{n \rightarrow \infty}  \fkeps\big(\Yi_{\eps}^{(k)} , \Xin^{(k)}  \big)$. 
			The application requires two conditions: That the mappings $\fkn, \fkeps:\mySet{R}^{2k} \mapsto \mySet{R}^+$ satisfy that for any convergent sequence ${\bf t}_n^{(2k)} \in \mySet{R}^{2k}$ with limit $ \mathop{\lim}\limits_{n \rightarrow \infty}{\bf t}_n^{(2k)} = {\bf t}^{(2k)}$, it holds that $\mathop{\lim}\limits_{n \rightarrow \infty} \fkn\big({\bf t}_n^{(2k)} \big)  =  \fkeps\big( {\bf t}^{(2k)}\big)$, and second, that the limit distribution is separable \cite[Pg. 101]{Kosorok:07}.
			%
			%then, for any sequence of random vectors ${\bf T}_n   \in \mySet{R}^{2k}$ such that  ${\bf T}_n \ConvDist{n \rightarrow \infty}  {\bf T}$, where ${\bf T}$ is separable, we have that  $\fkn\left( {\bf T}_n\right)  \ConvDist{n \rightarrow \infty}  \fkeps\left( {\bf T}\right) $  \cite[Thm. 7.24]{Kosorok:07}.
			%
			\ifFullVersion
			Specifically, we will show that the following two properties hold:
			\begin{enumerate}[label={\em P\arabic*}]
				\item \label{itm:Tight} The limiting distribution of $\left[\big( \Yi_{\eps}^{(k)}\big) ^T , \big( \Xin^{(k)}\big) ^T \right]$ is separable\footnote{By \cite[Pg. 101]{Kosorok:07}, an \ac{rv} $X \in \mySet{X}$ is separable  if $\forall \eta > 0$ there exists a compact set $\mySet{K}(\eta) \subset \mySet{X}$ such that $\Pr \left( X \in \mySet{K}(\eta)\right) \ge 1 - \eta$.}.
				\item \label{itm:Conv} For all convergent sequences $\yvec_n^{(k)}, \xvec_n^{(k)} \in \mySet{R}^{k}$ such that $\mathop{\lim}\limits_{n \rightarrow \infty}\yvec_n^{(k)} = \yvec^{(k)}$ and $\mathop{\lim}\limits_{n \rightarrow \infty}\xvec_n^{(k)} = \xvec^{(k)}$, we have that $\mathop{\lim}\limits_{n \rightarrow \infty} \fkn\big(\yvec_n^{(k)},\xvec_n^{(k)}\big)  =  \fkeps\left( \yvec^{(k)}, \xvec^{(k)}\right)$.
			\end{enumerate}
			
			To prove property \ref{itm:Tight}, we show that  ${\bf U}^{(k)} \triangleq \left[\big( \Yi_{\eps}^{(k)}\big)^T , \big( \Xin^{(k)}\big) ^T \right]$ is separable \cite[Pg. 101]{Kosorok:07}, i.e., that $\forall \eta > 0$, there exists $t > 0$ such that $\Pr\left(\|{\bf U}^{(k)}\|^2 > t \right) < \eta$.
			To that aim, recall first that by Markov's inequality \cite[Pg. 114]{Papoulis:91}, it follows that $\Pr\left(\right\|{\bf U}^{(k)}\left\|^2 > t \right) < \frac{1}{t}\E\left\{\left\|{\bf U}^{(k)}\right\|^2 \right\}$. From the input power constraint \eqref{eqn:AsyncConst1} it follows that $\E\left\{\left\|{\bf U}^{(k)}\right\|^2 \right\}$ is bounded, and thus for each $t > \frac{1}{\eta}\E\left\{\left\|{\bf U}^{(k)}\right\|^2 \right\}$ we have that $\Pr\left(\left\|{\bf U}^{(k)}\right\|^2 > t \right) < \eta$, and thus ${\bf U}^{(k)}$ is separable. 
			
			To prove property \ref{itm:Conv},  we note that by \cite[Eq. (8.39)]{Papoulis:91} 
			\begin{align}
			%	\pdf{\Ynvec^{(k)} | \Xnvec^{(k)}}\left( {\bf y} \big| {\bf x}\right)
			%	&= \int\limits_{\xz \in \mySet{R}^\Mem}  \pdf{\Ynvec^{(k)} | \Xnvec^{(k)}, \Xz}\left( {\bf y} \big| {\bf x}, \xz\right)\pdf{ \Xz|\Xnvec^{(k)}}\left( \xz \big| {\bf x}\right)d \xz \notag \\
			\pdf{\Yi_n^{(k)} | \Xin_n^{(k)} }\left( \yvec^{(k)}  \big| \xvec^{(k)} \right)
			%	&\stackrel{(a)}{=}  \int\limits_{\xz \in \mySet{R}^\Mem} 	 \pdf{\Wnvec^{(k)}}\left( {\bf y} - \myMat{H}_1 {\bf x} - \myMat{H}_0  \xz\right)\pdf{ \Xz}\left( \xz \right)d \xz,
			&\stackrel{(a)}{=} \pdf{\Wi_n^{(k)}}\left(  \yvec^{(k)} -\xvec^{(k)}\right),
			\label{eqn:ContFproof1}
			\end{align}
			where $(a)$ follows since $\Yi_n^{(k)} = \Xin_n^{(k)} + \Wi_n^{(k)}$ and since $\Wi_n^{(k)}$ and $\Xin_n^{(k)}$ are mutually independent.
			Similarly, we have that
			\begin{equation}
			\pdf{\Yi_{\eps}^{(k)} | \Xin^{(k)} }\left( \yvec ^{(k)}\big| \xvec^{(k)} \right)
			%=  \int\limits_{\xz \in \mySet{R}^\Mem} 	 \pdf{\Wepsvec^{(k)}}\left( {\bf y} - \myMat{H}_1 {\bf x} - \myMat{H}_0  \xz\right)\pdf{ \Xz}\left( \xz \right)d \xz,
			=\pdf{\Wi_{\eps}^{(k)}}\left( \yvec ^{(k)}-  \xvec^{(k)} \right).
			\label{eqn:ContFproof1a}
			\end{equation}
			Combining this with the fact that the  $ \pdf{\Wi_n^{(k)}}(\wvec^{(k)})$ is continuous  implies that  $	\pdf{\Yi_n^{(k)} | \Xin^{(k)} }\left( \yvec^{(k)} \big| \xvec^{(k)}\right) $ is continuous.
			Furthermore, we note that by Lemma \ref{Lem:PDF-convergence} it holds that    $\forall \eta > 0$ there exists $n_0(\eta) >0$ such that for all $n > n_0(\eta)$ we have that $\forall \wvec^{(k)} \in \mySet{R}^{k}$, $\big| \pdf{\Wi_n^{(k)}}\left( \wvec^{(k)}  \right)  -  \pdf{\Wi_{\eps}^{(k)}}\left(\wvec ^{(k)}\right) \big| < \eta$, for all sufficiently large $k \in \mySet{N}$. Consequently, for $n > n_0(\eta)$ and a sufficiently large $k \in \mySet{N}$, 
			\begin{align}
			\hspace{-1cm}\left|\pdf{\Yi_n^{(k)} | \Xin_n^{(k)} }\left( \yvec^{(k)} \big|  \xvec^{(k)} \right) - \pdf{\Yi_{\eps}^{(k)} | \Xin^{(k)}}\left( \yvec^{(k)} \big|  \xvec^{(k)} \right)  \right|  
			&=  \left|	\pdf{\Wi_{\eps}^{(k)}}\left( \yvec^{(k)} - \xvec^{(k)} \right) -  \pdf{\Wi_n^{(k)}}\left( \yvec^{(k)} -  \xvec^{(k)} \right)\right|  \notag \\
			& <  \eta,
			\label{eqn:ContFproof3}
			\end{align} 
			for all $\left( \yvec^{(k)} ,  \xvec^{(k)}\right)  \in \mySet{R}^{2k}$. 
			%		It thus follows that as $n \rightarrow \infty$, $\pdf{\Yi_n^{(k)} | \Xin_n^{(k)} }\left( \yvec ^{(k)}\big|  \xvec ^{(k)}\right) $ converges   to $\pdf{\Yi_{\eps}^{(k)} |\Xin^{(k)},  }\left( \yvec^{(k)} \big|  \xvec ^{(k)}\right) $, and that this convergence is uniform in $k\in\mN$ and in $\left( \yvec^{(k)} ,  \xvec^{(k)}\right)  \in \mySet{R}^{2k}$.
			It thus follows from Lemma \ref{lem:AsyncConvDist} that $\mathop{\lim}\limits_{n \rightarrow \infty}\pdf{\Yi_n^{(k)} | \Xin_n^{(k)} }\left( \yvec ^{(k)}\big|  \xvec ^{(k)}\right) =\pdf{\Yi_{\eps}^{(k)} |\Xin^{(k)},  }\left( \yvec^{(k)} \big|  \xvec ^{(k)}\right) $, and that this convergence is uniform in $k\in\mN$ and in $\left( \yvec^{(k)} ,  \xvec^{(k)}\right)  \in \mySet{R}^{2k}$.

			Next, we show that $\mathop{\lim}\limits_{n \rightarrow \infty} \pdf{\Yi_n^{(k)}   }\left( \yvec^{(k)}\right) = \pdf{\Yi_{\eps}^{(k)}  }\left( \yvec^{(k)}\right)$ uniformly with respect to $k$ and $\yvec^{(k)}\in\mySet{R}^{k}$. 
			Let  $\sigma_{\Yn}^2[i]$ and $\sigma_{\Yeps}^2[i]$ denote the variances of $ \Yn[i]$ and $\Yeps[i]$, respectively. Since $ \Xin_n^{(k)} $ and  $\Wi_{n}^{(k)}$ are zero-mean Gaussians with independent entries and are mutually independent, it holds that $\Yi_n^{(k)}$ is a zero-mean Gaussian with independent entries, and that the variance of each entry of $\Yi_n^{(k)}$ is given by the sum of the variances of the corresponding entries of  $ \Xin_n^{(k)} $ and  $\Wi_{n}^{(k)}$, $\sigma_{\Yn}^2[i] = \sigma_{X_n}^2[i] + \sigma_{\Wn}^2[i]$. Similarly, $\Yi_{\eps}^{(k)}$ is also zero-mean Gaussian with independent entries, and the variance of each entry of $\Yi_{\eps}^{(k)}$ is given by the sum of the variances of the corresponding entries of  $ \Xin^{(k)} $ and  $\Wi_{\eps}^{(k)}$, $\sigma_{Y_{\eps}}^2[i] = \sigma_{X}^2[i] + \sigma_{W_{\eps}}^2[i]$. Consequently, % as both  $\Wi_n^{(k)} \ConvDist{n \rightarrow \infty} \Wi_{\eps}^{(k)}$ and  $\Xin_n^{(k)} \ConvDist{n \rightarrow \infty} \Xin^{(k)}$ uniformly in $k$, 
			$\sigma_{\Yn}^2[i] \ge  \mathop{\min}\limits_{t \in \mySet{R}} \Cwc(t)$ and $\sigma_{\Yn}^2[i]$ converges to $\sigma_{\Yeps}^2[i]$ as $n \rightarrow \infty$ for each $i \in \mySet{Z}$. 
			%It therefore follows by repeating the proof of Lemma \ref{Lem:PDF-convergence} with $\Yi_n^{(k)}$ and $\Yi_{\eps}^{(k)} $ instead of $\Wi_n^{(k)}$ and $\Wi_{\eps}^{(k)} $ that  $\pdf{\Ynvec^{(k)} }\left( {\bf y}^{(k)}\right)$ is continuous and converges  to $\pdf{\Yepsvec^{(k)} }\left( {\bf y}^{(k)}\right) $ in the limit $n \rightarrow \infty$, and the convergence is uniform over ${\bf y}^{(k)} \in \mySet{R}^{k}$ and $k \in \mySet{N}$.
			It therefore follows by repeating the proof of Lemma \ref{Lem:PDF-convergence} with $\Yi_n^{(k)}$ and $\Yi_{\eps}^{(k)} $ instead of $\Wi_n^{(k)}$ and $\Wi_{\eps}^{(k)} $ that  $\pdf{\Ynvec^{(k)} }\left( {\bf y}^{(k)}\right)$ is continuous and $\mathop{\lim}\limits_{n \rightarrow \infty}\pdf{\Ynvec^{(k)} }\left( {\bf y}^{(k)}\right)=\pdf{\Yepsvec^{(k)} }\left( {\bf y}^{(k)}\right) $, where  convergence is uniform over ${\bf y}^{(k)} \in \mySet{R}^{k}$ and $k \in \mySet{N}$.    
			
			We  therefore conclude that $\fkn\left({\bf y}^{(k)}, {\bf x}^{(k)} \right)$ defined in \eqref{eqn:fkdef} is continuous\myFtn{The continuity of $\fkn\left({\bf y}^{(k)}, {\bf x}^{(k)} \right)$ follows as it is the ratio  of two continuous, positive, real functions.}  \cite[Thm. 4.9]{Rudin:76} and converges to $\fkeps\left({\bf y}^{(k)}, {\bf x}^{(k)} \right)$ for all ${\bf y}^{(k)}, {\bf x}^{(k)}  \in \mySet{R}^{2k}$   and for all $k$ sufficiently large in the limit $n \rightarrow \infty$ \cite[Thm. 3.3]{Rudin:76}.\myFtn{Note that since  $\pdf{\Yepsvec^{(k)} }\left( \yvec^{(k)}\right) =    \int\limits_{\xvec\in \mySet{R}^{k}} 	 \pdf{\Wepsvec^{(k)}}\left(\yvec ^{(k)}- \xvec^{(k)}\right) \pdf{ \Xin^{(k)}}\left( \xvec^{(k)} \right)d \xvec$, where $ \pdf{\Wepsvec^{(k)}}(\cdot)$ is the strictly positive \ac{pdf} of a Gaussian random vector, it follows that $\pdf{\Yepsvec^{(k)} }\left( {\bf y}^{(k)}\right) $ is also strictly positive. Consequently, \cite[Thm. 4.9]{Rudin:76} and \cite[Thm. 3.3]{Rudin:76}, which require the denominator of $\fkeps\left({\bf y}^{(k)}, {\bf x}^{(k)} \right)$ to be non-zero, both hold.}

			We can now prove 
			Property \ref{itm:Conv}   
			by considering an arbitrary pair of convergent sequences $\big\{{\bf y}_n^{(k)}\big\}_{n \in \mySet{N}}$, $\big\{{\bf x}_n^{(k)}\big\}_{n \in \mySet{N}}$, such that $\mathop{\lim}\limits_{n \rightarrow \infty}{\bf y}_n^{(k)} = {\bf y}^{(k)}$ and $\mathop{\lim}\limits_{n \rightarrow \infty}{\bf x}_n^{(k)} = {\bf x}^{(k)}$, for any $k \in \mySet{N}$, and letting  $\eta > 0$. 
			Since $\fkn\left({\bf y}^{(k)}, {\bf x}^{(k)} \right)$ is continuous, then $\exists \delta   > 0$ (which depends on $\eta$, ${\bf y}^{(k)}$, and ${\bf x}^{(k)}$) such that if $\Big\|\Big[ \big( {\bf y}_n^{(k)}\big) ^T,\big( {\bf x}_n^{(k)}\big) ^T\Big]^T - \Big[ \left( {\bf y}^{(k)}\right) ^T,\left( {\bf x}^{(k)}\right) ^T\Big]^T\Big\| < \delta$, then
			%\end{sloppypar} 
			\begin{equation}
			\label{eqn:ContFproof4}
			\Big| \fkn\left({\bf y}_n^{(k)}, {\bf x}_n ^{(k)}\right) - \fkn\left({\bf y}^{(k)}, {\bf x}^{(k)} \right) \Big| < \frac{\eta}{2}.
			\end{equation}  
			Since $\mathop{\lim}\limits_{n \rightarrow \infty}{\bf y}_n^{(k)} = {\bf y}^{(k)}$ and $\mathop{\lim}\limits_{n \rightarrow \infty}{\bf x}_n^{(k)} = {\bf x}^{(k)}$, then $\exists n_0(\delta) \in \mySet{N}$ such that $\forall n > n_0(\delta)$, it holds that $\Big\|\Big[ \big( {\bf y}_n^{(k)}\big) ^T,\big( {\bf x}_n^{(k)}\big) ^T\Big]^T \!-\! \Big[ \left( {\bf y}^{(k)}\right) ^T,\left( {\bf x}^{(k)}\right) ^T\Big]^T\Big\| < \delta$.
			Additionally, as $\fkn\left({\bf y}^{(k)}, {\bf x}^{(k)} \right)$ converges pointwise to $\fkeps\left({\bf y}^{(k)}, {\bf x}^{(k)} \right)$, then $\exists n_1 \left( \delta\right) \in \mySet{N}$ such that $\forall n > n_1(\delta)$ 
			\begin{equation}
			\label{eqn:ContFproof5}
			\Big| \fkn\left({\bf y}^{(k)}, {\bf x} ^{(k)}\right) - \fkeps\left({\bf y}^{(k)}, {\bf x}^{(k)} \right) \Big| < \frac{\eta}{2}.
			\end{equation}
			It follows from \eqref{eqn:ContFproof4}-\eqref{eqn:ContFproof5}  	that $\forall n > \max\big\{n_0(\delta),n_1(\delta)\big\}$, $\Big| \fkn\big({\bf y}_n^{(k)}, {\bf x}_n^{(k)} \big) - \fkeps\left({\bf y}^{(k)}, {\bf x}^{(k)} \right) \Big| < \eta$, proving	Property \ref{itm:Conv}. 
			As Properties and \ref{itm:Conv} and \ref{itm:Conv} are satisfied, then applying the extended CMT we obtain that $\fkn\big(\Ynvec^{(k)} , \Xnvec^{(k)} \big) \ConvDist{n \rightarrow \infty}  \fkeps\big(\Yepsvec^{(k)} , \Xin^{(k)} \big)$. 
			\fi
			As the  \acp{rv} $\zkn{k}{n}'\left( \cdf{\Xnvec}\right)$ and $\zkeps{k}'\left( \cdf{\Xin}\right)$, defined in \eqref{eqn:zkdefs}, are  continuous mappings of $\fkn\big(\Ynvec^{(k)} , \Xnvec^{(k)} \big)$ and $\fkeps\big(\Yepsvec^{(k)} , \Xin^{(k)} \big)$, respectively, it follows from  the \ac{cmt} \cite[Thm. 7.7]{Kosorok:07} that $\zkn{k}{n}'\left( \cdf{\Xnvec}\right) \ConvDist{n \rightarrow \infty} \zkeps{k}'\left( \cdf{\Xin}\right)$. 
			
			\smallskip

			Finally, we prove that the convergence $\zkn{k}{n}'\left( \cdf{\Xnvec}\right) \ConvDist{n \rightarrow \infty} \zkeps{k}'\left( \cdf{\Xin}\right)$ is uniform over $k \in \mySet{N}$. 
			To that aim, we show that  $k\cdot\zkn{k}{n}'\left( \cdf{\Xnvec}\right) \ConvDist{n \rightarrow \infty} k\cdot\zkeps{k}'\left( \cdf{\Xin}\right)$ uniformly over $k\in \mySet{N}$; 
			Since the \acp{cdf} of $k\cdot\zkn{k}{n}'\left( \cdf{\Xnvec}\right)$ and $k\cdot\zkeps{k}'\left( \cdf{\Xnvec}\right)$ evaluated at $\alpha \in \mySet{R}$ are equal to the \acp{cdf} of $ \zkn{k}{n}'\left( \cdf{\Xnvec}\right)$ and $ \zkeps{k}'\left( \cdf{\Xnvec}\right)$ evaluated at $\alpha/k \in \mySet{R}$, respectively, then when  $k\cdot\zkn{k}{n}'\left( \cdf{\Xnvec}\right) \ConvDist{n \rightarrow \infty} k\cdot\zkeps{k}'\left( \cdf{\Xin}\right)$ uniformly over $k\in \mySet{N}$ it holds that $ \zkn{k}{n}'\left( \cdf{\Xnvec}\right) \ConvDist{n \rightarrow \infty}  \zkeps{k}'\left( \cdf{\Xin}\right)$ uniformly over $k\in \mySet{N}$. 
			
			Let $\Charac_S(\cdot)$ denote the characteristic function of an \ac{rv} $S$, i.e., $\Charac_S(\alpha) \triangleq \E \{e^{j\cdot\alpha\cdot S}\}$. We prove that convergence is uniform over $k\in \mySet{N}$ by showing that  $\Charac_{k\cdot\zkn{k}{n}}(\cdot)$ converges to $\Charac_{k\cdot\zkeps{k}'}(\cdot)$ uniformly over $k\in \mySet{N}$.
			To that aim, we define the \acp{rv}
			\begin{equation}
			\Vin[i] \triangleq \log \frac{\pdf{\Yn[i] | \Xscal_n[i]}\left( \Yn[i] | \Xscal_n[i]\right) }{\pdf{\Yn[i]}\left( \Yn[i] \right)}; \qquad
			\Veps[i]\triangleq \log \frac{\pdf{\Yeps[i] | \Xscal[i]}\left( \Yeps[i] | \Xscal[i]\right) }{\pdf{\Yeps[i]}\left( \Yeps[i] \right)}.
			\label{eqn:VrvDef}
			\end{equation} 	
			As the random vectors $\Xnvec^{(k)}$ and $\Xin^{(k)}$ have independent entries, and since the channels \eqref{eqn:AsnycModel1} and \eqref{eqn:AsnycModel2}  are both memoryless, it holds  that the sequence of pairs of \acp{rv} $\{\Vin[i],\Veps[i] \}_{i\in \mySet{N}}$ are mutually independent over $i$, and that 
			\ifFullVersion
			\begin{equation}
			k \cdot \zkn{k}{n}'\left( \cdf{\Xnvec}\right)  =  \sum\limits_{i=1}^{k}\Vin[i]; \qquad	
			k \cdot\zkeps{k}'\left( \cdf{\Xin}\right)  =  \sum\limits_{i=1}^{k}\Veps[i].
			\label{eqn:ZkInd1}
			\end{equation} 
			We next compute the characteristic functions of $\Vin[i]$ and of $\Veps[i]$. It follows from the Gaussianity of $\Xscal_n[i]$ and $\Xscal[i]$ that  
			\begin{align*}
			\Vin[i] &= \frac{1}{2}\left(\frac{\Yn^2[i]}{\sigma_{\Yn}^2[i]} - \frac{\left( \Yn[i] - \Xscal_n[i]\right)^2 }{\sigma_{\Wn}^2[i]} \right) + \frac{1}{2}\log \left( \frac{\sigma_{\Yn}^2[i]}{\sigma_{\Wn}^2[i]}\right); 	\\
			\Veps[i] &= \frac{1}{2}\left(\frac{\Yeps^2[i]}{\sigma_{\Yeps}^2[i]} - \frac{\left( \Yeps[i] - \Xscal[i]\right)^2 }{\sigma_{\Weps}^2[i]} \right) + \frac{1}{2}\log \left( \frac{\sigma_{\Yeps}^2[i]}{\sigma_{\Weps}^2[i]}\right).
			\end{align*} 
			Defining the Gaussian \acp{rv} $A_n[i] \triangleq \frac{\Yn[i]}{\sigma_{\Yn}[i]} + \frac{\Yn[i] - \Xscal_n[i]}{\sigma_{\Wn}[i]}$,  $B_n[i] \triangleq  \frac{\Yn[i]}{\sigma_{\Yn}[i]} - \frac{\Yn[i] - \Xscal_n[i]}{\sigma_{\Wn}[i]}$, $A_\myEps[i] \triangleq \frac{\Yeps[i]}{\sigma_{\Yeps}[i]} +  \frac{\Yeps[i] - \Xscal[i]}{\sigma_{\Weps}[i]}$, and  $B_\myEps[i] \triangleq \frac{\Yeps[i]}{\sigma_{\Yeps}[i]} - \frac{\Yeps[i] - \Xscal[i]}{\sigma_{\Weps}[i]}$, and the deterministic quantities $\beta_n[i] \triangleq \log \Big( \frac{\sigma_{\Yn}^2[i]}{\sigma_{\Wn}^2[i]}\Big)$ and  $\beta_\myEps[i] \triangleq \log \Big( \frac{\sigma_{\Yeps}^2[i]}{\sigma_{\Weps}^2[i]}\Big)$, it follows that $\Vin[i]$ and $\Veps[i]$ can be expressed as  
			\begin{equation}
			\Vin[i]= \frac{1}{2} A_n[i] \cdot B_n[i] +\frac{1}{2}\beta_n[i]; \qquad 
			\Veps[i]= \frac{1}{2}A_\myEps[i] \cdot B_\myEps[i]+ \frac{1}{2}\beta_\myEps[i]. 
			\label{eqn:VrvDef2}
			\end{equation}
			Since $\Yn[i]$ and $\Xscal_n[i]$ are zero-mean and $\E \left\{\left( \frac{\Yn[i]}{\sigma_{\Yn}[i]}\right) ^2 \right\} = \E \left\{ \left( \frac{\Yn[i] - \Xscal_n[i]}{\sigma_{\Wn}[i]}\right) ^2 \right\} = 1$, we obtain that $A_n[i]$ and $B_n[i]$, as well as $A_\myEps[i]$ and $B_\myEps[i]$,  are each a pair of jointly Gaussian and uncorrelated  \acp{rv}, hence mutually independent. 
			Thus, denoting  the variances of $A_n[i]$, $B_n[i]$, $A_\myEps[i]$,  and $B_\myEps[i]$ by letting $\sigma_{A_n}^2[i]$, $\sigma_{B_n}^2[i]$, $\sigma_{A_{\myEps}}^2[i]$, and $\sigma_{B_{\myEps}}^2[i]$, respectively, it follows from \eqref{eqn:VrvDef2} that for any $\alpha \in \mySet{R}$
			\begin{align}
			\Charac_{V_n[i]}(\alpha) 
			&\stackrel{(a)}{=} e^{j\frac{1}{2}{\beta_n[i]}\alpha} \Charac_{A_n[i] \cdot B_n[i]}(\alpha/2) \notag \\
			&\stackrel{(b)}{=} e^{j\frac{1}{2}{\beta_n[i]}\alpha} \E\big\{\Charac_{A_n[i] }\left((\alpha/2)   B_n[i]\right)  \big\} \notag \\
			&\stackrel{(c)}{=} e^{j\frac{1}{2}{\beta_n[i]}\alpha} \E\big\{e^{-\frac{1}{8}\alpha^2   B_n^2[i] \sigma_{A_n}^2[i]} \big\} \notag \\
			&\stackrel{(d)}{=} e^{j\frac{1}{2}{\beta_n[i]}\alpha} \int\limits_{b=-\infty}^\infty e^{-\frac{1}{8}\alpha^2   b^2 \sigma_{A_n}^2[i]}\frac{1}{\sqrt{2\pi\sigma_{B_n}^2[i]}}e^{-\frac{b^2}{2\sigma_{B_n}^2[i]}}db \notag \\
			&= e^{j\frac{1}{2}{\beta_n[i]}\alpha} \frac{1}{\sqrt{\alpha^2   \sigma_{B_n}^2[i] \sigma_{A_n}^2[i]/4 + 1} } \int\limits_{b=-\infty}^\infty \frac{1}{\sqrt{2\pi\frac{\sigma_{B_n}^2[i]}{\alpha^2   \sigma_{B_n}^2[i] \sigma_{A_n}^2[i] + 4}}} e^{-\frac{b^2}{2\frac{\sigma_{B_n}^2[i]}{\alpha^2   \sigma_{B_n}^2[i] \sigma_{A_n}^2[i] + 4}}}db 
			\label{eqn:CharFuncV1} \\ 
			&\stackrel{(e)}{=}  e^{j\frac{1}{2}{\beta_n[i]}\alpha} \frac{1}{\sqrt{\alpha^2   \sigma_{B_n}^2[i] \sigma_{A_n}^2[i]/4 + 1} },
			\label{eqn:CharFuncV2}
			\end{align}
			where $(a)$ follows from substituting \eqref{eqn:VrvDef2} into the definition of the characteristic function; 
			$(b)$ follows from the law of total expectation \cite[Ch. 7.4]{Papoulis:91}; 
			$(c)$ follows from \cite[Ch. 5.5]{Papoulis:91} as $A_n[i]$ is a zero-mean Gaussian \ac{rv} with variance  $\sigma_{A_n}^2[i]$; 
			$(d)$ holds since $B_n[i]$ is a zero-mean Gaussian \ac{rv} with variance  $\sigma_{B_n}^2[i]$;	
			and $(e)$ follows since the integrand in \eqref{eqn:CharFuncV1} is a Gaussian \ac{pdf} with zero mean and variance $\frac{\sigma_{B_n}^2[i]}{\alpha^2   \sigma_{B_n}^2[i] \sigma_{A_n}^2[i] + 1}$. 	
			The characteristic function $\Charac_{V_\myEps[i]}(\alpha)$ can be obtained by repeating the above arguments, which results in 
			\begin{align}
			\Charac_{V_\myEps[i]}(\alpha) =  e^{j\frac{1}{2}{\beta_\myEps[i]}\alpha} \frac{1}{\sqrt{\alpha^2   \sigma_{B_\myEps}^2[i] \sigma_{A_\myEps}^2[i] + 1} }.
			\label{eqn:CharFuncV3}
			\end{align}	
			
			Now, for each $\alpha \in \mySet{R}$, it follows from \eqref{eqn:ZkInd1}  that the characteristic functions of $k \cdot \zkn{k}{n}'\left( \cdf{\Xnvec}\right)$ and $k \cdot \zkeps{k}'\left( \cdf{\Xin}\right) $ are  given by
			\cite[Ch. 8.2]{Papoulis:91}
			\begin{equation*}
			\Charac_{k \cdot \zkn{k}{n}'}(\alpha) = \prod\limits_{i=1}^{k}\Charac_{\Vin[i]}\left( \alpha\right); \qquad 
			\Charac_{k \cdot \zkeps{k}'}(\alpha) = \prod\limits_{i=1}^{k}\Charac_{\Veps[i]}\left( \alpha\right).
			\end{equation*}	
			Fix  $\eta > 0$ and $k_0 \in \mySet{N}$.  For any $k > k_0$ it holds that
			\begin{align}
			\left|\Charac_{k\cdot\zkn{k}{n}'}(\alpha)\! -\! 	\Charac_{k\cdot\zkeps{k}'}(\alpha)   \right| 
			&= \bigg|\Charac_{k_0\cdot\zkn{k_0}{n}'}\left( \alpha\right)  \prod\limits_{i=k_0 +1}^{k}\Charac_{\Vin[i]}\left( \alpha\right) \! - \! \Charac_{k_0\cdot\zkeps{k_0}'}\left( \alpha\right)  \prod\limits_{i=k_0 +1}^{k}\Charac_{\Veps[i]}\left( \alpha\right)  \bigg| \notag \\
			&= \bigg|\Charac_{k_0\cdot\zkn{k_0}{n}'}\left(\alpha\right)  \left( \prod\limits_{i=k_0 +1}^{k}\Charac_{\Vin[i]}\left( \alpha\right) \! - \!\prod\limits_{i=k_0 +1}^{k}\Charac_{\Veps[i]}\left( \alpha\right) \right)  \notag \\
			&\qquad + \left( \Charac_{k_0\cdot\zkn{k_0}{n}'}\left(\alpha\right) \! - \!\Charac_{k_0\cdot\zkeps{k_0}'}\left(\alpha\right) \right) \prod\limits_{i=k_0 +1}^{k}\Charac_{\Veps[i]}\left(\alpha\right) \bigg| \notag  	\\
			&\le \left|\Charac_{k_0\cdot\zkn{k_0}{n}'}\left( \alpha\right)\right|  \left|  \prod\limits_{i=k_0 +1}^{k}\Charac_{\Vin[i]}\left(\alpha\right) \! - \!\prod\limits_{i=k_0 +1}^{k}\Charac_{\Veps[i]}\left( \alpha\right)  \right|\notag \\
			&\qquad + \prod\limits_{i=k_0 +1}^{k}\left|\Charac_{\Veps[i]}\left( \alpha\right)\right| \left| \Charac_{k_0\cdot\zkn{k_0}{n}'}\left( \alpha\right) \! - \!\Charac_{k_0\cdot\zkeps{k_0}'}\left( \alpha\right) \right|.
			\label{eqn:CharBound1}
			\end{align}
			Next, we note that by \eqref{eqn:CharFuncV2}-\eqref{eqn:CharFuncV3} and \cite[Ch. 5.5]{Papoulis:91}, the characteristic functions are uniformly continuous  with magnitude smaller than one, except when evaluated at $\alpha = 0$, in which the function is equal to $1$. 
			Thus, by defining  $\theta_k(\alpha) \triangleq \prod\limits_{i=k_0 +1}^{k}\Charac_{\Vin[i]}\left(\alpha\right) \! - \!\prod\limits_{i=k_0 +1}^{k}\Charac_{\Veps[i]}\left( \alpha\right)$ for $k >k_0$, it holds that:
			\begin{enumerate}[label={\em Q\arabic*}]
				\item \label{itm:Pr1} The function $\theta_k(\alpha)$ is uniformly continuous and $\theta_k(0)=0$  for each $k > k_0$ .
				\item \label{itm:Pr2} Since $\big|\Charac_{\Vin[i]}\left(\alpha\right)\big| < 1$ and $\big|\Charac_{\Veps[i]}\left(\alpha\right)\big| < 1$ for each $\alpha >0$, it holds that $\mathop{\lim}\limits_{k \rightarrow \infty} \theta_k(\alpha) = 0$. 
			\end{enumerate}	
			We now define $\tilde{\alpha}_0(k, \eta)$ to be the smallest positive value such that for each $|\alpha|<\tilde{\alpha}_0(k, \eta)$ we have that $| \theta_k(\alpha)| < \frac{\eta}{2}$. It follows from \ref{itm:Pr1} that $\tilde{\alpha}_0(k, \eta) > 0$ for each $k>k_0$. Furthermore, since the magnitude of the characteristic function \eqref{eqn:CharFuncV3} is monotonically decreasing in $\alpha$,  it follows from \ref{itm:Pr2} that  $ \mathop{\lim}\limits_{k \rightarrow \infty}\tilde{\alpha}_0(k, \eta)  = \infty$. The fact that  $ \mathop{\lim}\limits_{k \rightarrow \infty}\tilde{\alpha}_0(k, \eta)  = \infty$ implies that there exists a finite $ k_1(\eta)$ such that  	$\alpha_0(\eta) \triangleq \mathop{\inf}\limits_{k > k_0}\tilde{\alpha}_0(k, \eta) =  \tilde{\alpha}_0(k_1(\eta), \eta)$. Consequently, $\alpha_0(\eta) $  is strictly larger than zero, and   $| \theta_k(\alpha)| = \Big|\prod\limits_{i=k_0 +1}^{k}\Charac_{\Vin[i]}\left(\alpha\right) \! - \!\prod\limits_{i=k_0 +1}^{k}\Charac_{\Veps[i]}\left( \alpha\right)\Big| < \frac{\eta}{2} $ for all $k > k_0$. For values of $\alpha \ge \alpha_0(\eta)$, it holds that $\big|\Charac_{\Vin[i]}\left(\alpha\right)\big| < 1$ and $\big|\Charac_{\Veps[i]}\left(\alpha\right)\big| < 1$, thus $\big|\theta_k(\alpha)\big| < \frac{\eta}{2} $ for all sufficiently large $k \in \mySet{N}$ and $\forall\alpha \in \mySet{R}$. 
			Furthermore, since $\zkn{k}{n}'\left( \cdf{\Xnvec}\right) \ConvDist{n \rightarrow \infty} \zkeps{k}'\left( \cdf{\Xin}\right)$, it follows from \cite[Ch. 8.8]{Lebanon:12} that $\exists n_0(\eta, k_0) \in \mySet{N} $ such that $\forall n > n_0(\eta, k_0)$, $\big|\Charac_{k_0\zkn{k_0}{n}'}(\alpha) - \Charac_{k_0\zkeps{k_0}'}(\alpha) \big| < \frac{\eta}{2}$ for all $\alpha \in \mySet{R}$. 
			Substituting this into \eqref{eqn:CharBound1} we obtain that  
			\begin{align}
			\left|\Charac_{k\cdot\zkn{k}{n}'}(\alpha)\! -\! 	\Charac_{k\cdot\zkeps{k}'}(\alpha)   \right| 
			&\le \left|\Charac_{k\cdot\zkn{k_0}{n}'}\left( \alpha\right)\right|   \cdot  \frac{\eta}{2} + \prod\limits_{i=k_0 +1}^{k}\left|\Charac_{\Veps[i]}\left( \alpha\right)\right|\cdot  \frac{\eta}{2}   
			\le \eta,
			\label{eqn:CharBound2}
			\end{align} 
			for all sufficiently large $k \in \mySet{N}$ and $\forall\alpha \in \mySet{R}$. 
			Eqn. \eqref{eqn:CharBound2} implies that for all $\eta>0$,  $	\left|\Charac_{k\cdot\zkn{k}{n}'}(\alpha) - 	\Charac_{k\cdot\zkeps{k}'}(\alpha)   \right| < \eta$ for all sufficiently large $k \in \mySet{N}$ and $\forall\alpha \in \mySet{R}$. Thus, by Levy's Theorem \cite[Prop. 8.8.1]{Lebanon:12} it follows that $k\cdot\zkn{k}{n}'\left( \cdf{\Xnvec}\right) \ConvDist{n \rightarrow \infty} k\cdot\zkeps{k}'\left( \cdf{\Xin}\right)$ and that this convergence is uniform over $k\in \mySet{N}$.   
			%%%%%%%%%%%%%%%%%%%%%%%%%%%%%%%%%
			\fi
		\end{IEEEproof}

		Before stating the next lemma, we recall that for a fixed $n \in \mySet{N}$,  $\cdf{\Xnvec}\opt$ is the optimal input distribution for the channel \eqref{eqn:AsnycModel2} subject to \eqref{eqn:AsyncConst1}.
		
		\begin{lemma}
			\label{lem:AsyncZk2}
			For any fixed $n$, every subsequence of $\big\{\zkn{k}{n}'\left( \cdf{\Xnvec}\opt\right)\big\}_{k \in \mySet{N}}$ converges in distribution in the limit $k \rightarrow \infty$ to a finite deterministic  scalar.
		\end{lemma}
		\begin{IEEEproof}
			Recall that the \acp{rv} $\zkn{k}{n}'\left( \cdf{\Xnvec}\right)$  represent the mutual information density rate between the input and the output of the channel defined in \eqref{eqn:AsnycModel2}, when the input is distributed according to $\cdf{\Xnvec}$. The channel \eqref{eqn:AsnycModel2} is a memoryless additive cyclostationary Gaussian noise channel, thus, by \cite{Shlezinger:16b}, it can be equivalently represented as a multivariate memoryless additive stationary Gaussian noise channel. The channel corresponding to the equivalent representation is {\em information stable} \cite[Sec. 1.5]{Dobrushin:63} (see \cite[Eq. (3.9.2)]{Han:03} for the definition of information stable channels). 
			%	Information stable channels can be roughly described as having the property that the capacity-achieving input  and its corresponding output behave	ergodically.
			For such channels, as noted in \cite[Remark 3.5.2]{Han:03}, when the input obeys the capacity-achieving distribution $\cdf{\Xnvec}\opt$, the mutual information density rate converges as $k$ increases almost surely to the finite and deterministic  mutual information rate. Since almost sure convergence implies convergence in distribution \cite[Lemma 7.21]{Kosorok:07}, this proves the lemma. 
		\end{IEEEproof}

		%-----------------------------------
		%	Proving the theorem
		%-----------------------------------
		\vspace{-0.2cm}
		\subsection[text]{Showing that $\Capacity_\myEps = \mathop{\lim \inf}\limits_{n \rightarrow \infty} \Capacity_n$}
		\label{app:proof2c}
		\vspace{-0.1cm}
		We are now ready to prove the conclusion of Theorem \ref{thm:AsycCap}.
		We first note that from \cite[Thm. 3.2.1]{Han:03} it follows that the capacities of the channels \eqref{eqn:AsnycModel2} and \eqref{eqn:AsnycModel1} are given by $\Capacity_n = \mathop{\max} \limits_{\cdf{\Xnvec}}\left\{ {\rm p-}\mathop{\lim \inf}\limits_{k \rightarrow \infty} \zkn{k}{n}'\left( \cdf{\Xnvec}\right)\right\}$ and  $\Capacity_\myEps = \mathop{\max}  \limits_{\cdf{\Xin}}\left\{{\rm p-}\mathop{\lim \inf}\limits_{k \rightarrow \infty} \zkeps{k}'\left( \cdf{\Xin}\right)\right\} $, respectively. The next lemma characterizes the capacity-achieving distribution  $\cdf{\Xnvec}\opt$: 
		
		\begin{lemma}
			\label{lem:tightness}
			The capacity achieving distribution for $\Xnvec^{(k)}$, $\cdf{\Xnvec}\opt$,  is Gaussian with independent entries and has a subsequence (in the index $n$) which converges in distribution to a multivariate Gaussian random vector uniformly with respect to $k\in \mySet{N}$.
		\end{lemma}
		
		\begin{IEEEproof}
			The Gaussianity of $\Xnvec^{(k)}$ follows from \cite[Thm. 1]{Shlezinger:16b}. For a fixed $k\in \mySet{N}$, every limit distribution of every convergent subsequence of $\Xnvec^{(k)}$ in the index $n$ is also Gaussian \cite[Ch. 4.3]{Neerven:16}.   
			\ifFullVersion
			In particular, for any fixed $n$, it follows from \cite[Thm. 1]{Shlezinger:16b} that the optimal input distribution for 
			a channel with additive zero-mean, memoryless, cyclostationary, Gaussian noise $W_n[i]$ with variance $\Sigwn[i]$,
			\else
			In particular, for a fixed $n$, it follows from \cite[Thm. 1]{Shlezinger:16b} that the optimal input distribution for 
			the channel \eqref{eqn:AsnycModel2} subject to \eqref{eqn:AsyncConst1} 
			\fi 
			is a temporally independent zero-mean Gaussian process with  variance $\sigma_{X_n}^2[i] \triangleq \E\left\{\big(X_n[i] \big)^2  \right\}$,  which satisfies:
			\ifFullVersion
			\begin{equation}
			\label{eqn:SigmaX}
			\sigma_{X_n}^2[i] = \left(\TDelta_n - \Sigwn[i] \right)^+,
			\end{equation}
			where $\TDelta_n > 0$  satisfies \eqref{eqn:MainThmCst}, namely, 
			\begin{equation*} 
			\frac{1}{  \Td_n}\sum\limits_{i=0}^{\Td_n -1}   \left(\TDelta_n - \Sigwn[i] \right)^+  =  \PCst.
			\vspace{-0.1cm}
			\end{equation*}	
			\else
			$	\sigma_{X_n}^2[i] = \left(\TDelta_n - \Sigwn[i] \right)^+$, 
			where $\TDelta_n > 0$  satisfies \eqref{eqn:MainThmCst}, namely, 
			$\frac{1}{  \Td_n}\sum\limits_{i=0}^{\Td_n -1}   \left(\TDelta_n - \Sigwn[i] \right)^+  =  \PCst$. 
			\fi
			Consequently, if  $\TDelta_n \ge \mathop{\max}\limits_{t \in \mySet{P}} \Cwc(t) \ge \mathop{\max}\limits_{i \in \mySet{N}} \Sigwn[i]$, then 
			\vspace{-0.25cm}
			\begin{align}
			\TDelta_n 
			&= \PCst + 	\frac{1}{  \Td_n}\sum\limits_{i=0}^{\Td_n -1}   \Sigwn[i]  
			%	\notag \\
			%	&
			\le \PCst +\mathop{\max}\limits_{i \in \mySet{N}} \Sigwn[i] 
			%	\notag \\
			%	&
			\le \PCst +\mathop{\max}\limits_{t \in \mySet{P}} \Cwc(t). 
			\label{eqn:TDeltaBound1}
			\vspace{-0.3cm}
			\end{align}
			
			\vspace{-0.25cm}
			\noindent
			It follows from \eqref{eqn:TDeltaBound1} that the sequence $\{\TDelta_n\}_{n \in \mySet{N}}$ is bounded in the interval $[0, \PCst +\mathop{\max}\limits_{t \in \mySet{P}} \Cwc(t)]$ for all $n \in \mySet{N}$. Thus, by the Bolzano-Weirstrass Theorem \cite[Thm. 2.42]{Rudin:76}, it has a convergent subsequence, and we let $n_1 < n_2 < \ldots$ denote the indexes of this convergent subsequence. %, implying that $\forall i \in \mySet{N}$,  $\{\sigma_{\Xnvec}^2[i]\}_{n \in \mySet{N}}$ has a convergent subsequence. 
			
			Next, we recall
			\ifFullVersion
			from \eqref{eqn:limcorr} 
			\fi
			% TODO NIR CONTINUE FROM HERE
			that the subsequence $\sigma_{w_{n_l}}^2[i]$ is pointwise convergent as $l \rightarrow \infty$. Consequently, the subsequence $\{\TDelta_{n_l} - \sigma_{w_{n_l}}^2[i]  \}_{l \in \mySet{N}}$ converges as $l \rightarrow \infty$, and thus, by the \ac{cmt} \cite[Thm. 7.7]{Kosorok:07}, $\{\sigma_{{X}_{n_l}}^2[i]\}_{l \in \mySet{N}}$ 
			\ifFullVersion
			given in \eqref{eqn:SigmaX} 
			\fi
			also converges as $l \rightarrow \infty$ for each $i \in \mySet{N}$.
			Since the optimal input has temporally independent elements, it follows from the above description that the elements of the length $k$ vector  $\Xnvec^{(k)}$ are mutually independent zero-mean Gaussian \acp{rv}, with variance $\E \Big\{ \big(\Xnvec^{(k)} \big)_i^2 \Big\} \equiv  \E\left\{\big(X_n[i] \big)^2  \right\}  =  \sigma_{X_{n}}^2[i]$ for each $i \in 1,2,\ldots, k$. It now follows from the proof of Lemma \ref{Lem:PDF-convergence} that convergence of the sequence $\big\{ \sigma_{X_{n_l}}^2[i]\big\}_{l \in \mySet{N}}$ as $l \rightarrow \infty$ for each $i\in \mySet{N}$ implies that the sequence of Gaussian   random vectors $\{\Xin_{n_l}^{(k)}\}_{l \in \mySet{N}}$ converges in distribution as $l \rightarrow \infty$ to a Gaussian distribution for any fixed  
			\ifFullVersion	
			$k \in \mySet{N}$.	
			
			Lastly, we show that this convergence is uniform with respect to $k\in\mySet{N}$. 
			Similarly to the proof\myFtn{The proof of Lemma \ref{Lem:PDF-convergence} shows when the variances of a Gaussian random vector with a non-singular diagonal covariance matrix converge uniformly, then it converges in distribution. Here,  $\sigma_{\Xin_{n}}^2[i]$ given in \eqref{eqn:SigmaX} can be zero, thus the covariance of $\Xnvec^{(k)}$ can be singular. Nonetheless, as detailed in \cite[Ch. 3.4.3]{Gallager:13}, the \ac{pdf} of Gaussian random vectors with singular covariance can be obtained by removing its deterministically dependent entries, resulting in an equivalent non-singular covariance matrix, and thus the uniform convergence of $\sigma_{\Xin_{n}}^2[i]$ implies that $\Xnvec^{(k)}$ converges in distribution.} of Lemma \ref{Lem:PDF-convergence}, we fix $\eta > 0$ and $k_0 \in \mySet{N}$, and show that $\exists l_0(\eta, k_0)$ such that for all $l,m > l_0(\eta, k_0)$ and sufficiently large $k \in \mySet{N}$, the \acp{cdf} of $\Xin_{n_l}^{(k)}$ and $\Xin_{n_m}^{(k)}$ satisfy
			$ |\cdf{\Xin_{n_l}^{(k)}}(\xvec^{(k_0)}) - \cdf{\Xin_{n_m}^{(k)}}(\xvec^{(k_0)})| < \eta $ for all $\xvec \in \mySet{R}^{k}$. 
			Define  the Gaussian function  $Q(\sigma^2, x) \triangleq \int\limits_{\alpha=-\infty}^{x}\frac{1}{\sqrt{2\pi \sigma^2}} e^{-\frac{\alpha^2}{2\sigma^2}d\alpha}$.
			Since the \ac{cdf} of $\Xin_{n_l}^{(k_0)}$ is pointwise convergent as $l \rightarrow \infty$, where convergence is uniform in ${\xvec}^{(k_0)} \in \mySet{R}^k$ as  $\Xin_{n_l}^{(k_0)}$ is Gaussian, it follows that $\exists l_0(\eta, k_0)$ such that for all $l,m > l_0(\eta, k_0)$ it holds that 
			\vspace{-0.1cm}
			\begin{equation}
			\mathop{\sup}\limits_{\xvec^{(k_0)} \in \mySet{R}^{k_0}}\left|\cdf{\Xin_{n_l}^{(k_0)}}\left( \xvec^{(k_0)}\right)  - \cdf{\Xin_{n_m}^{(k_0)}}\left( \xvec^{(k_0)}\right) \right| < \frac{\eta}{2}, \quad \forall{\xvec}^{(k_0)} \in \mySet{R}^k. 
			\label{eqn:UnifromConva}
			\vspace{-0.1cm}
			\end{equation} 
			Consequently, for all $k > k_0$, we have that 
			\vspace{-0.1cm}
			\begin{align}
			\left|\cdf{\Xin_{n_l}^{k}}\! \left( \xvec^{(k)}\right) \! -\! \cdf{\Xin_{n_m}^{k}} \!\left( \xvec^{(k)}\right) \right| 
			&= 	\left|\cdf{\Xin_{n_l}^{(k_0)}}\! \left( \xvec^{(k_0)}\right)\!\prod\limits_{i=k_0\!+\!1}^{k}\! Q(\sigma_{\Xin_{n_l}}^2[i], x_i)  \!-\! \cdf{\Xin_{n_m}^{(k_0)}} \!\left( \xvec^{(k_0)}\right)\!\prod\limits_{i=k_0\!+\!1}^{k}\! Q(\sigma_{\Xin_{n_m}}^2[i], x_i) \right|  \notag \\
			&\le \cdf{\Xin_{n_l}^{(k_0)}} \left( \xvec^{(k_0}\right) \left|\prod\limits_{i=k_0+1}^{k} Q(\sigma_{\Xin_{n_l}}^2[i], x_i)  \!-\! \prod\limits_{i=k_0+1}^{k} Q(\sigma_{\Xin_{n_m}}^2[i], x_i)    \right| \notag \\
			&\quad \!+\! 	
			\prod\limits_{i=k_0+1}^{k} Q(\sigma_{\Xin_{n_m}}^2[i], x_i) \left|\cdf{\Xin_{n_l}^{(k_0)}} \left( \xvec^{(k_0)}\right) \!-\!\cdf{\Xin_{n_m}^{(k_0)}} \left( \xvec^{(k_0)}\right) \right|.
			\label{eqn:UnifromConv1}
			\vspace{-0.1cm}
			\end{align}
			Next, we note that while, in general, $\sigma_{\Xin_{n}}^2[i]$ can be zero,  each period of the optimal cyclostationary input must include at least a single index $i$ for which  $\sigma_{\Xin_{n}}^2[i]>0$. Consequently, for sufficiently large $k$, $0 <\prod\limits_{i=k_0+1}^{k} Q(\sigma_{\Xin_{n_l}}^2[i], x_i) < \frac{\eta}{2}$ and $0 <\prod\limits_{i=k_0+1}^{k} Q(\sigma_{\Xin_{n_m}}^2[i], x_i) < \frac{\eta}{2}$. Substituting these two inequalities and the fact that $\cdf{\Xin_{n_l}^{(k_0)}} <1$  into \eqref{eqn:UnifromConv1} results in 
			\vspace{-0.1cm}
			\begin{align}
			\left|\cdf{\Xin_{n_l}^{(k)}} \left( \xvec^{(k)}\right)  - \cdf{\Xin_{n_m}^{(k)}} \left( \xvec^{(k)}\right) \right| 	 
			&\le \frac{\eta}{2} + \frac{\eta}{2} \left|\cdf{\Xin_{n_l}^{(k_0)}} \left( \xvec^{(k_0)}\right) -\cdf{\Xin_{n_m}^{(k_0)}} \left( \xvec^{(k_0)}\right) \right| 
			%	 \notag \\
			%	&
			\stackrel{(a)}{\le} \frac{\eta}{2}\left(1 + \frac{\eta}{2} \right), 
			\label{eqn:UnifromConv2}
			\vspace{-0.1cm}
			\end{align}
			where $(a)$ follows from \eqref{eqn:UnifromConva}. Eqn. \eqref{eqn:UnifromConv2} implies that for all sufficiently small $\eta<1$, we have that $	\left|\cdf{\Xin_{n_l}^{(k)}} \left( \xvec^{(k)}\right)  - \cdf{\Xin_{n_m}^{(k)}} \left( \xvec^{(k)}\right) \right|  < \eta$ for all sufficiently large $k$, thus concluding the proof. 
			Note that differently from Lemma \ref{lem:AsyncConvDist}, here we did not need an assumption on the variance to show uniform convergence in $k \in \mySet{N}$, as here we consider convergence in distribution (of the \acp{cdf}) while in Lemma \ref{lem:AsyncConvDist} we considered convergence of the \acp{pdf}. 
			\else 
			$k \in \mySet{N}$, and  that this convergence is uniform in $k \in\mySet{N}$. 
			\fi
		\end{IEEEproof}

		\begin{lemma}
			\label{lem:CapLowBound}
			$\Capacity_\myEps \ge \mathop{\lim \inf}\limits_{n \rightarrow \infty} \Capacity_n$, and a rate of $\mathop{\lim \inf}\limits_{n \rightarrow \infty} \Capacity_n$ is achievable for the channel \eqref{eqn:AsnycModel1} when the input obeys a Gaussian distribution.
		\end{lemma}
		\begin{IEEEproof}
			From Lemma \ref{lem:tightness} it follows that the sequence of distributions with independent entries $\{\cdf{\Xnvec}\opt \}_{n \in \mySet{N}}$  has a convergent subsequence, i.e., there exists a set of indexes $n_1 < n_2 < \ldots$ such that the sequence of distributions with independent entries $\{\cdf{\Xnvec[n_l]}\opt \}_{l \in \mySet{N}}$ converges in the limit $l \rightarrow \infty$ to a   Gaussian distribution $\cdf{\Xin}'$ with independent entries.
			\ifFullVersion	
			By lemma  \ref{lem:tightness}, each $k \times 1$ vector $\Xnvec[{n_l}]^{(k)}$ distributed via $\cdf{\Xnvec[n_l]}\opt$ converges to the $k \times 1$ vector $\Xin^{(k)}$ which is distributed via  $\cdf{\Xin}'$, and this convergence is uniform in $k \in \mySet{N}$.  Therefore, $\{\Xnvec[{n_l}]^{(k)}\}_{l \in \mySet{N}}$ and $\Xin^{(k)}$ satisfy the conditions on Lemma \ref{lem:AsyncZk}, as $\{\Xnvec[{n_l}]^{(k)}\}_{l \in \mySet{N}}$ is a sequence of $k \times 1$ zero-mean Gaussian random vectors with independent entries which converges in distribution as $l\rightarrow \infty$ uniformly in $k \in \mySet{N}$ to the zero-mean Gaussian random vectors with independent entries   $\Xin^{(k)}$. 
			\fi
			With this input distribution,  
			it follows from Lemma \ref{lem:AsyncZk}  that $\zkn{k}{n_l}'\big( \cdf{\Xnvec[n_l]}\opt\big)  \ConvDist{l \rightarrow \infty} \zkeps{k}'\left( \cdf{\Xin}'\right)$ uniformly with respect to $k \in \mySet{N}$. 
			By Lemma \ref{lem:AsyncZk2}, every subsequence of $\big\{\zkn{k}{n_l}'\big( \cdf{\Xnvec[n_l]}\opt\big)\big\}_{l \in \mySet{N}}$ converges in distribution to a finite deterministic scalar for $k \rightarrow \infty$.
			Thus, by Theorem \ref{thm:plim} we have that
			\vspace{-0.1cm}
			\begin{align}
			\mathop{\lim}\limits_{l \rightarrow \infty}\left( {\rm p-}\mathop{\lim \inf}\limits_{k \rightarrow \infty} \zkn{k}{n_l}'\left( \cdf{\Xnvec[n_l]}\opt\right) \right)
			&= {\rm p-}\mathop{\lim \inf}\limits_{k \rightarrow \infty} \zkeps{k}'\left( \cdf{\Xin}'\right) \notag \\
			& \le \mathop{\max} \limits_{\cdf{\Xin}}\left\{  {\rm p-}\mathop{\lim \inf}\limits_{k \rightarrow \infty} \zkeps{k}'\left( \cdf{\Xin}\right)\right\} 
			%		\notag \\
			%		&
			= \Capacity_\myEps.
			\label{eqn:CapLowBound1}
			\vspace{-0.1cm}
			\end{align}
			Noting that 
			%\ifFullVersion	
			by definition of $\Capacity_n$ we have that 
			%\fi
			$\Capacity_n =  {\rm p-}\mathop{\lim \inf}\limits_{k \rightarrow \infty} \zkn{k}{n}'\left( \cdf{\Xnvec}\opt\right)$, then from \eqref{eqn:CapLowBound1} it follows that
			\vspace{-0.1cm}
			\begin{align}
			\Capacity_\myEps
			&\ge \mathop{\lim}\limits_{l \rightarrow \infty}\Capacity_{n_l} 
			%	\notag \\
			%	&
			\stackrel{(a)}{\ge} \mathop{\lim \inf}\limits_{n \rightarrow \infty} \Capacity_n,
			\label{eqn:CapLowBound2}
			\vspace{-0.1cm}
			\end{align}
			where $(a)$ follows since, by definition, the limit of every subsequence is not smaller than the  limit inferior \cite[Pg. 56]{Rudin:76}. 
			Noting that $\cdf{\Xin}'$ is Gaussian by Lemma \ref{lem:tightness} concludes the proof. 
			%	
			%	Lastly, 	Since $\{\cdf{\Xnvec[n_l]}\opt \}_{l \in \mySet{N}}$ are Gaussian distributions, it follows that $\cdf{\Xin}'$ is also a Gaussian distribution \cite[Ch. 4.3]{Neerven:16}.
		\end{IEEEproof}

		\begin{lemma}
			\label{lem:CapUpBound}
			$\Capacity_\myEps \le \mathop{\lim \inf}\limits_{n \rightarrow \infty} \Capacity_n$.%\footnote{Here I cannot assume only Gaussian inputs and state that this holds. What I can say is, by combining this lemma with the previous one, that the maximum achievable rate of the async channel with Gaussian inputs equals $\mathop{\lim \inf}\limits_{n \rightarrow \infty} \Capacity_n$.  }
		\end{lemma}
		
		\begin{IEEEproof}
			To prove the lemma, we note that by the general formula of \cite[Eq. (1.3)]{Verdu:94}:
			\vspace{-0.1cm}
			\begin{align}
			\Capacity_\myEps &\le \mathop{\lim \inf}\limits_{k \rightarrow \infty}  \mathop{\sup}\limits_{ \cdf{\Xin^{(k)}}}\frac{1}{k} I\left(\Xin^{(k)} ; \Yi_{\myEps}^{(k)} \right).  
			\label{eqn:CapBound1}
			\vspace{-0.1cm}
			\end{align}	
			Let $\cdf{\Xin}\opt$ be the distribution which achieves the right hand side of \eqref{eqn:CapBound1}\myFtn{As noted in \cite[Remark 3.2.2]{Han:03}, there exists an optimal input distribution which maximizes the mutual information, thus the $\sup$ statement in \eqref{eqn:CapBound1} can be replaced with $\max$. }. Note that for each channel input distribution $\cdf{\Xin}$, 
			\ifFullVersion
			\vspace{-0.1cm}
			\begin{align}
			I\left(\Xin^{(k)} ; \Yi_{\myEps}^{(k)} \right) 
			&= h\left( \Yi_{\myEps}^{(k)}\right) - h\left(\Wepsvec^{(k)} \right) 
			%\notag \\
			%&
			\stackrel{(a)}{=}h\left( \Yi_{\myEps}^{(k)}\right) - \sum\limits_{i=1}^{k}h\big(\Weps[i]\big)  \notag \\
			&\stackrel{(b)}{\le} \sum\limits_{i=1}^{k} \Big( h\left( \Yeps[i]\right) -h\big(\Weps[i]\big)\Big) , 
			\label{eqn:CapBound1a}
			\vspace{-0.1cm}
			\end{align}
			\else
			\vspace{-0.1cm}
			\begin{align}
			I\left(\Xin^{(k)} ; \Yi_{\myEps}^{(k)} \right) 
			%&= h\left( \Yi_{\myEps}^{(k)}\right) - h\left(\Wepsvec^{(k)} \right) 
			%\notag \\
			%&
			\stackrel{(a)}{=}h\left( \Yi_{\myEps}^{(k)}\right) - \sum\limits_{i=1}^{k}h\big(\Weps[i]\big)  
			%\notag \\
			%&
			\stackrel{(b)}{\le} \sum\limits_{i=1}^{k}\Big( h\left( \Yeps[i]\right) -h\big(\Weps[i]\big)\Big), 
			\label{eqn:CapBound1a}
			\vspace{-0.1cm}
			\end{align}
			\fi
			where $(a)$ holds since the noise $\Weps[i]$ is memoryless, and $(b)$ follows from \cite[Thm. 8.62]{Cover:06}. 
			\ifFullVersion
			We note that the inequality  in \eqref{eqn:CapBound1a} is achievable  with equality when  $\cdf{\Xin}\opt$ is 
			memoryless \cite[Ch. 9.2]{Cover:06}.
			\else
			We note that the rightmost term in \eqref{eqn:CapBound1a} is achievable  with equality when  $\cdf{\Xin}\opt$ is 
			memoryless. 
			\fi 
			Furthermore, by \cite[Thm. 8.65]{Cover:06}, for any coice of $\big\{\E \{X^2[i]\} \big\}_{i \in \mySet{N}}$, \eqref{eqn:CapBound1a} is maximized when each $X[i]$ is Gaussian. Since \eqref{eqn:CapBound1a} holds  $\forall k \in \mySet{N}$, it follows that the optimal input distribution $\cdf{\Xin}\opt$ is memoryless and Gaussian. 
			
			%Since the channel \eqref{eqn:AsnycModel1} is memoryless, it holds that $\cdf{\Xin}\opt$ defines a temporally independent source\footnote{The fact that memoryless sources are optimal for memoryless channels follows from \cite[Pg. 187]{Han:03}, which is stated for channels with finite-alphabet inputs. However, following the rationale in \cite{Wyner:78}, this statement also holds for continuous-alphabets.}.

			Recalling that $\zkeps{k}' \left(\cdf{\Xin}\right)$, defined in \eqref{eqn:zkdefs}, is the mutual information density rate for input distribution $\cdf{\Xin}$, whose expected value is the  mutual information \cite[Ch. 3.3]{Han:03}, it follows   that \eqref{eqn:CapBound1} can be equivalently stated as
			\vspace{-0.1cm}
			\begin{align}
			\Capacity_\myEps &\le \mathop{\lim \inf}\limits_{k \rightarrow \infty} \E \left\{\zkeps{k}' \left(\cdf{\Xin}\opt \right)  \right\}.
			\label{eqn:CapBound2}
			\vspace{-0.1cm}
			\end{align}
			Next, 
			\ifFullVersion
			consider $ \mathop{\lim \inf}\limits_{k \rightarrow \infty} \E \left\{\zkeps{k}' \left(\cdf{\Xin}\opt \right)  \right\}$: Let
			\else 
			we let
			\fi
			$k_1 < k_2 < \ldots$ be the set of indexes of the subsequence of $\E \left\{\zkeps{k}' \left(\cdf{\Xin}\opt \right)  \right\}$ whose limit is equal to the limit inferior, i.e., $ \mathop{\lim}\limits_{l \rightarrow \infty} \E \left\{\zkeps{k_l}' \left(\cdf{\Xin}\opt \right)  \right\} = \mathop{\lim \inf}\limits_{k \rightarrow \infty} \E \left\{\zkeps{k}' \left(\cdf{\Xin}\opt \right)  \right\}$. Since by Lemma \ref{lem:AsyncZk}, the sequence of non-negative \acp{rv} $\left\{\zkn{k_l}{n}' \left(\cdf{\Xin}\opt \right)  \right\}_{n \in \mySet{N}}$ convergences in {distribution\myFtn{Note that the conditions of Lemma \ref{lem:AsyncZk} hold as both the input distribution is the same for both $\zkn{k_l}{n}' \left(\cdf{\Xin}\opt \right)  $ and $\zkeps{k_l}' \left(\cdf{\Xin}\opt \right)$.}} to $\zkeps{k_l}' \left(\cdf{\Xin}\opt \right)$ as $n \rightarrow \infty$ uniformly in $k \in \mySet{N}$, it follows from\myFtn{\cite[Thm. 3.4]{Billingsley:99} states that if $X_n \ConvDist{n \rightarrow\infty} X$ then $\E\{|X|\} \le \mathop{\lim\inf}\limits_{n \rightarrow \infty} \E\{|X_n |\}$. } \cite[Thm. 3.4]{Billingsley:99} that 
			$\E \left\{\zkeps{k_l}' \left(\cdf{\Xin}\opt \right)  \right\} \le  \mathop{\lim\inf}\limits_{n \rightarrow \infty} \E \left\{\zkn{k_l}{n}' \left(\cdf{\Xin}\opt \right)  \right\}$. Consequently, 
			Eq. \eqref{eqn:CapBound2} can now be written as 
			\vspace{-0.1cm}
			\begin{align}
			\Capacity_\myEps 
			&\le \mathop{\lim}\limits_{l \rightarrow \infty} \E \left\{\zkeps{k_l}' \left(\cdf{\Xin}\opt \right)  \right\} 
			%\notag \\
			%&
			\le \mathop{\lim}\limits_{l \rightarrow \infty} \mathop{\lim\inf}\limits_{n \rightarrow \infty} \E \left\{\zkn{k_l}{n}' \left(\cdf{\Xin}\opt \right)  \right\} 
			%\notag \\
			%&
			\stackrel{(a)}{=} \mathop{\lim\inf}\limits_{n \rightarrow \infty} \mathop{\lim}\limits_{l \rightarrow \infty} \E \left\{\zkn{k_l}{n}' \left(\cdf{\Xin}\opt \right)  \right\}
			\notag \\ 
			&
			\le  \mathop{\lim \inf}\limits_{n \rightarrow \infty} \mathop{\lim}\limits_{l \rightarrow \infty}  \mathop{\sup}\limits_{\cdf{\Xin}} \E \left\{\zkn{k_l}{n}' \left(\cdf{\Xin}  \right)  \right\}
			% \notag \\
			%&
			\stackrel{(b)}{=}  \mathop{\lim \inf}\limits_{n \rightarrow \infty} \mathop{\lim}\limits_{l \rightarrow \infty}  \mathop{\sup}\limits_{ \cdf{\Xin^{(k_l)}}}\frac{1}{k_l} I\left(\Xin^{(k_l)} ; \Yi_n^{(k_l)} \right),
			\label{eqn:CapBound3}
			\vspace{-0.1cm}
			\end{align}	
			%where $(a)$ follows by applying Lemma \ref{lem:AsyncZk} with the sequence of distributions  $\cdf{\Xnvec} = \cdf{\Xin}\opt$, i.e., the input distribution does not vary with $n$; 
			where $(a)$ follows since the convergence  $\zkn{k}{n}' \left(\cdf{\Xin}\opt \right)\ConvDist{n\rightarrow \infty}\zkeps{k_l}' \left(\cdf{\Xin}\opt \right)$ is uniform with respect to $k$, thus the limits are interchangeable \cite[Thm. 7.11]{Rudin:76}; and $(b)$ holds since mutual information is the expected value of the mutual information density rate \cite[Ch. 2.3]{Cover:06}. 
			%and $(c)$ follows since, by writing the real valued sequence $\{\alpha_n\}_{n \in \mySet{N}}$ with entries $\alpha_n \triangleq  \mathop{\lim}\limits_{l \rightarrow \infty} \E \left\{\zkn{k_l}{n} \left(\cdf{\Xin}\opt \right)  \right\}$, when $\mathop{\lim}\limits_{n \rightarrow \infty} \alpha_n$ exists, it holds that  $\mathop{\lim \inf}\limits_{n \rightarrow \infty}\alpha_n  = \mathop{\lim}\limits_{n \rightarrow \infty} \alpha_n $ \cite[Example 3.18]{Rudin:76}. 
			%
			Lastly, recall that in the proof of Lemma \ref{lem:AsyncZk2} it was established that the channel \eqref{eqn:AsnycModel2} is information stable. For such channels, we have from \cite{Dobrushin:63} that $\Capacity_n =  \mathop{\lim}\limits_{k \rightarrow \infty}\mathop{\sup}\limits_{ \cdf{\Xin^{(k)}}} \frac{1}{k}I\left(\Xin^{(k)} ; \Yi_{n}^{(k)}\right) $, and the limit exists. Substituting this in \eqref{eqn:CapBound3} results in 
			\ifFullVersion
			\begin{equation*}
			\Capacity_\myEps \le \mathop{\lim \inf}\limits_{n \rightarrow \infty} \Capacity_n,
			\end{equation*}
			\else 
			$\Capacity_\myEps \le \mathop{\lim \inf}\limits_{n \rightarrow \infty} \Capacity_n$,
			\fi
			thus proving the lemma. 
		\end{IEEEproof}

		Combining Lemma \ref{lem:CapLowBound} and \ref{lem:CapUpBound}  proves that $\Capacity_\myEps = \mathop{\lim \inf}\limits_{n \rightarrow \infty} \Capacity_n$, and by Lemma \ref{lem:CapLowBound}, this rate is achievable with Gaussian inputs, thus proving the theorem.
		% the theorem\footnote{If I also manage to prove Lemma \ref{lem:AsyncZk2} then I can conclude that Gaussian inputs are optimal.}.
		\qed

	\end{appendices}
	%----------------------------------------------------------------------------------------
	%	BIBLIOGRAPHY
	%----------------------------------------------------------------------------------------

\end{document}